\pdfoutput=1
\documentclass[multphys,vecphys]{svmult}

\usepackage{makeidx}         % allows index generation
\usepackage{graphicx}        % standard LaTeX graphics tool
\usepackage{multicol}        % used for the two-column index
\usepackage{cite}            % adjusts the "syntax" of the refs in the
\usepackage[bottom]{footmisc}% places footnotes at page bottom
\makeindex             % used for the subject index
\font\tenmsam=msam8
\def\EDT{{(EDT-TTF-CONMe$_2$)$_2$AsF$_6$}}
\def\Bech{(TMTSF)$_2$X}
\def\SePF{(TMTSF)$_2$PF$_6$}

\def\SeClO{(TMTSF)$_2$ClO$_4$}

\def\SeAsF{(TMTSF)$_2$AsF$_6$}
\def\Fabre{(TMTTF)$_2$X}
\def\SPF{(TMTTF)$_2$PF$_6$}
\def\SBF{(TMTTF)$_2$BF$_4$}
\def\SBr{(TMTTF)$_2$Br}
\def\SAsF{(TMTTF)$_2$AsF$_6$}
\def\TM{(TM)$_2$X}

\def\Trho{$T_\rho$}
\def\TrhoD{T_\rho}
\def\TSP{$T_{\rm SP}$}
\def\TSPD{T_{\rm SP}}
\def\TSPO{$T^0_{\rm SP}$}
\def\TSPOD{T^0_{\rm SP}}
\def\dm{{1\over 2}}
\def\gaeq{{\hbox{\tenmsam\symbol{"26}}}}

\def\tmpf{$\mathrm{(TMTSF)_{2}PF_{6}}$\,}
\def\tmas{$\mathrm{(TMTSF)_{2}AsF_{6}}$}
\def\tmc{$\mathrm{(TMTSF)_{2}ClO_{4}}$}
\,
\def\tms{$\mathrm{(TMTTF)_{2}PF_{6}}$\,}
\def\tmbr{$\mathrm{(TMTTF)_{2}Br}$\,}

\def\tmttfbf{$\mathrm{(TMTTF)_{2}BF_{4}}$\,}

\def\tmx{$\mathrm{(TM)_{2}X}$\,}
\def\tq{$\mathrm{(TTF-TCNQ)}$\,}
\,
\,
\def\R{$\mathrm{ReO_{4}^{-}}$}

\def\tsm{$\mathrm{TMTSF}$}
\def\tst{$\mathrm{TMTTF}$}
\def\tmsx{$\mathrm{(TMTSF)_{2}X}$}

\def\sb{$\mathrm{SbF_{6}}$}
\def\pf{$\mathrm{PF_{6}}$}
\def\re{$\mathrm{ReO_{4}}$}

\def\cl{$\mathrm{ClO_{4}}$}
\def\fb{$\mathrm{BF_{4}^{}}$}
\def\scn{$\mathrm{SCN^{}}$}
\def\TTdm{(TTDM-TTF)$_2$Au(mnt)$_2$}
\def\edt{$\mathrm{(EDT-TTF-CONMe_{2})_{2}AsF_{6}}$}
\def\tfx{$\mathrm{(TMTTF)_{2}X}$}
\def\tsx{$\mathrm{(TMTSF)_{2}X}$\,}
\def\tmdm{$\mathrm{TMTSF-DMTCNQ}$}
\def\tms{$\mathrm{(TMTTF)_{2}PF_{6}}$}

\def\tmcrx{$\mathrm{(TMTSF)_{2}ClO_{4(1-x)}ReO_{4x}}$}\,
\,
\def\tm{$\mathrm{(TMTSF)_{2}X}$}\,
\,
\begin{document}

\title{Interacting electrons in quasi-one-dimensional organic superconductors}
\author{ C. Bourbonnais \and D. J\'erome}
\institute{R\'eseau Qu\'ebecois sur les Mat\'eriaux de Pointe (RQMP), D\'epartement de physique, Universit\'e de Sherbrooke, Sherbrooke, Qu\'ebec, Canada J1K-2R1 \texttt{cbourbon@physique.usherb.ca}
\and Laboratoire de Physique des Solides, UMR 8502, Universit\'e de Paris-sud, 91405 Orsay, France 
\texttt{jerome@lps.u-psud.fr}}
% Use the package "url.sty" to avoid problems with special characters
% used in your e-mail or web address.
% Addresses should be removed from contribution and entered into
% blist.tex" (by the compiler).

\maketitle
%Your text goes here. Separate text sections with the standard \LaTeX\
%sectioning commands.
\section{Introduction}
\label{Intro}

 Superconductivity in organic materials has emerged in 1979 from an important background of preexisting knowledge and experimental techniques. All previous studies undertaken since 1973, which had been mostly performed on the  \tq series of  charge transfer organic conductors had failed to reveal superconductivity using chemistry and (or) pressure to suppress   the density-wave or the so-called Peierls instability inherent to one-dimensional conductors. A breakthrough, which contributed to the discovery of organic superconductivity, has been the synthesis of the molecule \tsm\ by K. Bechgaard and coworkers\cite{Bechgaard74}. 
 
Actually, in the early 70's   leading ideas governing the search for new materials likely to exhibit good
metallicity and possibly superconductivity were driven by the possibility to minimize the role of electron-electron repulsions and at the
same time to increase the electron-phonon interaction, while keeping the overlap between conducting stacks as large as possible. 
This led to the
synthesis of the new electron donating molecule \tsm,  presenting much analogy with the previously known fulvalene donors in which the redox potential $(\Delta E)_{1/2}$ can be minimized 
\cite{Garito74,Engler77}, by utilizing selenium instead of sulfur as hetero-atoms\cite{Andersen78}. The next step was quite encouraging since the use of a high pressure has allowed to remove the instability due to the divergence of the Peierls channel down to the lowest temperatures in the two-chain conductor \tmdm \cite{Andrieux79a}. A lucky situation has also been the  synthesis of a series of 1D organic salts based on  the  radical cationic  molecule TMTTF (the sulfur analog of the TMTSF molecule) and on a variety of inorganic
monoanions such as \cl , \fb\ or \scn \cite{Brun77,Coulon82}. A conducting character could thus be anticipated from the intermolecular overlap of partly filled highest molecular orbitals (HOMO) of  individual molecules. 
%%%%%%%%%%%%%%%%%%Figure Structure des stacks
\begin{figure}[htbp]			
\centerline{\includegraphics[width=0.8\hsize]{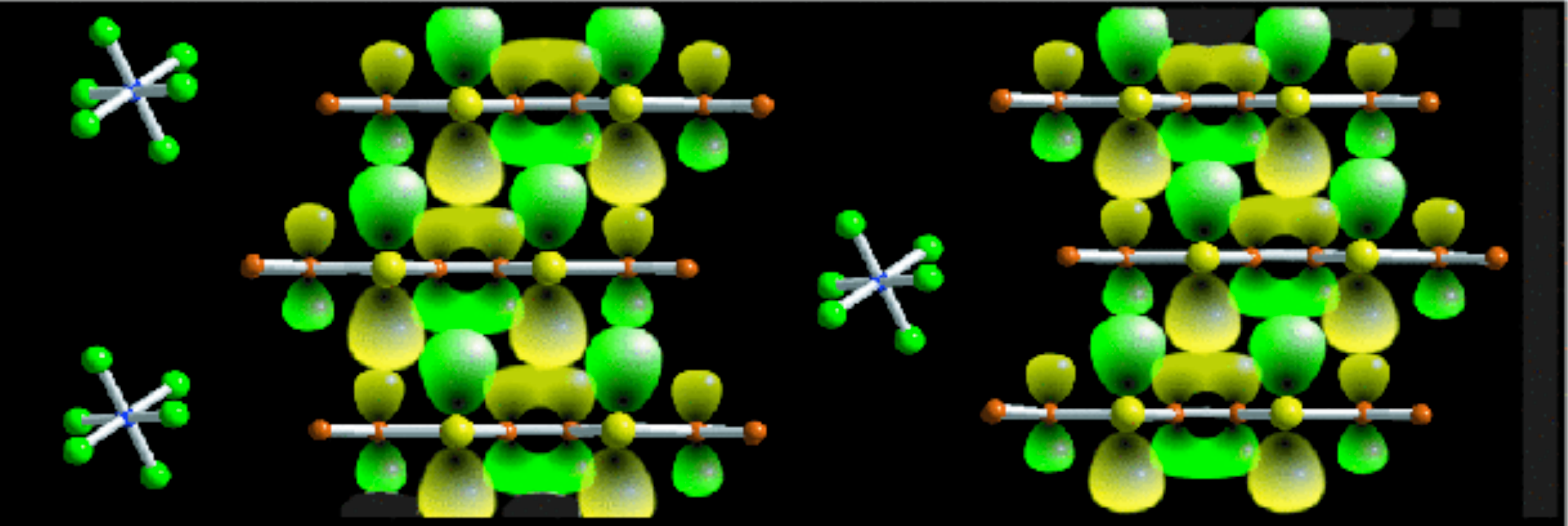}}
\caption{\TM, view of the cationic and anionic stacking perpendicular to the stacking axis, \textit{courtesy} of J. Ch. Ricquier, IMN, Nantes}
\label{stacks.eps}
\end{figure}
%%%%%%%%%%%%%%%%%%%%
The compounds, \tfx, were all insulating at ambient pressure but their crystal structure  is the prototype of the \tsx\  series in which superconductivity has been subsequently discovered. In the rest of this article when we mention a $\mathrm{(TM)_{2}X}$ compound, this means that the organic molecule can be either \tsm\ or \tst. The structure exhibits a face to face packing of  flat molecular units along the $a$ direction and the formation of molecular layers in the $a-b$ planes separated by anions stacks (Fig. \ref{stacks.eps}). The overall symmetry is triclinic, not too far from orthorhombic often  taken as the approximate structure by theoreticians. In addition, the structure reveals an important peculiarity namely, the   anions are located in centrosymmetrical cavities lie slightly above or below each molecular plane with a zigzag stacking of the molecules along the $a$ direction. This
structure  leads in turn to a weak alternation of the interplanar distance (dimerization and a concomitant splitting of
the HOMO conduction band into a filled lower sub-band separated from a half-filled upper (hole-like) band by a
gap  $\Delta_D$  at $\pm 2k_F$, called the dimerization gap
in the new Brillouin zone.  However, on account of the finite transverse
dispersion, this dimerization gap does not lead to a genuine gap in the middle of the  density of states as given from
the extended-H\"uckel band calculation. The only claim that can be made is that
these conductors show commensurate band filling (three-quarter filled with electrons or one quarter-empty with holes). This originates from the 2:1 stoichiometry. Consequently, according to a non interacting particle band calculation, all compounds in the
\tfx\ series should be found conducting.
%%%%%%%%%Table
\begin{table}
\begin{center}\caption{Calculated band parameters for three representative members of the  $\mathrm{(TM)_{2}X}$ series according to the room
temperature crystallographic data in Ref. \cite{Ducasse86}. The average intra and interstack interactions
are given in lines 3 and 5, respectively. The bond dimerization is shown in line 4. All energies are in
meV}
\begin{tabular}{llll}
\hline  & $\mathrm{(TMTTF)_{2}PF_{6}}$ & $\mathrm{(TMTSF)_{2}PF_{6}}$ & $\mathrm{(TMTSF)_{2}ClO_{4}}$ \\  
\hline
 $t_1$& \hfil 137\hfil &\hfil 252\hfil &\hfil 258\hfil \\
$t_2$&\hfil 93\hfil&\hfil209\hfil &\hfil 221\hfil\\  
 $\overline{t}$&\hfil 115\hfil &\hfil 230\hfil &\hfil 239\hfil\\
 $\frac{\Delta t}{\overline{t}}$&\hfil 0.38\hfil &\hfil 0.187\hfil &\hfil 0.155\hfil \\  
 $\overline{t}_{\perp b} $&\hfil 13\hfil &\hfil 58\hfil &\hfil 44\hfil \\  
\hline
\end{tabular}
\end{center}
\label{table1} 
\end{table}

\begin{figure}[htbp]			
\centerline{\includegraphics[width=0.6\hsize]{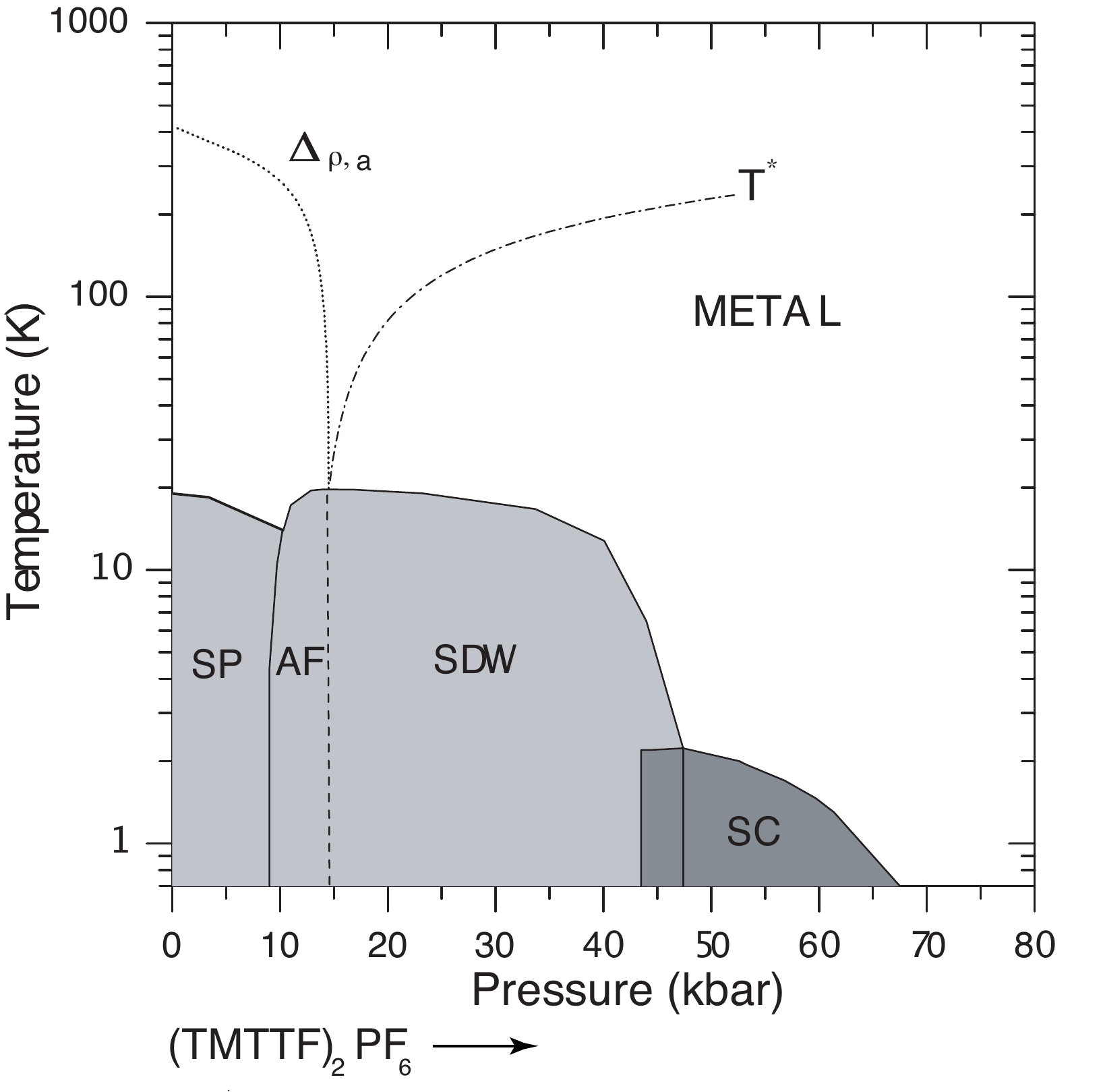}}
\caption{Phase diagram of \SPF, as determined from transport measurements. After Ref.  \cite{Auban04,Jaccard01}}
\label{PF6generic.eps}
\end{figure}

In Table (\ref{table1}) we report the band parameters of
different members of the \tsx series as computed from crystallographic data \cite{Ducasse86}.
The sulfur compounds exhibit bands that are significantly narrower and their crystallographic structure is more dimerized than those of the selenide
compounds. $\mathrm{(TMTTF)_{2}Br}$ (not listed in Table (\ref{table1})) is, however, an exception among the sulfur
compounds with a dimerization of 0.13, which is smaller than the value calculated for
$\mathrm{(TMTSF)_{2}ClO_{4}}$. This might be due to the calculation of electronic bands based on
rather old and less accurate  crystallographic data  than those used for the other compounds \cite{Behnia95}. All \tmx compounds with diversified anions can be gathered on a generic phase diagram displaying a wealth of different physical properties\cite{Jerome91}.
The gross features of  the $\mathrm{(TM)_{2}X}$ phase diagram are shown in Figures \ref{PF6generic.eps} and \ref{fig:1}. Compounds on the left hand side of the phase diagram, such as $\mathrm{(TMTTF)_{2}PF_{6}}$, are insulators below room temperature with a
broad metal to insulator transition, while those on the right hand side of 
$\mathrm{(TMTTF)_{2}Br}$ exhibit an extended temperature  regime with a metallic behavior and a sharp transition towards
an insulating ground state. Therefore, the cause of the insulating nature of some 
members of the
\tfx\  series should be determined in relation to the role of e-e repulsion and low dimensionality as we shall show
later.
Although the most extensive  pressure studies have been performed on $\mathrm{(TMTSF)_{2}PF_{6}}$  and $\mathrm{(TMTTF)_{2}PF_{6}}$,  recent studies 
of other compounds of the \tfx \ series with X= \re, \fb\  and $\mathrm{Br}$ have shown that the main features observed under pressure in
$\mathrm{(TMTTF)_{2}PF_{6}}$  or in $\mathrm{(TMTSF)_{2}PF_{6}}$   are also observed in other systems \cite{Auban04}.
At this stage we may emphasize that  the band filling of these materials is commensurate and in addition the existence of a  dimerization in the crystal structure of the \tmx\  series raises quite a challenging  problem for  the physics of one dimensional conductors since with the axial dimerization the conduction band becomes half-filled while it is originally quarter-filled from stoichiometry considerations.
The commensurate band filling opens a new scattering channel  for the carriers between both
sides of the Fermi surface as then the total momentum transfer for two (four) electrons from one side of the 1-D  Fermi surface
to the other is equal to  a  reciprocal lattice vector (Umklapp scattering for half (quarter)-filled bands). 
%%%%%%%%%%%%%%%%%%Figure diagram PF6
%%%%%%%%%%%%%%%%%%%%
\section{Elements of theory for interacting electrons in low  dimension}
\label{theory}
In this section we shall depict some of  the main  results of the theory of  quasi-one-dimensional metals.  Given the pronounced one-dimensional anisotropy of the compounds,   it is natural to first consider  the 1D limit.  To this end,  the study of susceptibilities of non interacting electrons is particularly revealing  of the natural instabilities that can take place in one dimension. Take for example the bare {\it Peierls} susceptibility of the system, which is the response to the formation of $2k_F$ electron-hole pairs, the constituents of charge-density-wave (CDW) and spin-density-wave (SDW) correlations (here $k_F$ being the Fermi wave vector).  In one dimension, the energies  of an electron (hole)  state at $k$ and a hole (electron ) state  at
$k-2k_F$ are  connected  through  the  {\it nesting} relation $\epsilon(k)=-\epsilon(k-2k_F)$
of the electron spectrum $\epsilon(k)$ close to the Fermi level. The summation over  a
macroscopic number of  intermediate electron-hole states linked by  this relation   leads to an infrared logarithmic singularity of the form $\chi^0_P(2k_F,T)\sim (\pi v_F)^{-1}\ln(E_F/T)$.
Similarly, the {\it Cooper} susceptibility $\chi^0_C$, which probes the formation  of pairs (two holes or two electrons) of particles of total momentum zero that are connected through the inversion property of the spectrum $\epsilon(k)=\epsilon(-k)$, also gives rise to a  logarithmic divergence  $\chi^0_C(T) \sim   (\pi v_F)^{-1}\ln(E_F/T)$ -- a singularity that is actually found in any dimension. 

What thus really makes one dimension so peculiar resides in the fact
that both singularities refer to the same set of electronic states and will then interfere  one another\cite{Bychkov66}. In the presence of interactions, the interference is found to  all order of  perturbation theory  for the   scattering  amplitudes of electrons with opposite  Fermi velocities and it modifies the nature of the electron system in a essential way.
 In the framework of the 1D electron gas model, the selected emphasis put by these infrared singularities on  electronic states  close to the Fermi level allows  us to define  interactions with respect to the Fermi points $\pm k_F$\cite{Dzyaloshinskii72,Solyom79}. Thus for a rotationally invariant system of length $L$, the Hamiltonian of the electron gas model can be written in the form
\begin{eqnarray}
   H &=& \sum_{k,p,\sigma} \epsilon_p(k) a^\dagger_{p,k,\sigma} a_{p,k,\sigma}\cr
 &+& {1\over L}\sum_{\lbrace k,q,\sigma \rbrace}g_1\
a^{\dagger}_{+,k_1+2k_F+q,\sigma}a^{\dagger}_{-,k_2-2k_F-q,\sigma'}
a_{+,k_2,\sigma'}a_{-,k_1,\sigma} \cr
&+& {1\over L}\sum_{\lbrace k,q,\sigma \rbrace}g_2\
a^{\dagger}_{+,k_1+q,\sigma}a^{\dagger}_{-,k_2-q,\sigma'}
a_{-,k_2,\sigma'}a_{+,k_1,\sigma} \cr
& + &{1\over2 L}\sum_{\lbrace p,k, q,\sigma \rbrace}g_3\  a^{\dagger}_{p,k_1+p2k_F +q,\sigma}a^{\dagger}_{p,k_2-p2k_F -q+pG,\sigma'}
a_{-p,k_2,\sigma'}a_{-p,k_1,\sigma}\cr 
&+& {1\over L}\sum_{\lbrace p,k, q,\sigma \rbrace}g_4\  a^{\dagger}_{p,k_1+q,\sigma}a^{\dagger}_{p,k_2-q,-\sigma}
a_{p,k_2,-\sigma}a_{-p,k_1,\sigma}
\label{hamiltonian}
\end{eqnarray}
where $\epsilon_p(k) = v_F(pk-k_F)$ is the electron spectrum energy for right ($p=+$) and left ($p=-$) going electrons, $g_1$ and $g_2$
are the back and forward scattering amplitudes, respectively, whereas  $g_3$   corresponds to the Umklapp
scattering, a process  made
possible at half-filling, where the reciprocal lattice vector
$G=4k_F=2\pi/a $ enters in the momentum conservation law; finally, one
has the coupling $g_4$ by which two electrons near $k_F$ (resp.
$-k_F$) experience a small momentum transfer which keeps them on the
same branch \cite{Solyom79}.

In the one-loop perturbation theory, the electron scattering amplitudes are corrected by the aforementioned Cooper and Peierls logarithmic singularities. These logarithms are scale invariant quantities as a function of energy or temperature,  which allow us to write down scaling or flow equations for the various scattering amplitudes. After  all cancellations due to Cooper-Peierls interference being made, we have \cite{Dzyaloshinskii72,Solyom79}, 
\begin{eqnarray}
&& {\tilde{g}'_1}   = 
-\tilde{g}^{2}_1 \,+ \ldots\cr
 &&(2\tilde{g}'_2 -\tilde{g}'_1)   =    \ \tilde{g}^{2}_3 \,+ \ldots\cr
&&{\tilde{g}'_3}  =     \ \tilde{g}_3
(2\tilde{g}_2 -\tilde{g}_1)\, + \ldots,
\label{flow1D}
\end{eqnarray}
where $\tilde{g}_i'= (d/d\ell) \tilde{g}_i$ and $\ell=\ln E_F/T$ is the logarithmic -- loop --   variable. The long wavelength spin excitations are governed by the $\tilde{g}_1\equiv g_1/\pi v_\sigma $ ($v_\sigma = v_F(1-g_4/2\pi v_F)$ coupling, which is decoupled  from  both $\tilde{g}_3\equiv g_3/\pi v_\rho$ ($v_\rho=v_F(1+g_4/2\pi v_F)$) and the combination $2\tilde{g}_2-\tilde{g}_1\equiv (2g_2-g_1)/\pi v_\rho$ connected to charge excitations. In  the physically relevant repulsive sector for systems like \TM\ where $g_{1,2}>0$, and owing to the existence of a small dimerization gap $\Delta_D\ll E_F$ of organic stacks (Table\,\ref{table1}),  weak half-filled Umklapp scattering $g_3 \approx g_1 \Delta_D/E_F$ is present\cite{Barisic81,Emery82}. Thus  for $g_1-2g_2 < \mid g_3\mid$, both $2g_2- g_1$ and $ g_3$ are relevant variables for the charge and scale to  the strong coupling sector, where a charge gap $\Delta_\rho $ is found below the temperature scale $T_\rho (\sim \Delta_\rho/2)$. This can be seen as a $4k_F$ charge localization responsible for  a  Mott insulating (MI) state. On the other hand, the solution $\tilde{g}_1(T)= \tilde{g}_1/(1+\tilde{g}_1\ln E_F/T)$ for the $g_1$ coupling, which follows from Eq. (\ref{flow1D}), is marginally irrelevant and scales to zero, leaving the  spins degrees of freedom gapless as shown by the calculation  of the uniform spin susceptibility\cite{Bourbon93,Dzyaloshinskii72},
\begin{equation}
\label{spinchi}
\chi_\sigma(T) = {2\mu_B^2(\pi v_\sigma)^{-1}\over 1- {1\over 2} \tilde{g}_1(T)}.
\end{equation}
 The spin susceptibility  decreases   monotonically as a function of temperature and is  unaffected by the occurrence of a charge gap. The electron system develops singularities, however,  for  staggered density-wave   response. Thus  the $2k_F$ SDW or antiferromagnetic response, which is governed by the  combination of couplings $\tilde{g}_2(\ell)+ \tilde{g}_3(\ell)$ that flows to strong coupling, develops a  power law singularity of the form 
\begin{equation}
\label{AF}
\chi_{\rm AF}(2k_F,T) \propto  (\pi v_F)^{-1} (T/\Delta_\rho)^{-\gamma},
\end{equation} 
where the power law exponent $\gamma =  \tilde{g}_2(T_\rho) +\tilde{g}_3(T_\rho)\sim 1$. The response for the $2k_F$ charge-density-wave `on bonds', called the bond-order-wave (BOW) response, which is governed by the combination of couplings $ \tilde{g}_2(\ell)+ \tilde{g}_3(\ell)- 2\tilde{g}_1(\ell)$, also develops a power law singularity in temperature
\begin{equation}
\label{BOW}
\chi_{\rm BOW}(2k_F,T) \propto  (\pi v_F)^{-1} (T/\Delta_\rho)^{-\gamma_{\rm BOW}}.
\end{equation} 
Here the exponent $\gamma_{\rm BOW}\sim 1$ is essentially the same as the one of AF response -- the amplitude of the latter being larger, however\cite{Kimura75}. When $2k_F$ phonons are included, their coupling to singular BOW correlations   yields a lattice instability of the spin-Peierls (SP) type.  It is worthwhile to note that all the above properties of the 1D electron gas model  find some echo in the phase diagram of \TM\  (Fig. \ref{fig:1}).

\subsection{Some results of the bosonization picture \cite{Giamarchi04}}
\label{bosons}
We now turn to the description of the one-dimensional electron gas using the bosonization method.  A major property of interacting electrons  in one dimension is that long wavelength
 charge or spin-density-wave oscillations constructed by the combination
of electron-hole pair excitations at low energy form  extremely stable
excitations \cite{Giamarchi04,Voit95}.  Quasi-particles excitations like those taking place in a Fermi liquid (FL) for example,  are absent at low energy and are replaced by  these collective  
acoustic excitations for both spin ($\sigma$) and charge ($\rho $) degrees and freedom, thus allowing the construction of a phase representation  of  the electron gas Hamiltonian.   The  Fermi field
$$
\psi_{p,\sigma}(x)  = ÊL^{-\dm} \sum_k a_{p,k,\sigma}\ e^{ikx}
$$
\begin{equation}
\sim \lim_{\alpha_0 \to 0} {e^{ipk_Fx}\over \sqrt{2\pi\alpha_0}} \exp\Bigl(-{i\over
\sqrt{2}}[p(\phi_\rho + \sigma \phi_\sigma) + (\theta_\rho + \sigma\theta_\sigma)]\Bigr),
\end{equation}
can be expressed in terms of the spin and charge
phase fields $\phi_{\nu=\rho,\sigma}$  \cite{Schulz98,Giamarchi04} ($\alpha_0 $ is a short distance cut-off). These satisfy the
commutation relations
\begin{equation}
[\Pi_{\nu^\prime}(x^\prime),\phi_\nu(x)]= -i\delta_{\nu\nu'} \delta(x-x'),
\end{equation}
where   $\Pi_{\nu}(x)$ is   the momentum conjugate to
$\phi_\nu(x)$ and is defined by $\theta_\nu(x)= \pi\int \Pi_\nu(x') dx'$.
In this   phase variable
representation  the  full  electron gas Hamiltonian  takes the form
   \begin{eqnarray}
   \label{PhaseH}
H & = & \sum_{\nu =\rho,\sigma}\dm \int  \left[ \pi u_\nu K_\nu \Pi_\nu^2 + u_\nu (\pi K_\nu)^{-1}
\left({\partial
\phi_\nu \over \partial x}\right)^2 \right] dx  \cr
&+ & {2g_1\over (2\pi\alpha_0)^2} \int  \cos(\sqrt{8}\phi_\sigma) \ dx
+ \  {2g_3 \over (2\pi\alpha_0)^2}\int  \cos(\sqrt{8}\phi_\rho) \ dx.
\end{eqnarray}
The harmonic part of the phase Hamiltonian on the first line corresponds to the Tomonaga-Luttinger model,  which is exactly solvable. The spectrum
shows no quasi-particles but only collective excitations and all the properties of the model
then become entirely governed by the velocity $u_\nu
$ and the `stiffness constant'
$K_\nu$ of  acoustic  excitations, which depend  on interactions. This corresponds to the physics of the so-called  Luttinger (LL) or Tomonaga-Luttinger liquid.   In the Tomonaga-Luttinger limit the power law singularity of the AF spin response $\chi_{\rm AF}(2k_F,T)\sim T^{-\gamma}$ is confirmed  and the exponent
\begin{equation}
\label{gammaAF}
\gamma= 1-K_\rho
\end{equation}
is expressed in terms of the charge stiffness constant $K_\rho$. The absence of quasi-particle  excitations  is   captured  by the power law decay of the density of states at the Fermi  level
\begin{equation}
\label{density}
N(0) \sim (\pi v_F)^{-1}\Big({T\over E_F}\Big)^\alpha,
\end{equation}
with the exponent 
\begin{equation}
\label{alpha}
\alpha= {1\over 4} (K_\rho + 1/K_\rho -2).
\end{equation}
The quasi-particle weigth at the Fermi level $z(T) \sim T^{\alpha}$  follows a similar power law decrease.

In the presence of the sine Gordon terms due to the backscattering and Umklapp couplings in the phase Hamiltonian Eq. (\ref{PhaseH}), an exact solution cannot be found in the general case.
However,  a perturbative scaling procedure can be used for the various parameters that define the Hamiltonian\cite{Giamarchi04}. For rotationally invariant repulsive couplings, $g_1$ is marginally irrelevant as found in the  many-body description (\S\ref{theory}),  and only the flow in the charge sector  essentially  matters. In low order, one can write the following flow equations
\begin{eqnarray}
{dK_\rho\over dl}= &&  - \dm K_\rho^2 \tilde{g}_3^2,\cr
{d\tilde{g}_3\over dl} = && \tilde{g}_3 (2-2K_\rho).
\label{Flowbose}
\end{eqnarray}
For repulsive couplings, the bare  $K_\rho=K_\rho(g_4,2g_2-g_1)< 1$, $g_3$ is relevant and scales to strong coupling,  as found in the previous fermion scaling description in Eqs. (\ref{flow1D}), while $K^*_\rho (l\gg 1)\to 0$. An expression for the charge gap can be found   
\begin{equation}
\Delta_\rho \sim W \tilde{g}_U^{1/ [2(1-n^2K_\rho)]},
\label{Gap}
\end{equation}
where for half-filling Umklapp $n=1$ and $\tilde{g}_U=\tilde{g}_3$\cite{Giamarchi04}.

A charge gap is not limited to half-filling but may be present for other commensurabilities
too \cite{Giamarchi97}. At quarter-filling for example, the transfer of four particles from one side of the Fermi
surface to the other leads  to the Umklapp coupling
\begin{equation}
H_{1/4} \simeq   {2g_{1/4}\over (2\pi\alpha_0)^2} \int dx \cos(2\sqrt{8}\phi_\rho).
\end{equation}
The phase argument of this  term differs and leads to a distinct flow equation  

\begin{equation}
{d\tilde{g}_{1/4}\over dl} =  (2-8K_\rho)\tilde{g}_{1/4},
\label{Uquart}
\end{equation}
which goes to strong coupling if $K_\rho< 1/4$, namely for sizable long-range Coulomb interaction\cite{Mila93}.  The value of the insulating gap is given by (\ref{Gap}) by taking  
$n=2$  and $\tilde{g}_U=\tilde{g}_{1/4}$ at quarter-filling ($\tilde{g}_{1/4}\sim (U/W)^3$ in the Hubbard limit)\cite{Schulz94,Mila93}. It worth noting that in the  special situation where the
quarter-filled chains are weakly dimerized, both half-filling and quarter-filling Umklapp couplings are  present in practice and should interfere one another \cite{Tsuchiizu01}. In effect for materials like \TM, stoichiometry imposes
half a carrier (hole) per TM molecule, a concentration that cannot be modified by applying
pressure. Consequently,  uniformly spaced molecules along the stacking axis should lead to a situation where a unit
cell contains 1/2 carriers,
\textit{i.e.}, the conduction band is quarter-filled.  However, the non-uniformity of molecular packing has been observed in early structural studies of the \tfx\  crystals \cite{Ducasse86}. The dimerization of the overlap between
molecules occurs along the stacks, a situation that is pronounced in the sulfur series, although it is also
encountered in some members of the  
%:
$\mathrm{(TMTSF)_{2}X}$ series (Table 1). The impact of such a dimerization on the electronic structure is
generally quantified by a modulation of the intra-stack overlap integral, because both longitudinal and transverse
molecular displacements can contribute to the  intermolecular overlap and could
 make them half-filled band compounds.

\subsection{The role of interchain coupling}
\label{interchain}

Electronic materials like  \TM\  can  be only considered
as close realizations of 1D   
interacting fermion systems so that   interchain coupling, though small,  must be
taken into account in their description. A non zero intermolecular overlap perpendicular to the chains yields finite interchain  hopping integrals $t_{\perp b}$ and $t_{\perp c}$ along the $b$ and $c$ directions, respectively. These play an essential role either in the restoration of Fermi liquid (FL) quasi-particles or in establishing long-range order.  Considering a square lattice of $N_\perp$ chains, the electron spectrum takes the form
\begin{equation}
E_p({\bf k})= \epsilon_p(k) -2t_{\perp b} \cos k_{\perp b}-2t_{\perp c}\cos k_{\perp c},
\label{3Dspect}
\end{equation}
where ${\bf k}=(k,k_{\perp b},k_{\perp c})$ and $t_{\perp c}\ll t_{\perp b}\ll E_F$. Owing to the strong anisotropy in the transfer integrals, a one-electron coherent   motion in the transverse direction is not present  at all temperatures. In the  non  interacting case for example, the temperature scale below which thermal fluctuations no longer blur the transverse quantum mechanical   coherence for the electron is simply $T_{x^1}\sim t_{\perp b}$, which can be seen as a one-particle dimensionality crossover. In the presence of interactions, however,   the quasi-particle weight $z(T)\sim (T/E_F)^\alpha$  being reduced by the LL behavior, the condition   becomes $T_{x^1}\sim z(T_{x^1})t_{\perp b}$ \cite{Bourbon84}, namely 
\begin{equation}
\label{cross1}
T_{x^1}\sim t_{\perp b}\Big({t_{\perp b}\over E_F}\Big)^{\alpha / 1-\alpha}.
\end{equation}
\begin{figure}[t]
\centering
\includegraphics*[width=.7\textwidth]{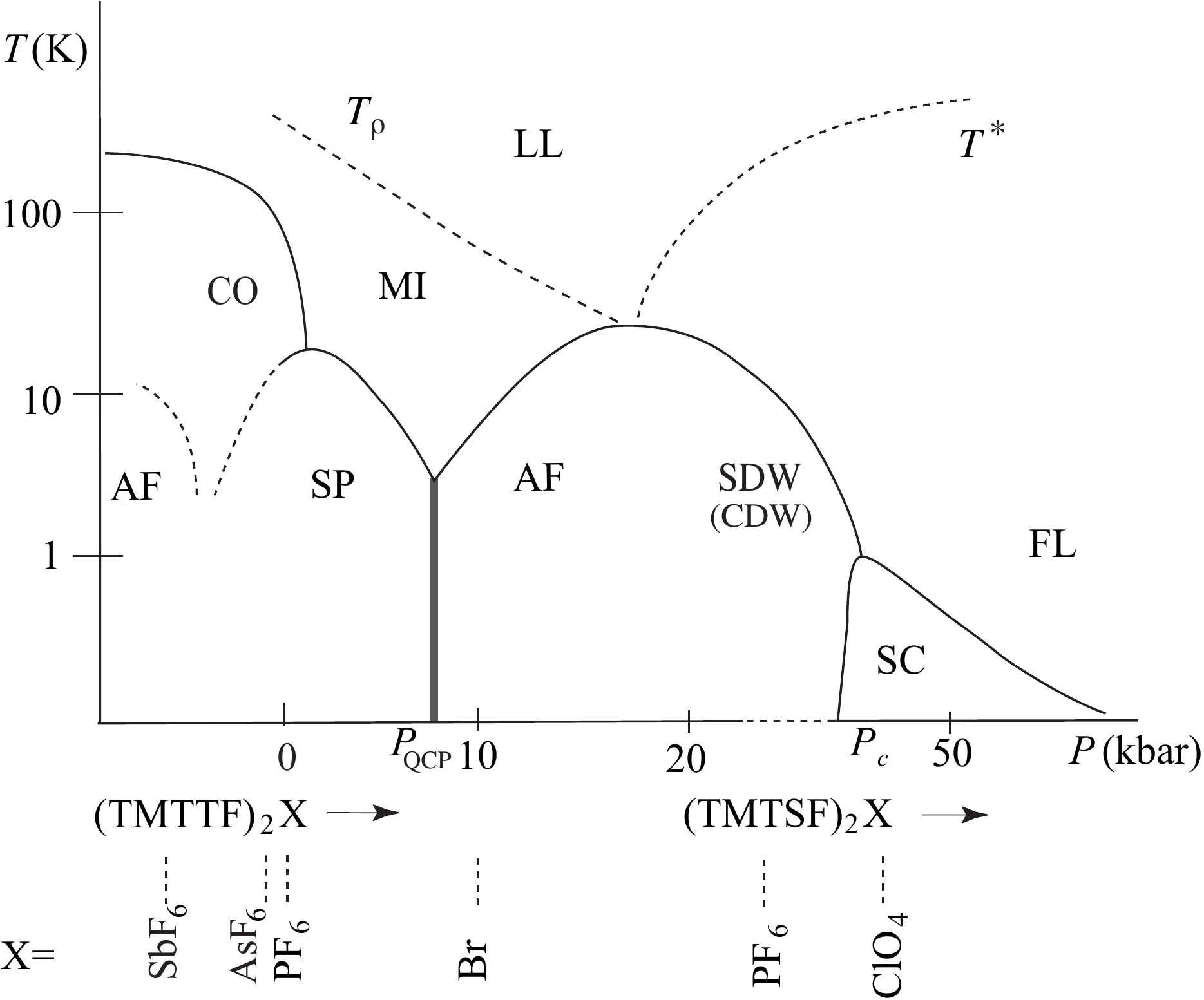}
\caption[]{The generic phase diagram of (TM)$_2$X.}
\label{fig:1}       % Give a unique label
\end{figure}
The one-particle crossover scale $T_{x^1}$ then decreases in the presence of interactions. When a Mott insulating phase takes place in the 1D domain, $\alpha$ reaches unity, $T_{x^1}$  vanishes, and the  transverse band motion is not possible  the
single-particle coherence becomes spatially confined along the
stacks. 

\begin{figure}[t]
\centering
\includegraphics*[width=.55\textwidth]{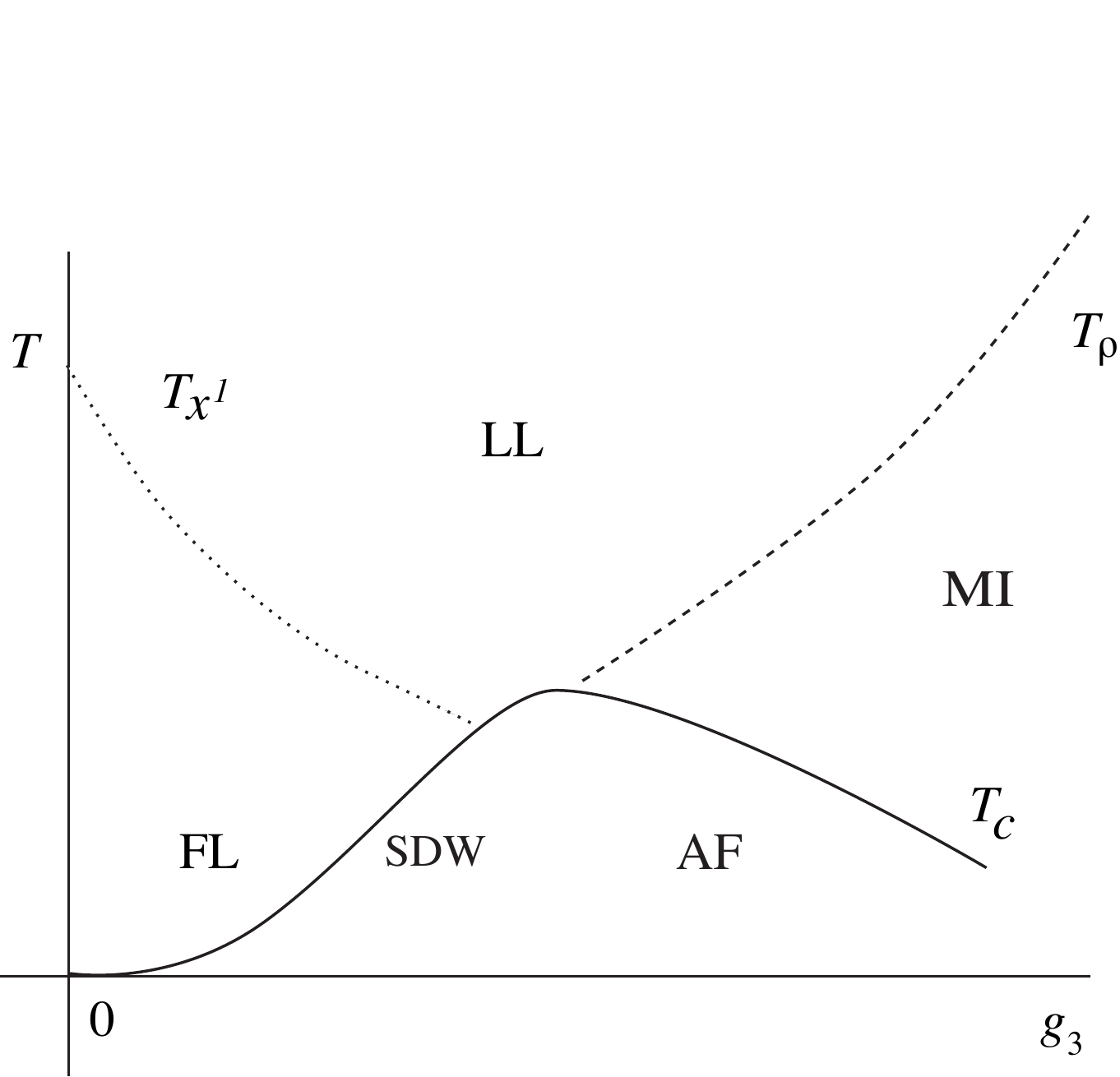}
\caption[]{Schematic phase diagram of a  quasi-one-dimensional electron system at repulsive coupling calculated from the scaling theory. Here $g_3\propto g_{1,2}$ is the running Umklapp parameter for the amplitude of interaction. }
\label{Tctheo}       % Give a unique label
\end{figure}

Transverse coherence is nevertheless possible but it is achieved through two-particle pair hopping, a mechanism  for long-range order that is not present in the Hamiltonian at the start but
which emerges
when  interactions along
the stacks  combine  with   
$t_{\perp, b,c}$ in the one-dimensional region\cite{Bourbon86,Brazovskii85b,Bourbon91,Bourbon88}. For repulsive interactions and commensurate fillings, the most important pair hopping processes that gradually emerges as a function of energy is an interchain antiferromagnetic   exchange, 
$$
\delta H_\perp = \sum_{i,j} \int dx \,J_{\perp i,j}\, {\bf S}_i(x)\cdot {\bf S}_j(x).
$$ 
  In the one-dimensional
regime,  it is  governed at the one-loop level  by the flow equation in the Fourier space
\begin{equation}
\tilde{J}'_{\perp}({\bf q}_0) = \tilde{f}(\ell)
+ \tilde{J}_{\perp}({\bf q}_0)\gamma(\ell)
- {1 \over 2} (\tilde{J}_{\perp}({\bf q}_0))^2,
\label{VRPA}
\end{equation}
where $\tilde{f}(\ell) \simeq -2\sum_{i=b,c}[(\tilde{g}_2(\ell) + \tilde{g}_3(\ell))t_{\perp,i}/W]^2
e^{(2-2\alpha(\ell))\ell}$ and ${\bf q}_0=(2k_F,\pi,\pi)$ is the modulation wave vector of AF order. Here  
$\alpha(\ell)$ and $\gamma(\ell)$ are    scale dependent power law exponents  of the quasi-particle weight (Eq. (\ref{alpha})) and  antiferromagnetic susceptibility (Eq.~(\ref{AF})), respectively.
Depending on the sign of 
$2-2\alpha(\ell)-\gamma(\ell)$, different cases can be considered. Thus when a Mott  gap is formed at $\ell_\rho=\ln E_F/T_\rho$, $2-2\alpha(\ell)-\gamma(\ell)$ becomes negative  above $\ell_\rho$ and the solution of (\ref{VRPA}) can be put in the following simple Stoner form at temperature $T$
\begin{equation}
\label{StonerJ}
\tilde{J}({\bf q}_0) \approx {\tilde{J}_{\perp b} + \tilde{J}_{\perp c}\over 1- ({J}_{\perp b} + {J}_{\perp c}) \chi_{\rm AF}(2k_F,T)},
\end{equation}
where the low temperature 1D AF susceptibility $\chi_{\rm AF}(2k_F,T)$ is given by Eq. (\ref{AF}), for $\gamma=1$ ($K_\rho^*=0$), and
\begin{equation}
\label{ExchangeA}
J_{\perp b,c} \sim \pi v_F{t_{\perp b,c}^{*2}\over \Delta_\rho^2}.
\end{equation}  
is the effective exchange in the $b$ and $c$ directions at the scale $T_\rho$. From (\ref{StonerJ}), a singularity will then occur at  
\begin{equation}
\label{TcNeel}
T_c \sim  {(t_{\perp b}^{*2} + t_{\perp c}^{*2})\over \Delta_\rho}, 
\end{equation} 
signaling a transition to a N\'eel ordered state. This result indicates that $T_c$ -- essentially dominated by the exchange in the $b$ direction --  increases as the Mott temperature $T_\rho$ decreases (Fig. \ref{Tctheo}). When the commensurability effects are decreasing  and $T_\rho$ eventually merges with the critical behavior of the transition,   antiferromagnetism becomes itinerant and corresponds to a SDW state. The interchain exchange enters in the weak coupling domain where $2-2\alpha(\ell)-\gamma(\ell)$ is small but positive. This  modifies the critical temperature, which reads
\begin{equation}
\label{Tcweak}
T_c \sim  \tilde{g}^{*2} t_{\perp b}^{*},
\end{equation}   
where $\tilde{g}^*=\tilde{g}_2^* +\tilde{g}_3^*$ and  $t_{\perp b}^*=zt_{\perp b}$ are respectively the  effective amplitude of electron-electron coupling and transverse hopping close to the transition. The calculations show  that $T_c$ starts to decrease in this domain giving rise to a maximum in $T_c$\cite{Bourbon91,Bourbon95,Bourbon02}. When the strength of 1D correlations becomes weaker one quickly arrives at the regime where $T_{x^1}$ becomes larger than  the above $T_c$ so that  the one-electron motion is no longer confined along the stacks and interchain coherence develops  before the onset of criticality linked to the transition.   

The temperature $T_{x^1}$ then becomes the  scale  below which the nesting of the whole Fermi surface becomes coherent. 
By looking at the effective
spectrum  Eq.(\ref{3Dspect}) in which, 
$t_{\perp b,c} 
\to t_{\perp b,c}^*=z(T_{x^1})t_{\perp b,c}$, perfect nesting  occurs at ${\bf q}_0=(2k_F,\pi,\pi)$ where the electron-hole symmetry $E_-^*({\bf k })=-E_+^*({\bf k}+{\bf q}_0)$ holds. 
It follows that electron-hole excitations within the   energy shell $\sim
T_{x^1}$ above and below the coherent --
warped -- Fermi surface lead to a logarithmically singular response 
$\chi^0({\bf q}_0,T)\sim (\pi v_F)^{-1}\ln T_{x^1}/T$ in the Peierls channel. This singularity is also found in the perturbation
theory of  the scattering amplitudes and 
for repulsive
interactions,  it yields an instability of the normal state towards SDW long-range order. When perfect nesting prevails, a 
not   too bad
approximation consists of neglecting the
interference between  the Cooper and Peierls channels (we shall
revert to the problem  of interference
below $T_{x^1}$ later in \S~\ref{Supra}).
This corresponds to the ladder diagrammatic summation. In the scaling theory language, the ladder approximation  corresponds to the  flow equation 
\begin{equation}
{d\tilde{J}\over d\ell} = \dm \tilde{J}^2 + ....,
\end{equation}
for an effective coupling constant $\tilde{J} = \tilde{g}_2 +\tilde{g}_3 -
\tilde{J}_\perp$
that defines the net attraction between an
electron and a hole separated by ${\bf q}_0$. The integration leads to the Stoner expression
\begin{equation}
\tilde{J}(T) ={\tilde{J}^*\over 1-
\dm {J}^*\chi^0({\bf q}_0,T) },
\label{ladder}
\end{equation}
where
$\tilde{J}^*$ is the effective SDW coupling
obtained from Eqs. (\ref{VRPA}) and (\ref{flow1D}) at $\ell_{x^1}= \ln E_F/T_{x^1}$, and $\chi^0({\bf q}_0,T)= (\pi v_F)^{-1} \ln T_{x^1}/T$.  The above expression leads to a BCS
singularity at the SDW critical temperature
\begin{equation}
T_c= T_{x^1} \,e^{-2/J^*},
\label{BCSTc}
\end{equation}
which decreases as the interactions decrease. Nesting
frustration is  required to suppress the transition\cite{Horovitz82,Yamaji82}.
When nesting deviations are
sufficiently strong,  however,     the FL remains unstable. Actually, when   the partial but finite interference between the Peierls and
   the Cooper channels is restored, the system turns out to develop a
  superconducting instability.  We shall
return to this  in \S\ \ref{Supra}.

%%%%%%%%%%%% Fabre series high temperature
\section{The Fabre salts series}
\subsection{The generic \tmx phase diagram }
\label{Fabre}
\begin{figure}[h]			
\centerline{\includegraphics[width=1\hsize]{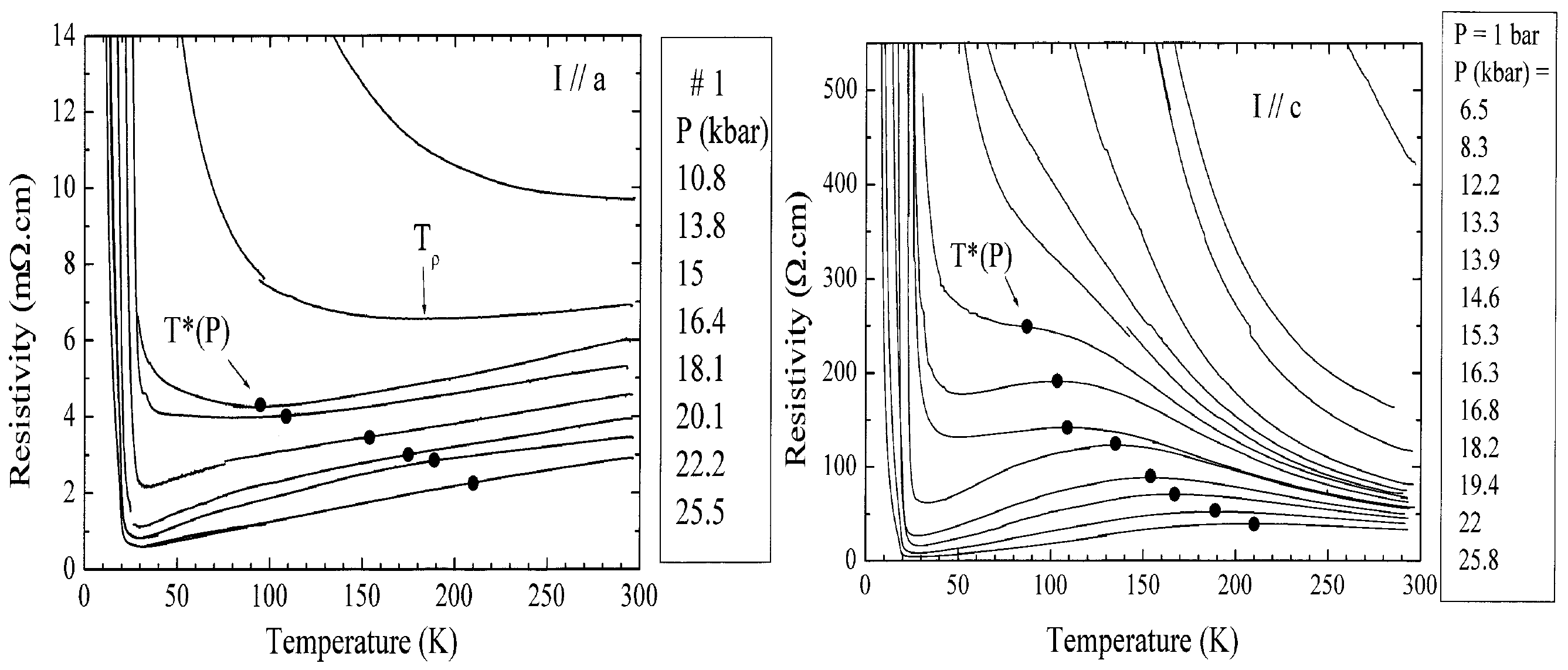}}
\caption{\tms, longitudinal (left) and transverse (right) resistances versus temperature at different pressures. After \cite{Auban04}}
\label{tmttf2pf6rhosousP.eps}
\end{figure}

Although the \tmx generic phase diagram can be established by measuring the transport properties of various compounds, it
has been most rewarding to use a single compound, namely \tms, and the help of a high pressure to span the generic diagram in Figures \ref{PF6generic.eps} and \ref{fig:1}. The study of this strongly
insulating system
\tms\  under high pressure has been very useful not only because it has led to the stabilization of superconductivity in
a strongly insulating sulfur compound \cite{Jaccard01}, but also because its location at the left end of the 
phase diagram has allowed several key properties of quasi 1-D conductors to be carefully monitored under pressure. For instance,
the longitudinal or 
 transverse transports activation and the one-dimensional deconfinement arising under pressure \cite{Auban04}, which will be discussed later in \S  \ref{Deconf}. 
The phase diagram in Fig. \ref{PF6generic.eps} where \tms \, is the reference compound at ambient pressure 
has  been obtained from the temperature dependences of
$\rho_c (T)$ and $\rho_a (T)$ to be discussed below. In the low pressure region (see Fig. \ref{tmttf2pf6rhosousP.eps}, $P< $10 kbar) $\rho_c (T)$ and $\rho_a (T)$ are both 
activated, with $\Delta_{\rho, c}$ being about 30\% larger than $\Delta_{\rho,a}$. At higher
pressures, the activation of $\rho_c $ follows a gentle decrease, while
$\Delta_{\rho,a}$ collapses abruptly at a pressure of about 14 kbar, which also marks the onset of  a non-monotonous temperature dependence  of  $\rho_c $ with a maximum arising at the temperature denoted $T^\star$.
Above 14 kbar, $\rho_a (T)$ displays a metallic
temperature dependence down to the sharp metal-insulator transition below 20 K, while $\rho_c (T)$ remains indicative of a weakly
insulating state above the temperature $T^\star$,  which increases under pressure. The different pressure dependences observed for longitudinal and transverse transport will be considered in the context of the pressure-induced deconfinement in \S \ref{Deconf}.

\subsection{Longitudinal transport}
A major problem encountered with the \tmx materials (as well as in most organic conductors) is the very strong
pressure (or volume) dependence of their electronic properties, particularly in   transport measurements \cite{Cooper79,Jerome94,Auban99}. This strong volume dependence is going along a particularly large 
 thermal expansion. Hence, the only temperature dependence that can be compared with the
prediction of the theory is the one measured at a constant volume. As all temperature dependences are obtained experimentally 
under constant pressure, a transformation to the constant volume $ T$-dependence must   be performed\cite{Jerome94,Auban99}, in order to make a significant comparison with the theory. Fortunately, this correction is not that relevant for the case of \tms\ under high pressure since the compressibility of these materials is known to decrease under pressure\cite{Gallois87}. This is no longer the case when the metallic phase is already stable at ambient pressure as we shall see later  for the case of \tmpf (\S\  \ref{Metallic}).
%%%%%%%%%%%Figure infra rouge des isolants

\begin{figure}[h]			
\centerline{\includegraphics[width=0.55\hsize]{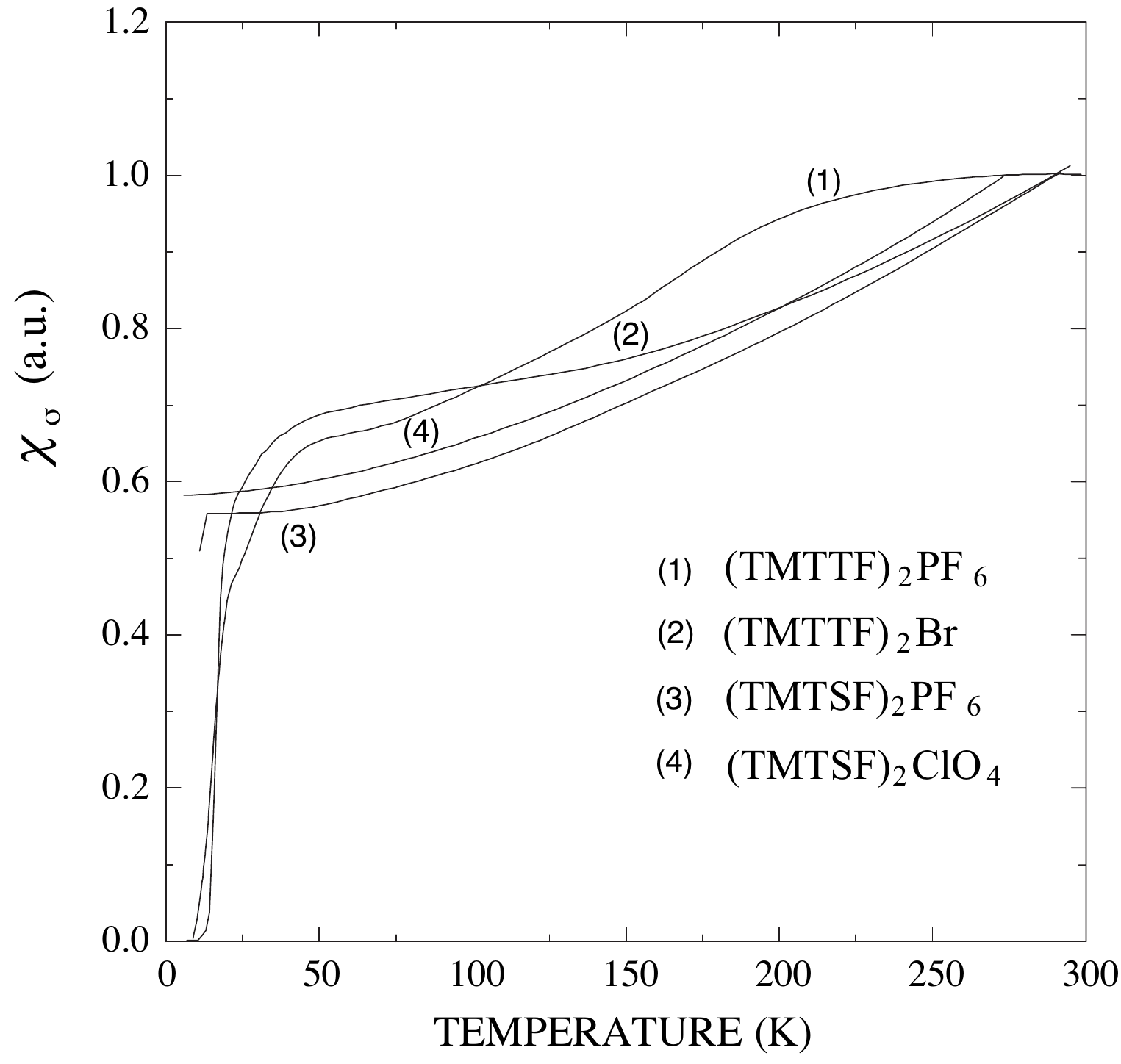}}
\caption{Temperature dependence of the uniform spin susceptibility $\chi_\sigma$ of  \TM\ compounds at ambient pressure. After References  \cite{Creuzet87b} [\SPF]; \cite{Wzietek93} [\SBr]; \cite{Miljak83} [\SePF]; \cite{Miljak85} [\SeClO]}
\label{chiT}
\end{figure}

%%%%%%%%%%%
The insulating character of these materials
with partly filled conduction bands is not expected in the framework of non interacting electrons. The reason for this must be the existence of strong repulsive interactions between 1D carriers. One explanation for 
the different values and pressure
dependences of the activation energies can be taken as an evidence for in-chain conduction made possible by thermally
excited 1-D objects similar to solitons in conducting polymers \cite{Brazovskii80}, whereas transverse
transport requires the excitation of real quasi-particles (QP's) through a  Mott-Hubbard gap larger than the soliton gap as we shall see. Contrasting with the strongly pressure dependence of the transport properties, the susceptibility is hardly sensitive to the location of a specific compound in the generic phase diagram. The spin susceptibility is dropping by about 40\% between room and low temperature for all materials, although the actual magnitude and the low temperature behaviour depend on the compound (see Fig. \ref{chiT}). According to the 1D result Eq. \ref{spinchi} in the high temperature domain, a value of $\tilde{g}_1\sim 0.5$ can reasonably account for the temperature dependence of the magnetic susceptibility.
%%%%%%%%%%%%%%%%%Figure resistance a et c PF6
%%%%%%%%%%%%%%%%%
%%%%%%%%%%%%%%%%%Figure resistance a et c PF6
%%%%%%%%%%%%%%%%%

Although the insulating behavior of \tfx\  salts can be clearly ascribed to their commensurate band filling, a closer
examination is needed to determine which of the Umklapp scattering channels 1/2- or 1/4-filled is the most active.
In the presence of such a gap, the transport is activated at low temperature $\rho_{a} \propto e^{\Delta_{\rho, a} / T}$
but is expected to vary according to the power
law
\begin{eqnarray}
\rho_{a}\propto T^{4{n}^{2}K_{\rho}-3}
\end{eqnarray}
  in the high temperature regime, \textit{i.e.}, 
$T>\Delta_{\rho, a}$ \cite{Giamarchi97}. The material resembles a metal at high temperature along the longitudinal
direction. In the high $T$  1-D regime ($T>t_{\perp b}$), the picture of non-coupled chains is approached.  Therefore, the density of
quasiparticle states should resemble the situation that prevails in a Luttinger liquid namely,
$N(\omega)\sim \mid \omega \mid^{\alpha}$, where $\alpha$ is related to a bare $K_{\rho}$ through Eq. (\ref{alpha}), forgetting about the influence of the Mott gap (supposedly smaller
than the temperature).

\subsection{ Transverse transport and deconfinement}
\label{Deconf}
\begin{figure}[htbp]			
\centerline{\includegraphics[width=0.6\hsize]{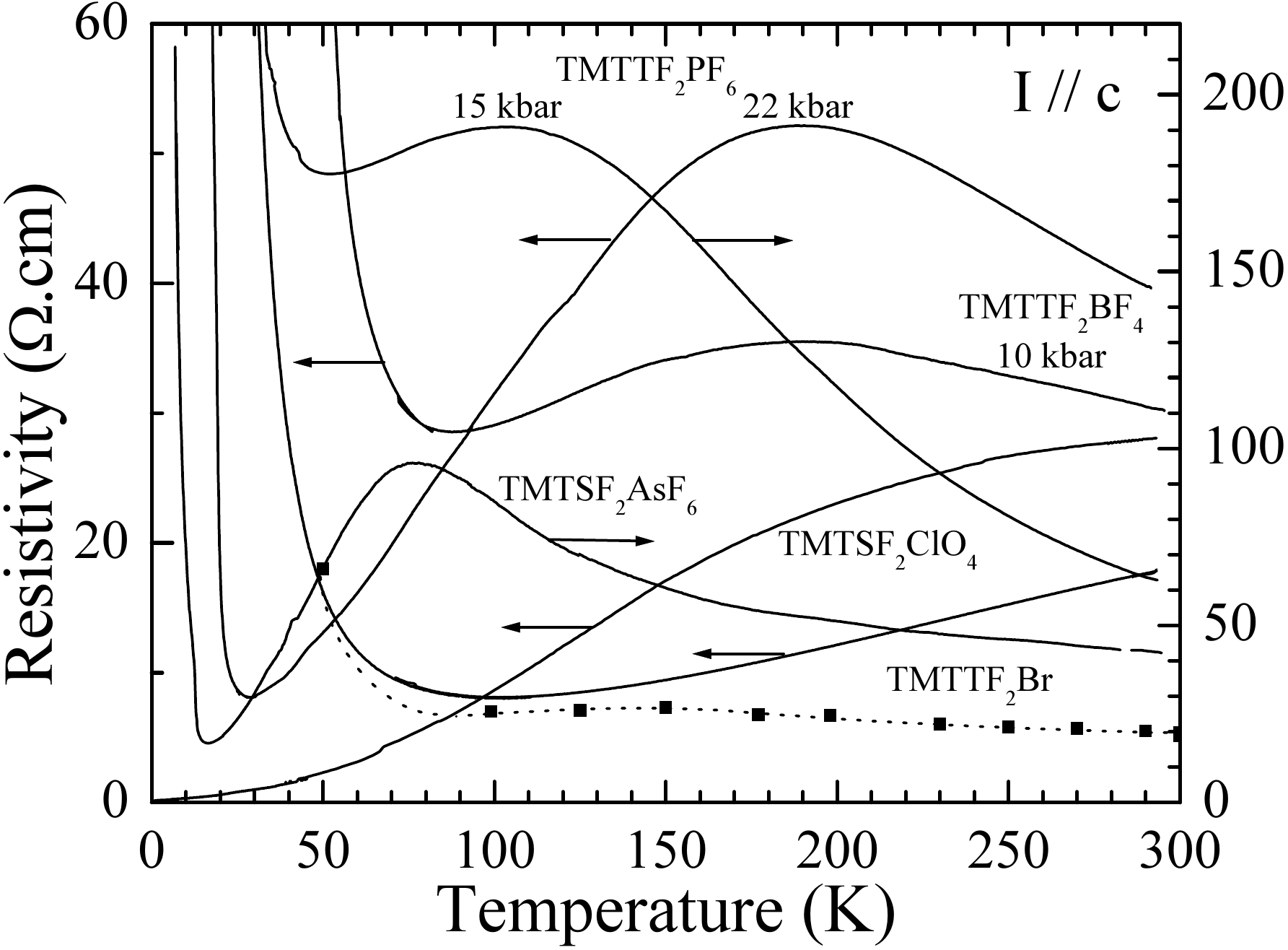}}
\caption{Temperature dependence of the transverse transport along the c-axis in several compounds belonging to the (TM)$_2$X series studied under pressure. The temperature of the maximum of resistivity shows the location of the dimensional crossover $T^{\star}$, which is strongly pressure dependent for each compound. The resistivity upturn at low temperatures represents the stabilization of the insulating SDW phase. In case of the selenium compound (TMTSF)$_2$ClO$_4$, the crossover lies around room temperature whereas for the sulfur compound (TMTTF)$_2$Br a resistivity maximum is seen only after the constant volume correction is taken into account. After Ref. \cite{Auban05} }
\label{rhocunderpressure.eps}
\end{figure}

Under a pressure higher than 14 kbar
the behavior of the \tms\ resistance along the direction of the weakest coupling, \textit{i.e.}, along the $c$-axis,
displays an insulating character with a maximum around 80-100 K and becomes metallic at lower temperatures,
although remaining several orders of magnitude above the  Mott-Ioffe critical value, which is considered as the limit
between metal and insulating-like transport \cite{Mott74}. Figure~\ref{rhocunderpressure.eps} displays the temperature dependence of $\rho_c$ in \tms, and also for other members of the \tmx\ family with different anions such as \tmttfbf
and \tmbr.
%%%%%%%%%%%%%%%%Figure Rc sous pression
%%%%%%%%%%%%%%%%%%%

The insulating character of the  transverse transport has been interpreted as the signature of a non
Fermi liquid  behavior for carriers within planes (chains) \cite{Moser98}. When transverse transport along the
$c$-direction is incoherent, transverse conductivity probes the physics of the $a-b$ planes and
conductivity in terms of the transverse coupling
$t_{\perp c}$ is expressed in the tunneling approximation as\\
\begin{eqnarray}
&&\sigma_{c}(\omega , T)\! \propto \ \!\!t_{\perp c}^{2 }\int \!\! dx \!\!
\int \!\! d\omega^{' }   A_{1D}(x,\omega^{'}) A_{1D}(x,\omega + \omega^{'})
\frac{f(\omega^{'})-f(\omega^{' }+\omega )}{\omega },\cr
&&
\label{sigmaperp}
\end{eqnarray} where $A_{1D}(x,\omega)$ is the one-electron spectral function of a single chain and $f(\omega)$ the Fermi-Dirac function. For
$a-b$ planes   made of an array of weakly interacting Luttinger chains, Eq. (\ref{sigmaperp}) leads to a power law
temperature dependence for the $c$-axis conduction.
The temperature at which the $c$-axis transport switches from an insulating
to a metallic temperature dependence corresponds to a crossover between two regimes: a high temperature regime
with no QP weight at the Fermi energy (possibly a TL liquid in the 1D case) and another regime in which the QP
weight increases with decreasing temperature. This interpretation does not  imply that the transport along
the $c$-direction must also become coherent below the cross-over. The $c$-axis transport may well remain
incoherent with a FL  being established in the $a-b$ plane at temperatures below
$T^\star$.   
The temperature dependence  of the resistivity along the least conducting direction is thus expressed as \cite{Georges00}: 
\begin{eqnarray}
\rho_{c}(T) \propto T^{1-2\alpha}.
\label{5}
\end{eqnarray}
Consequently, the temperature dependence of transport properties along the \textit{a} and \textit{c}-axes above
$T^\star$ should possibly lead to a consistent determination of
$K_{\rho}$. 

Now, regarding
\tms, we are facing a very interesting system, since the evolution from a Mott insulator to a metal can be
carefully studied under pressure in a single sample and a decrease in
compressibility under pressure makes constant volume correction less significant for  temperature dependences measured at high pressures. Turning to the evaluation of the correlation coefficient from the temperature
dependence of
$\rho_{c}$,  we
end up fitting the data for \tms\ in the pressure domain around 12 kbar, Fig.
\ref{tmttf2pf6rhosousP.eps},  with a very small value of $K_{\rho}$ (or large values of $\alpha$) which is
not  compatible with the  value 
$K_{\rho}=0.23$ derived from the far infrared (FIR, see below) and NMR data \cite{Wzietek93}. Consequently,  
the  Mott gap seems to be important in this temperature regime governing  the excitation for the motion of single
particles  along  $c$. Tentatively, one can expect a
transverse resistivity behaving according to \cite{Georges00},
\begin{eqnarray}
\rho_{c}(T)\propto  T^{1-2\alpha} e^{\Delta_{\rho,c}/T}.
\end{eqnarray}

\begin{figure}[htbp]			
\centerline{\includegraphics[width=0.65\hsize]{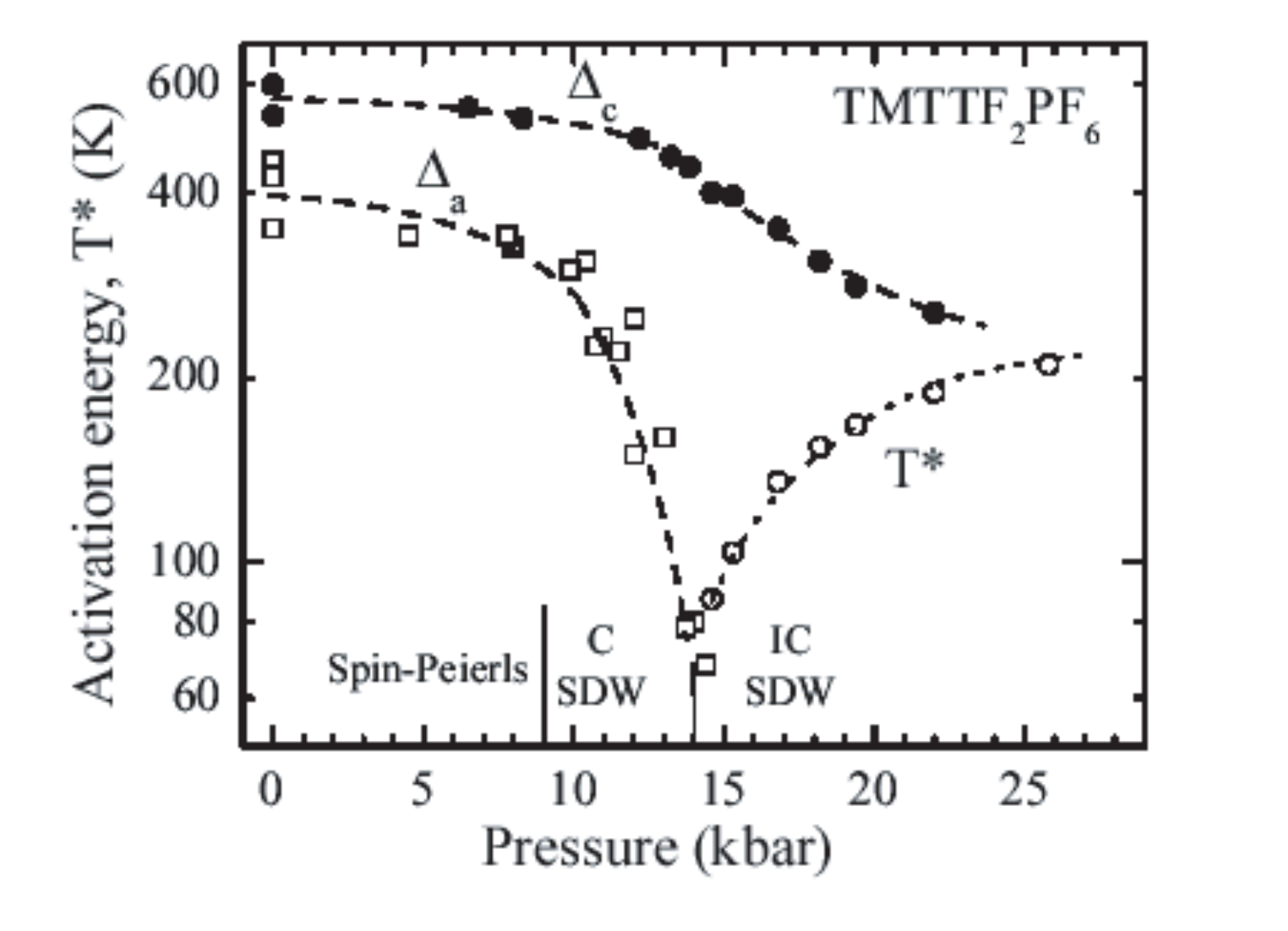}}
\caption{Pressure dependence of the transport activation in \tms . The activation for the
$c$ axis transport ($\Delta_c\equiv \Delta_{\rho,c}$, in the text) although decreasing under pressure survives up to high pressures, while the longitudinal transport ($\Delta_a\equiv \Delta_{\rho,a}$, in the text) is no
longer activated  above 
14 kbar when the dimensional cross-over arises at the finite temperature $T^\star$. After Ref.\cite{Auban04} }
\label{Deltarho.eps}
\end{figure}

Since the Mott -Hubbard gap varies as a power of $K_{\rho}$, even a small variation in the ratio between
the Coulomb interaction and the bandwidth under pressure can explain a significant decrease of all gaps
  moving from the left to the right in the generic phase  diagram. To summarize it is interesting to have a look at the data of longitudinal and transverse transports obtained in \tms\  under pressure  displayed on Fig. \ref{Deltarho.eps}. The transverse transport along $c$ is due to the hopping of quasi-particle and therefore requires an activation through a gap, which is the remanence of the Mott-Hubbard gap. It survives the onset of the dimensional cross-over. Thus we make the important identification 
  \begin{equation}
\label{Tstar}
T^\star \equiv T_{x^1}
\end{equation}
between the temperature at which the metallic behavior in $\rho_c$ is restored and the single-particle dimensionality crossover Eq. (\ref{cross1}). From Fig. \ref{Deltarho.eps}, we see that $T^\star$ goes to zero when the insulating behavior in the longitudinal transport  is restored, which is at 14 kbar for a system like \SPF, namely  at the onset of electron confinement where the renormalization of $t_{\perp b, c}^*$ is strong (Eq. (\ref{cross1})).  
 As for the longitudinal transport, it proceeds via the thermal excitation of 1-D objects similar to the solitons in conducting polymers through a gap smaller than the quasi-particle gap\cite{Brazovskii80}. They loose their one-dimensional character and thus acquire a metallic power law temperature dependence when the transverse coupling becomes pertinent under pressure (\textit{i.e.} above 14 kbar).
 %%%%%%Figure de delta c et delta a dans TMTTF2PF6
%%%%%%%%%%

\subsection{Far infrared response in the \tmx series}
\begin{figure}[htbp]			
\centerline{\includegraphics[width=0.5\hsize]{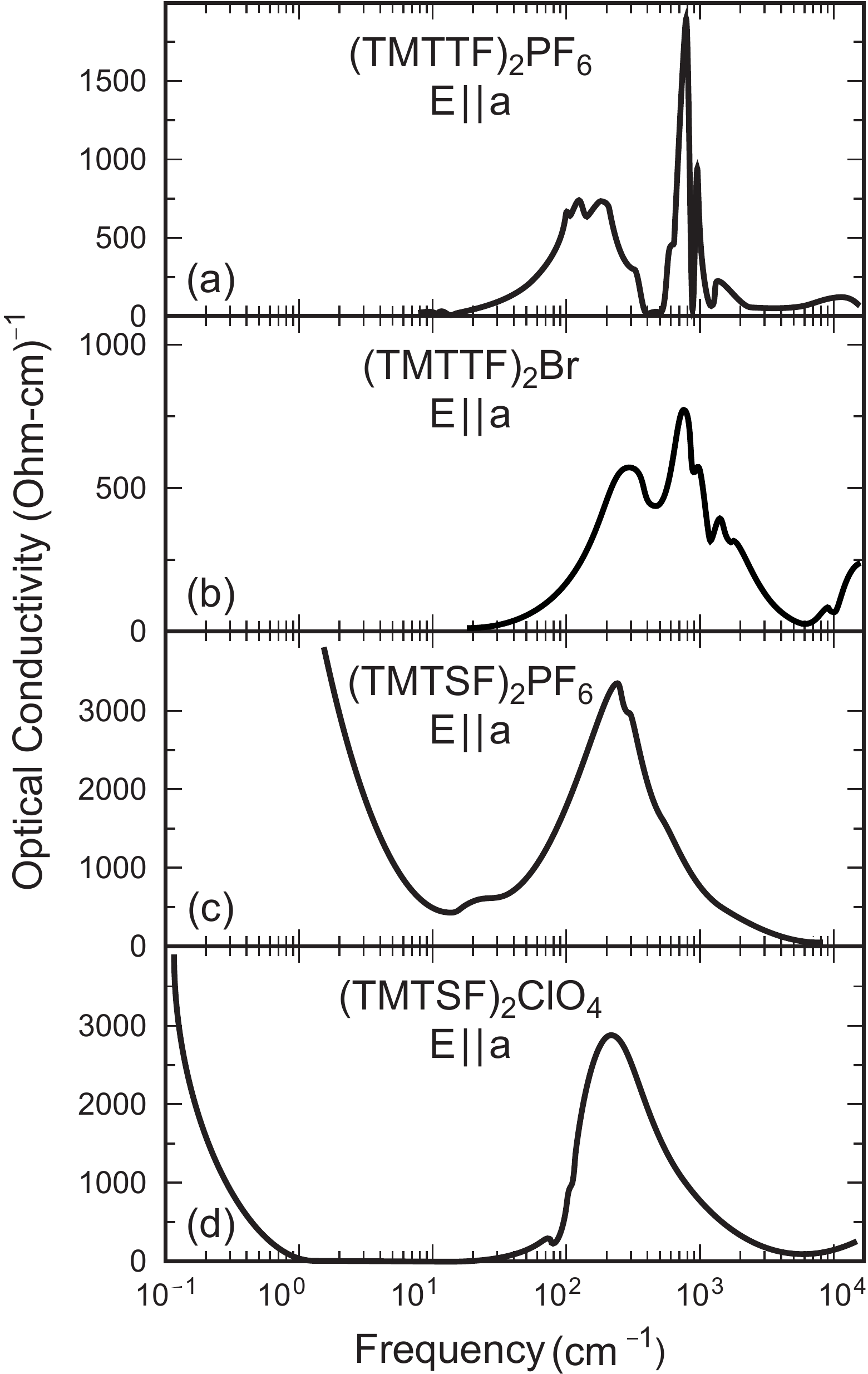}}
\caption{Frequency dependence of the optical conductivity of several \tmx compounds for the light polarized along the $a$ axis at ambient pressure and $T\approx 20$K. After \cite{Dressel03}}
\label{FIRTMTTFBrPF6.eps}
\end{figure}

  An other signature of the Mott-Hubbard gap has been given by the  frequency-dependent conductivity $\sigma(\omega)$ measured in various salts of the \tmx\ series exhibiting very different values of the conductivity at room temperature\cite{Jacobsen83}(see Fig.~\ref{FIRTMTTFBrPF6.eps}). The peak of the conductivity at a frequency $\omega_{0}$ correlates with the magnitude of the room temperature conductivity, namely both sulfur salts   \tms\ and \tmbr , which are insulating,  exhibit a conductivity peak around \hbox{$\omega_{0} \approx 1000$~cm$^{-1}$}. In \tmpf the peak occurs around 200~cm$^{-1}$  that is, very close to the zero frequency.
  %%%%%%%%%%%%%%Figure Infra rouge dans les isolants et les conducteurs Dressel
%%%%%%%%%%%%%%
  Both optical  and transport
data give
$2\Delta_{\rho,a}= 800-- 1000$ K  in 
$\mathrm{(TMTTF)_{2}PF_{6}}$ at ambient pressure \cite{Auban04}. The difference between
$K_{\rho}$  for selenium and sulfur compounds  ($K_{\rho}  = 0.18$ for the latter material) is a result of the
difference between their bare bandwidths, since on-site repulsion, being a molecular property, is likely to be less sensitive
to pressure than the intermolecular overlap along the stacking axis.

\subsubsection{Nuclear Magnetic Resonance}
\label{NMR}
  The  measurement of the temperature dependent nuclear spin-lattice
relaxation rate  in NMR denoted by $T_1^{-1}$  is another tool that has played a quite  important role in the description  of low energy electron  spin correlations in \TM \cite{Wzietek93}.  The connection between nuclear relaxation and the electron spin dynamics   is given by the Moriya  $T_1^{-1}$  expression
\begin{equation}
T_1^{-1} = \, \mid\! A\mid^2 T \int {\chi^{\prime\prime}(\vec{q},
\omega)\over \omega}\, d^Dq,
\end{equation}
which is taken in the zero Larmor frequency limit ($\omega\to 0$) and where $A$ is proportional to the
hyperfine matrix element. The relaxation of nuclear spins   
 gives  relevant information about the static, dynamics and dimensionality
$D$ of electronic spin correlations. In $D=1$, this expression gives  a relatively easy access to the interaction  parameter  $K_\rho$ that enters in most power laws expressions in one dimension 
\cite{Bourbon84,Bourbon89,Bourbon93}.
According to Eqns.~(\ref{spinchi}) and (\ref{AF}), the enhancement  of the imaginary part of the spin susceptibility 
$\chi^{\prime\prime}$ occurs  at
$q\sim 0 $ and    $q\sim 2k_F$, which yields
\begin{equation}
T_1^{-1} \simeq C_0T\chi_\sigma^2(T) + C_1T^{K_\rho}.
\label{Relaxation}
\end{equation}
Here  $C_0$ and $C_1$ contains weak logarithmic corrections in temperature. As a function of
temperature, two different behaviors can
  be singled out. At   high temperature, where uniform spin correlations
dominate and those at $2k_F$  are
small, the relaxation rate is   governed by the $T \chi_\sigma^2(T)$ term. In the low
temperature domain, however, $2k_F$ spin correlations
 are singularly enhanced, while  uniform correlations remain  finite  so that $T_1^{-1} \sim
T^{K_\rho}$.

\begin{figure}[t]
\centering
\includegraphics*[width=.6\textwidth]{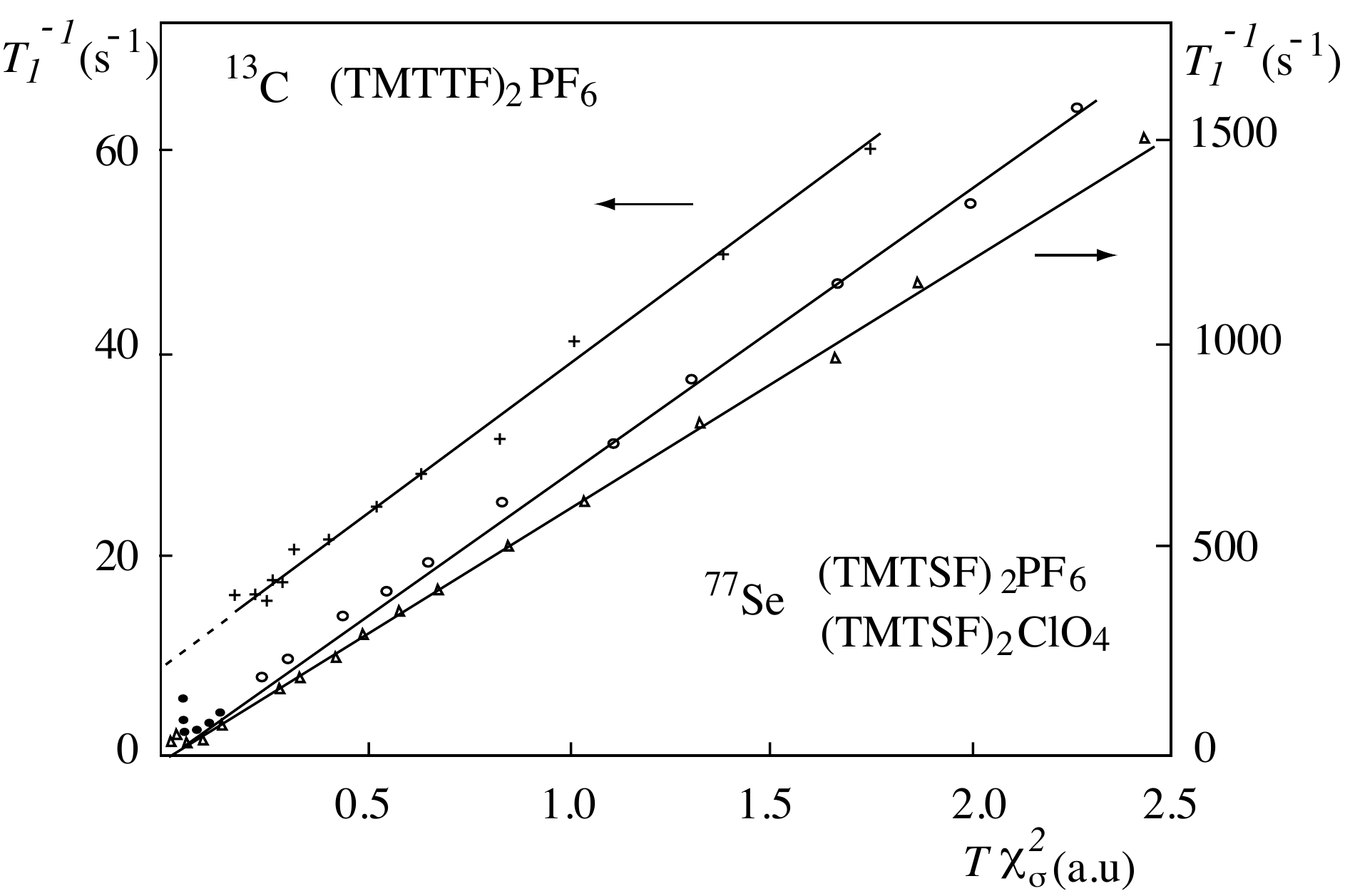}
\caption{Temperature dependence of NMR  nuclear relaxation rate  plotted as $T_1^{-1}$ {\it vs}   $T\chi_\sigma^2$ for $^{13}$C \SPF (crosses), $^{77}$Se \SePF (open circles) and  $^{77}$Se \SeClO (open triangles).  After Ref.\cite{Bourbon89} }
\label{T1SeS}       
\end{figure}

Consider for example the insulating compounds
   (TMTTF)$_2$X, we have seen in \S\ \ref{bosons} 
that the renormalized charge
stiffness \hbox{$K^*_\rho\to 0$}  essentially vanishes below the Mott scale.
 The resulting behavior for the relaxation rate becomes 
\begin{equation}
   T_1^{-1}\sim C_1
+ C_0T\chi_\sigma^2.
\end{equation}
As shown in Fig.~\ref{T1SeS}, this behavior   indeed emerges for
(TMTTF)$_2$PF$_6$ salt when the  relaxation rate
is combined to the spin susceptibility data ($T\chi_\sigma^2)$ in the MI
phase above three-dimensional ordering \cite{Bourbon89,Creuzet87}. A similar linear behavior of $ T_1^{-1}$ {\it vs}  $T\chi_\sigma^2$, with a finite intercept confirming the $K_\rho^*=0$ value of the charge stiffness, 
 is  invariably found in all  insulating 
materials down to the low
temperature  domain that surrounds the  three-dimensional
magnetic or lattice distorted long-range order \cite{Wzietek93,Jerome94,Gotschy92}.

As one moves to the right-hand-side in  the phase diagram of Fig. \ref{fig:1}, we see that the $ T_1^{-1}$ {\it vs}  $T\chi_\sigma^2$ law is relatively well obeyed over a large temperature domain in the normal phase.  If one takes   \SePF\ for example, deviations are seen only below $150$K or so  indicating that $K_\rho$ would be  finite over the whole temperature range\cite{Wzietek93}. For \SeClO,  $T_1^{-1}$ enhancement coming from antiferromagnetic correlations emerges  at even lower temperature ($\sim 30$K)\cite{Bourbon84}. 

%%%%%%%%%%%%%%%%%%%%%%%%%%%%%%%%%%%
%%%%%%%%%%%%%%%%%%%%%%%%%%%%%%%

\subsection{The ordered states at low temperature}
\subsubsection{The spin-Peierls instability of the sulfur compounds} 
 The two compounds \SPF\ and  \SAsF\ of the Fabre series develop an instability of the  MI phase that involves both spins and lattice degrees of freedom.  X-ray diffuse scattering measurements of Pouget {\it et al.} \cite{Pouget82}, on both compounds have soon  revealed  the existence of   diffuse scattering lines at the wave vector 2$k_F$ showing the onset of  1D lattice softening below the temperature scale  $T_{SP}^0\approx$  60K for \SPF\ and 45K for \SAsF. These lines condense into satellite reflections at  the  transition temperature $T_{SP}\approx$ 19K  for \SPF\ salt and 15K, for \SAsF, where a static lattice distortion takes place at the wave vector ${\bf q}_0=(2k_F,\pi,\pi)$. Recent elastic neutron scattering experiments did confirmed the existence of such a static distortion at $\mathbf{q}_0$ for  \SPF\ below \TSP
\cite{Foury04}. Since both the softening and the transition occur in the Mott insulating state where only spin excitations are gapless, one thus deals with a  spin-Peierls (SP) transition  with a pronounced quasi-1D character. The expected non magnetic nature of both SP fluctuations and long-range order has been confirmed by the temperature dependence  of the spin susceptibility (Fig. \ref{chiT}) and nuclear spin-lattice relaxation rate \cite{Wzietek93,Dumm00};  these quantities are reduced in the fluctuation  regime and show thermal activation below \TSP.

 Following the example of the Peierls transition, the  SP instability proceeds from the coupling  of singular 1D bond-order-wave (BOW) electronic correlations  to acoustic phonons at $2k_F$\cite{Bourbon96,Caron87,Riera00}. Unlike the Peierls case, however, the coupling becomes singular in the MI state instead of the metallic phase. The microscopic theory predicts a power law singularity in the  1D electronic BOW response below the Mott scale (Eq. \ref{BOW}), where for  weakly dimerized chains systems like \Fabre, the power law exponent $\gamma_{\rm BOW}=1 \,(K^*_\rho=0)$, as also found for the antiferromagnetic spin response Eq. (\ref{gammaAF}). The enhancement of the electron-phonon interaction at $2k_F$ by BOW correlations can be worked out by perturbation theory\cite{Bourbon96,Caron87}. In the random phase approximation, the static temperature electron-electron vertex part induced by the exchange of $2k_F$ phonons takes the form
\begin{equation}
\label{RPASP}
\Gamma_{\rm ph}(2k_F,T) = {g^0_{\rm ph} \over 1-{g}^0_{\rm ph}\chi_{\rm BOW}(2k_F,T)}, 
\end{equation}
where $g^0_{\rm ph}$ is the square of the bare $2k_F$ electron-phonon matrix element. A singularity of the temperature vertex part will then develop   at the mean field  temperature 
\begin{equation}
\label{TSPMF}
T_{\rm SP}^0= c\, \tilde{g}^0_{\rm ph} \TrhoD,
\end{equation}      
 where $ c\, \gaeq \,1$. Since there is no phase transition in one dimension, $T_{\rm SP}^0$ is not a true transition temperature but a temperature scale of lattice fluctuations that can be identified with the softening temperature seen in X-ray experiments. Although  higher order fluctuations corrections will bring back  the transition at $T=0$, $T_{SP}^0$ remains the right  temperature scale for the onset of short-range fluctuations \cite{Bourbon96}.

 As for long-range SP order, it is driven by an interchain interaction which we denote $V_\perp $, whose contributions combine Coulomb, interchain hopping and three-dimensional phonons \cite{Bourbon95,Caron87}. These coupling favor staggered bond order transversally to the chains at the wave vector $\mathbf{q}_0$.  A molecular field treatment of interchain coupling, which takes into account one-dimensional fluctuations rigorously, leads to  the  mean-field expression for the spin-Peierls susceptibility
  \begin{equation}
\label{MFSP}
\chi_{\rm SP}(\mathbf{q}_0,T) = {\chi_{\rm 1D, SP}(2k_F,T)\over 1- V_\perp \chi_{\rm 1D, SP}(2k_F,T)}
\end{equation} 
where $\chi_{\rm 1D, SP}$ is the 1D spin-Peierls fluctuations susceptibility. The singularity in $\chi_{\rm SP}$ 
occurs at 
\begin{eqnarray}
 \TSPD& = & \TSPOD f(V_\perp/\TSPOD),
\end{eqnarray} 
where from the singular behavior of $\chi_{\rm 1D, SP}$ at low temperature \cite{Bourbon96}, it is found that $f(V_\perp/\TSPOD)\sim 1/3$ for $V_{\perp}/T_{\rm SP}^0\ll 1$\cite{Caron87}. The resulting estimation $\TSPD \sim T_{\rm SP}^0/3$ apparently holds   for most   electronically driven quasi-1D structural transitions in the adiabatic limit\cite{Pouget87}. The observed values of  the  ratio $T_{\rm SP}/T_{\rm SP}^0$ in \SPF\ and \SAsF\  are  compatible with this estimation. 
 
%%%%%%%%%%%%%%%%%%%%%%%%%%%%%%%%%%%
\begin{figure}[t]
\centering
\includegraphics*[width=.5\textwidth]{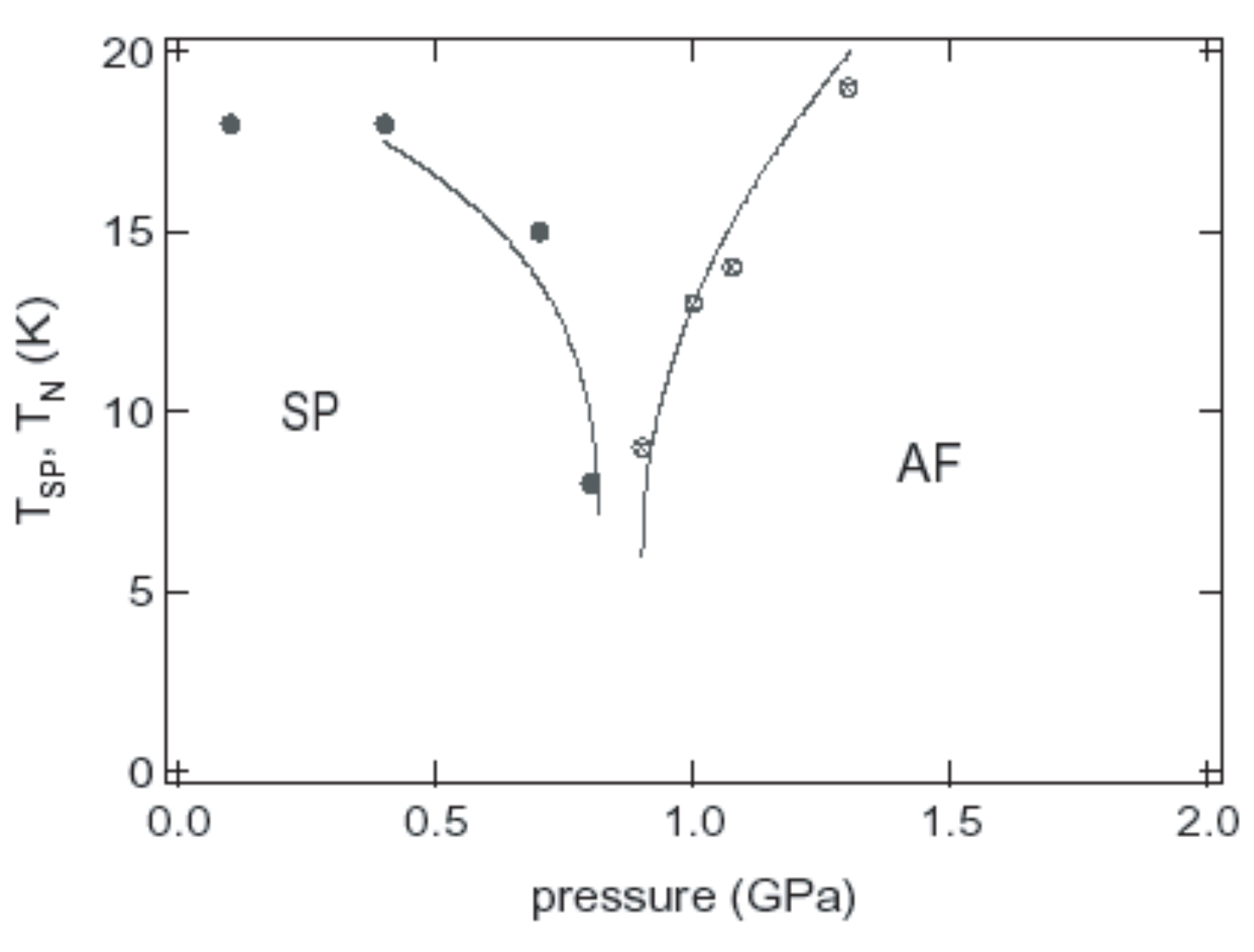}
\caption[]{The phase diagram of \SPF\ near the quantum critical point at $P_{\rm QCP}$. After Ref.\cite{Chow98} }
\label{QCP}       
\end{figure}
%%%%%%%%%%%%%%%%%%%%%%%%%%%%%%%   

  The respective $T_{\rm SP}$ of \SAsF\  and \SPF\  evolve  differently under low pressure. For \SAsF, \TSP\  first increases under pressure and reaches a maximum when the charge ordering temperature $T_{\rm CO}$ (see below) merges with \TSP\  at 1.5 kbar, and finally decrease at higher pressure\cite{Zamborszky02}. Charge order  breaks the inversion symmetry in the unit cell and acts as a $4k_F$ charge-density-wave potential on molecular sites.  The potential reduces the strength of BOW correlations \cite{Ilakovac94}, and  in turn the amplitude of both \TSPO\ and \TSP. For \SPF, however, $T_{\rm CO}$ occurs at much lower temperature \cite{Chow00}, and \TSP\ shows essentially a constant decrease under pressure.   It finally goes down rapidly,  extrapolating to zero at $P_{\rm QCP}\approx$ 9kbar, where it competes with a N\'eel state\cite{Chow98,Caron88,Creuzet87}. This critical pressure corresponding to the zero temperature extrapolation of the critical lines can be put in the category of a quantum critical point (Fig. \ref{QCP})\cite{Chow98}.

 The decrease of \TSPO\ and  \TSP\ under pressure can be qualitatively understood if one considers that the drop in \Trho\   weakens electronic BOW correlations, which according to  Eq. (\ref{TSPMF}), reduces \TSPO\, (Figs. \ref{QCP} and \ref{PF6generic.eps}).  As \TSPO\ carries on decreasing with pressure, it will reach   values that become small compared to the typical energy $\omega_D$ of $2k_F$ phonons (\hbox{$\omega_D \sim $} 100K in these materials \cite{Pouget76}), where quantum effects enter into play \cite{Caron84,Caron96,Bakrim05}.  In effect, the above expression for \TSPO\  has been obtained for  `static' phonons, in the so-called adiabatic approximation where the molecules are supposed to have an infinite mass. Adiabaticity is a reasonable assumption provided that $\omega_D/2\pi T_{\rm SP}^0\ll 1$\cite{Caron84}. As \TSPO\ decreases and $\omega_D$ increases under pressure, however, the adiabatic condition will not be satisfied any more and  the lattice softening will be reduced  by the zero point motion of the molecular lattice. It has been shown that a  quantum-classical crossover is expected at $\omega_D/\pi T_{\rm SP}^0(P) \approx 1 $, where quantum corrections completely suppress \TSPO\  and in turn $T_{\rm SP}$\cite{Caron84,Caron96}. For a system like \SPF, this would take place after a reduction of \Trho\  by a factor of two or so, which according to Fig. \ref{fig:1}, corresponds  to a pressure of the order of $P_{\rm QCP}$.

\paragraph{Charge ordering \cite{NoteCO}}

The MI state of most members of the Fabre series is characterized by  another temperature scale connected to a different type of long-range order. The study of the  temperature dependence of  electrical permittivity for \SPF\ and \SAsF\  has indeed revealed the existence of a singularity in the dielectric constant at 70K  and 100K, respectively \cite{Nad00a,Nad00b}. This singularity,  not seen in the magnetic susceptibility (Fig. \ref{chiT}), is associated with an instability in the charge sector.  The  nature of this state was clarified at  the same time by Chow {\it et al.} \cite{Chow00},   who showed from NMR that the  instability is actually a continuous phase transition towards  a charge disproportionation  in the unit cell. NMR does not tell, however, at which wave vector this charge ordered (CO)  state takes place. In this regard, Monceau {\it et al.}\cite{Monceau01}, suggested that the anion lattice  may undergo a uniform displacement when coupled to the $4k_F$ electron charge instability along the stacks. In this picture, the   CO instability in  \Fabre\  would be  akin in most cases  to a ferroelectric  phase transition with a divergent dielectric constant.  

The experimental  identification of a charge-ordered state in \Fabre\  salts lifted a sizable part of the veil surrounding the   nature of the so-called `structureless' phase transition that was detected much earlier from transport measurements  for several members of the \Fabre \ series \cite{Coulon85,Laversanne84}. It also puts an additional scale in the generic phase diagram of Fig. \ref{fig:1}. Moreover, for a compound like (TMTTF)$_2$SbF$_6$ with a larger  centro symmetrical anion, it was shown from NMR under pressure that the existence of both a CO transition at $T_{\rm CO} \simeq 150$K and a  N\'eel state at $T_c \simeq 7$K    forces to extend  the pressure scale of Fig. \ref{fig:1}, further on  the left\cite{Yu04}, where a   N\'eel rather than a spin-Peierls state is stable. This N\'eel state is suppressed under pressure and   replaced by a non magnetic phase, presumably of the spin-Peierls type such as  found in \SPF\ and \SAsF \cite{Yu04}, and which has been discussed above. At ambient pressure, (TMTTF)$_2$SbF$_6$ would then be found at the {\it left} of \SPF, which defines the origin on the  pressure scale of Fig. \ref{fig:1}. Other compounds like the 7K antiferromagnet (TMTTF)$_2$SCN would also be located in the same region of this extended phase diagram.

 It is not clear, however, how far  on the right-hand-side of    Fig \ref{fig:1}   CO ordering is found. As mentioned above, it is known to be rapidly suppressed under pressure for compounds like \SAsF\  and \SPF. However, it has been claimed to be present in compound like \SBr\, at ambient pressure\cite{Nad00a}.

The possibility for a CO order state, as a $4k_F$ charge instability, is predicted to take place in purely quarter-filled one-dimensional system for small charge stiffness  $K_\rho < 1/4$  (Eq. (\ref{Uquart})),  namely for sizable long-range Coulomb interaction where it coincides with an insulating state\cite{Mila93,Yoshioka01}. It has been found also for models of interacting electrons in weakly dimerized chains with and without anions displacements\cite{Tsuchiizu01,Riera01},   in the framework of mean-field theory \cite{Seo97} and numerical calculations\cite{Clay03}.

\subsubsection{The N\'eel order}

 Sufficiently above $P_{\rm QCP}$ in Fig. \ref{fig:1},  antiferromagnetic correlations within the MI state are much less affected by lattice SP fluctuations, which  are sizably weaker   in this pressure range. This is shown in the phase diagram by the absence of a spin pseudo gap from the temperature dependent nuclear relaxation rate    in \SPF\ at 13 kbar and in \SBr\  at 1 bar  \cite{Wzietek93}.  This is also confirmed in the case of \SBr\ by X-ray diffuse scattering experiments at ambient pressure\cite{Pouget97}. The observation in these conditions of a temperature independent  nuclear relaxation rate for both materials,  indicates that the power law exponent  of the singularity in the AF response in Eq.(\ref{AF})  below the  Mott  scale $T_\rho$  is  $\gamma=1  (K^*_\rho=0)$\cite{Wzietek93}. We have seen in \S\, \ref{interchain} that in  the presence of a Mott gap, $\Delta_\rho > t_{\perp b}$ (here $\Delta_\rho\equiv \Delta_{\rho,a}$), electron-hole bound pairs are formed and a coherent  electron band motion in the transverse directions cannot take place. The propagation of order in the transverse directions leading to a N\'eel ordered state  is provided by the antiferromagnetic interchain exchange $J_{\perp b,c}$ given by Eq. (\ref{ExchangeA}). 
  We have seen that the temperature scale for the  N\'eel ordering is determined by the singularity of the exchange coupling  at the one-loop level (Eq. (\ref{VRPA})), which leads to $T_c \propto 1/\Delta_\rho$ (Eq. (\ref{TcNeel})).

This result indicates that $T_c$ -- essentially dominated by the exchange in the $b$ direction --  increases as the Mott gap $\Delta_\rho$ decreases, a feature commonly observed in the Fabre series (Figs. \ref{fig:1} and \ref{PF6generic.eps}) \cite{
Klemme95,Brown97}. It is worth noting that in the quarter-filled Mott insulator compound   \EDT,  the interchain exchange is also the driving force of antiferromagnetic long-range order.  In these systems too, $T_c$ is  found to increase as $\Delta_\rho$ is decreasing, a behavior that proved to be independent of the commensurability of Umklapp scattering processes   behind the insulating gap (Fig. \ref{Tctheo}). 

\begin{figure}[t]
\centering
\includegraphics*[width=.5\textwidth]{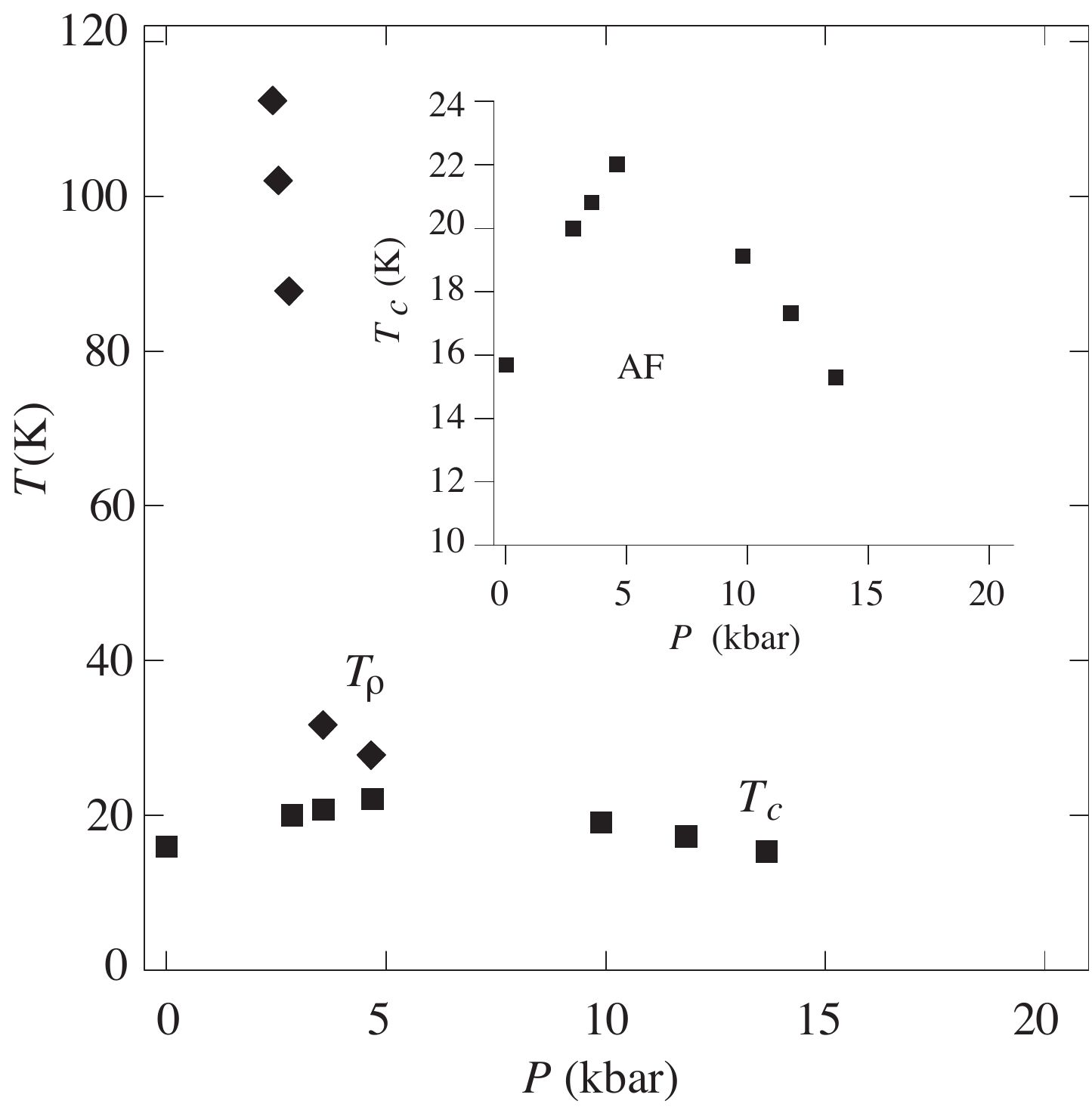}
\caption[]{Variation of the critical AF critical temperature as a function of hydrostatic pressure in \SBr. In the inset, a zoom of the maximum of $T_c$. After Ref.\cite{Klemme95}}
\label{TNBR}       
\end{figure}

When the pressure is further increased, the Mott insulating and N\'eel critical scales meet and then the spins   order themselves directly from the metallic state.  Antiferromagnetism becomes itinerant in character and corresponds to a SDW state. The interchain exchange enters in the weak coupling sector and continues to be  active, albeit on a relatively small pressure range with $T_c$ given by Eq (\ref{Tcweak}).  The calculations show  that $T_c$ starts to decrease in this restricted pressure domain giving rise to a maximum in $T_c$ seen in experiments for \SBr \ (Fig. \ref{TNBR}) \cite{Klemme95,Brown97}, \SPF\ \cite{Jaccard01} (Fig. \ref{PF6generic.eps}) and mixed selenium-sulfur  compounds  (TMDMTSF)$_2$PF$_6$ \cite{Auban89,Auban91}. This weak coupling domain coupling quickly evolves to a regime where $t_{\perp b}^*$ and then the single electron transverse coherence length along the $b$ direction  is increasing rapidly  under pressure, signaling the beginning of a coherent band motion perpendicular to the chains. This yields the onset of electronic deconfinement and  coherent nesting of the  Fermi surface at $T^\star$ \cite{Bourbon98}. The value of $T_c$ in this latter domain also decreases when the couplings decrease under pressure (Fig. \ref{Tctheo}).
%%%%%%%%%%%%%%%%%%%%%%%%%%%%%%%%%%%
%%%%%%%%%%%%%%%%%%%%%%%%%%%%%%%

\section{The Bechgaard salts}

\subsection{The metallic phase}
\label{Metallic}

\begin{figure}[htbp]			
\centerline{\includegraphics[width=0.65\hsize]{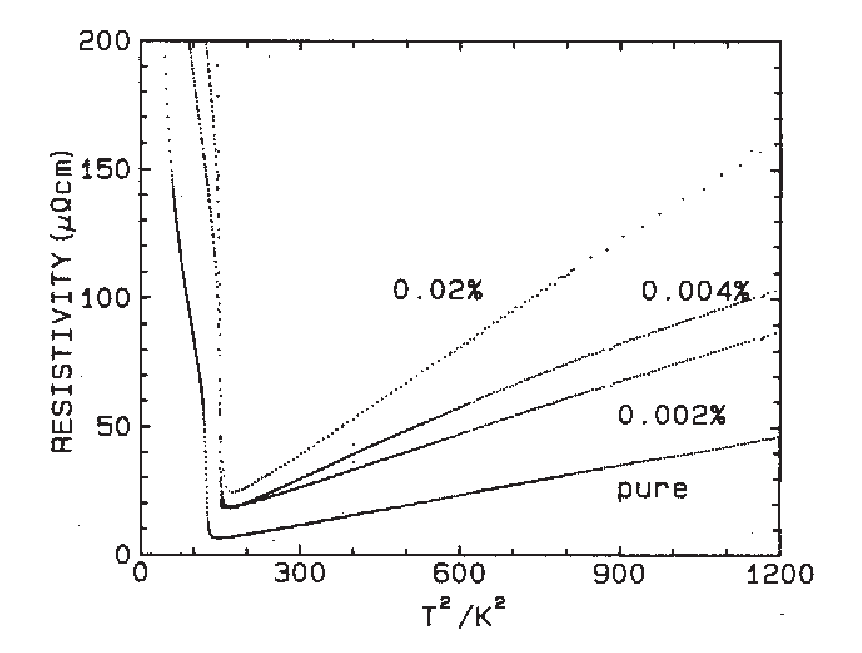}}
\caption{Temperature dependence of the \tmpf  longitudinal  resistivity plotted versus $T^2$ for pure and irradiated samples. After \cite{Tomic91}}
\label{T2.eps}
\end{figure}

The strongly metallic character of the \tmpf salt has been one of the highlights in the search for organic conductors \cite{Bechgaard80}. The temperature dependence of the longitudinal resistivity follows a power law $T^{1.4}$ from 300K down to about 100K. Below 35K, the resistivity of \tmpf or \tmas\  is quadratic in temperature $\rho_a (T) = \rho_0+ A T^2$, which is valid down to the metal-insulator transition due to the onset of an itinerant antiferromagnetic state at 12K (Fig. \ref{T2.eps}). 

%%%%%%%%%Figure rho_a T dans PF6
%%%%%%%%%%%%%%%
%%%%%%%%%%%%%%%%

\begin{figure}[htbp]			
\centerline{\includegraphics[width=0.5\hsize]{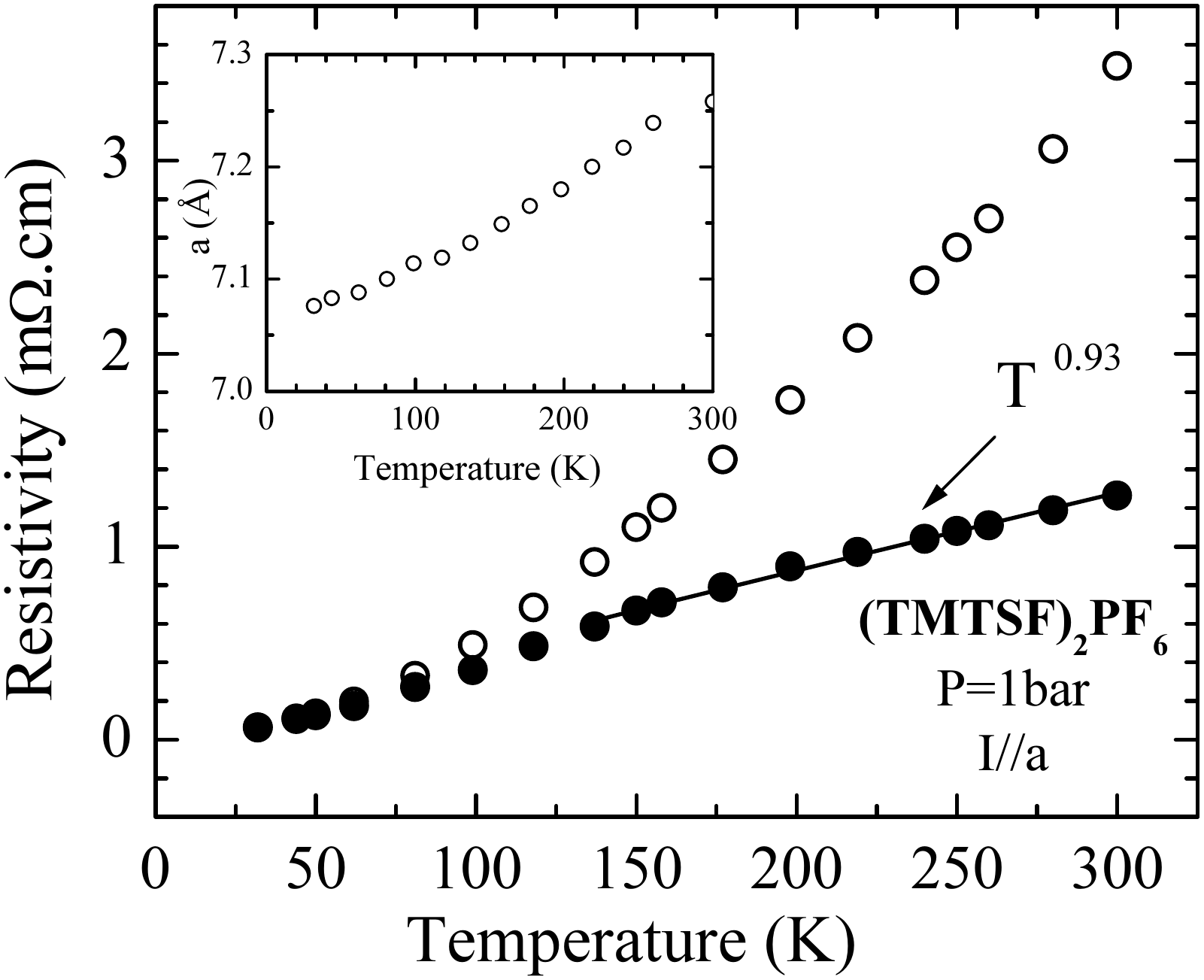}}
\caption{Temperature dependence of the \tmpf  longitudinal  resistivity at constant volume showing quasi linear $T$-dependence, with the thermal
dependence of the lattice
parameter $a$ displayed in the inset. After \cite{Auban99}}
\label{rhoconstantvolumePF6.eps}
\end{figure}

For high quality samples the resistance ratio $\rho_a (300K) /\rho_0 $ can reach  values as large as 800\cite{Tomic91} (Fig.\ref{T2.eps}).
Furthermore, an interesting behaviour encountered in  \tmpf materials (and also in most organic conductors) is the very strong
pressure (or volume) dependence of their electronic properties, particularly the transport property \cite{Cooper79,Jerome94,Auban99}. 
In addition, the thermal expansion of
these materials is particularly large. Hence, the only temperature dependence that can be compared with the
prediction of the theory is the one measured at a constant volume. As all temperature dependences are obtained
under constant pressure, a constant volume transformation must be performed. An example is given in Fig. \ref{rhoconstantvolumePF6.eps}
%%%%%%%%%%%%%%%
%%%%%%%%%%%%%%%%
by the longitudinal transport of \tmpf behaving at  high temperature similar to $T^2$ under ambient pressure but varying
sublinearly ($\sim T^{0.93}$) from 300 to 150 K, once the volume correction is taken into account \cite{Auban99}.
The experimental power law of longitudinal resistivity leads in turn to
${n}^{2}K_{\rho}$=0.98  according to the theory of resistivity \cite{Giamarchi97}. Note that a similar power law for the 
temperature dependence of longitudinal transport ($\sim  T^{0.93}$) can also be observed for two 
sulfur-compounds,
\fb\  and \pf\    under high enough pressure when the correction to constant volume becomes negligible.

In the early days of the \tm\ compound, the lattice dimerization was believed  to govern entirely the  amplitude of
the Mott-Hubbard gap \cite{Barisic81,Emery82}.
When  the   half-filled  scenario is privileged, namely ($ {n}=1$). Hence,  $ {n}^{2}K_{\rho}$=0.98 leads to  a bare value of  $K_{\rho}$  close to unity  implying
a weakly coupled electron gas. This  situation of a very weak coupling  is difficult to reconcile with an enhancement of the
spin susceptibility and the characteristic enhancement of the nuclear spin relaxation rate  \cite{Bourbon99,Wzietek93} but an additional  argument against this weak coupling is made possible by the
unusual behavior of transverse transport (Eq. (\ref{5})) \cite{Georges00}. 
The weak coupling value for $K_{\rho} \approx 1$ derived from the temperature dependence of
longitudinal transport and the optical data (see below) would imply $\alpha \approx 0$ and consequently a metal-like
temperature dependence for
$\rho_{c}(T)$  which is \textit{at variance} with the data. 

More recently, an alternative interpretation based on new
experimental results has been proposed  assuming that the 1/4-filled scattering could justify the existence
of the Mott gap in the entire \tmx  series \cite{Giamarchi97}. With such a hypothesis (${n}=2$), the fit of the  experimental
data would  thus  lead to
$K_{\rho}=0.23$ and $\alpha = 0.64$ (see below, discussion on optical response) 
\cite{Schwartz98}. This bare value for $K_{\rho}$   agrees fairly well in the 1/4-filled scenario at the very least at high temperature where the influence of half-filling Umklapp should be weak \cite{Bourbon99,Tsuchiizu01}.  This value that would imply $U/W = 0.7$ for the Hubbard parameter is compatible with plasma edge measurements  and the enhancement  in the spin susceptibility  \cite{Bourbon99,Wzietek93}.  Such a strong coupling value for the bare  $K_{\rho}$ implies that a system such as \tmpf lies
at the border between a 1D Mott insulator and a Luttinger liquid, though slightly on the insulating side. \tmbr\  is another particularly interesting system, in which the pressure coefficient of the resistivity is
very large. Hence, it is the correction to  constant volume which makes the maximum in $\rho_{c}$ to appear around 150K, while this maximum is
absent in the
 constant pressure runs, (see Fig. \ref{rhocunderpressure.eps}).  
 
\begin{figure}[htbp]			
\centerline{\includegraphics[width=0.9\hsize]{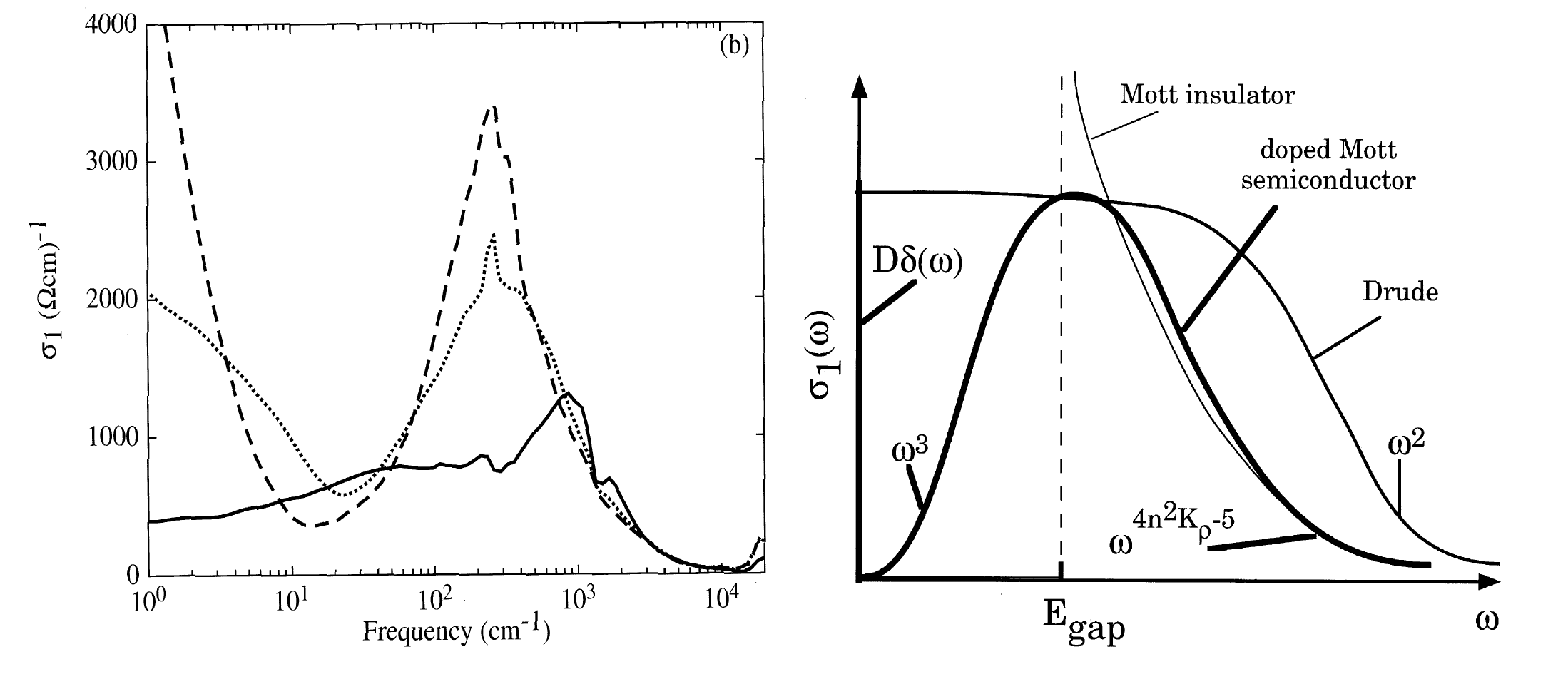}}
\caption{Far infra red optical conductivity (${\bf E}\| a$) data of \tmpf, experiment\cite{Schwartz98} for $T=300, 100, 20$K(left) and theory for a doped Mott insulator \cite{Giamarchi97} (right)}
\label{optiqexpertheoPF6.eps}
\end{figure}

 The other  approach to  the correlation coefficient is given by the  far infrared optical studies of
$\mathrm{(TMTSF)_{2}PF_{6}}$, which have been very helpful for the determination of  $K_{\rho}$ since  the FIR gap of
about 
$\Delta_{\rho, a}= 200$~cm$^{-1}$  in
$\mathrm{(TMTSF)_{2}PF_{6}}$  has been attributed to the signature of the Mott-Hubbard gap
\cite{Schwartz98}. Consequently,  the frequency dependence of the conductivity above the Mott gap is closely linked to the dynamics of the excited carriers in the 1D regime. The theory predicts a power law dependence for the optical conductivity at frequencies larger
than the Mott gap\cite{Giamarchi97} namely, 
\begin{eqnarray}
\sigma_{1,a}(\omega)\sim
\omega^{4n^{2}K_{\rho}-5},
\end{eqnarray}
 at $\omega>2\Delta_{\rho,a}$ see Fig.(\ref{optiqexpertheoPF6.eps}).
 %%%%%%%%%%%%%%%%%%%%Figure de l'optique sur TMTSF2PF6
%%%%%%%%%%%%%%%%%%%%%%Optique experimentale
\begin{figure}[htbp]			
\centerline{\includegraphics[width=0.5\hsize]{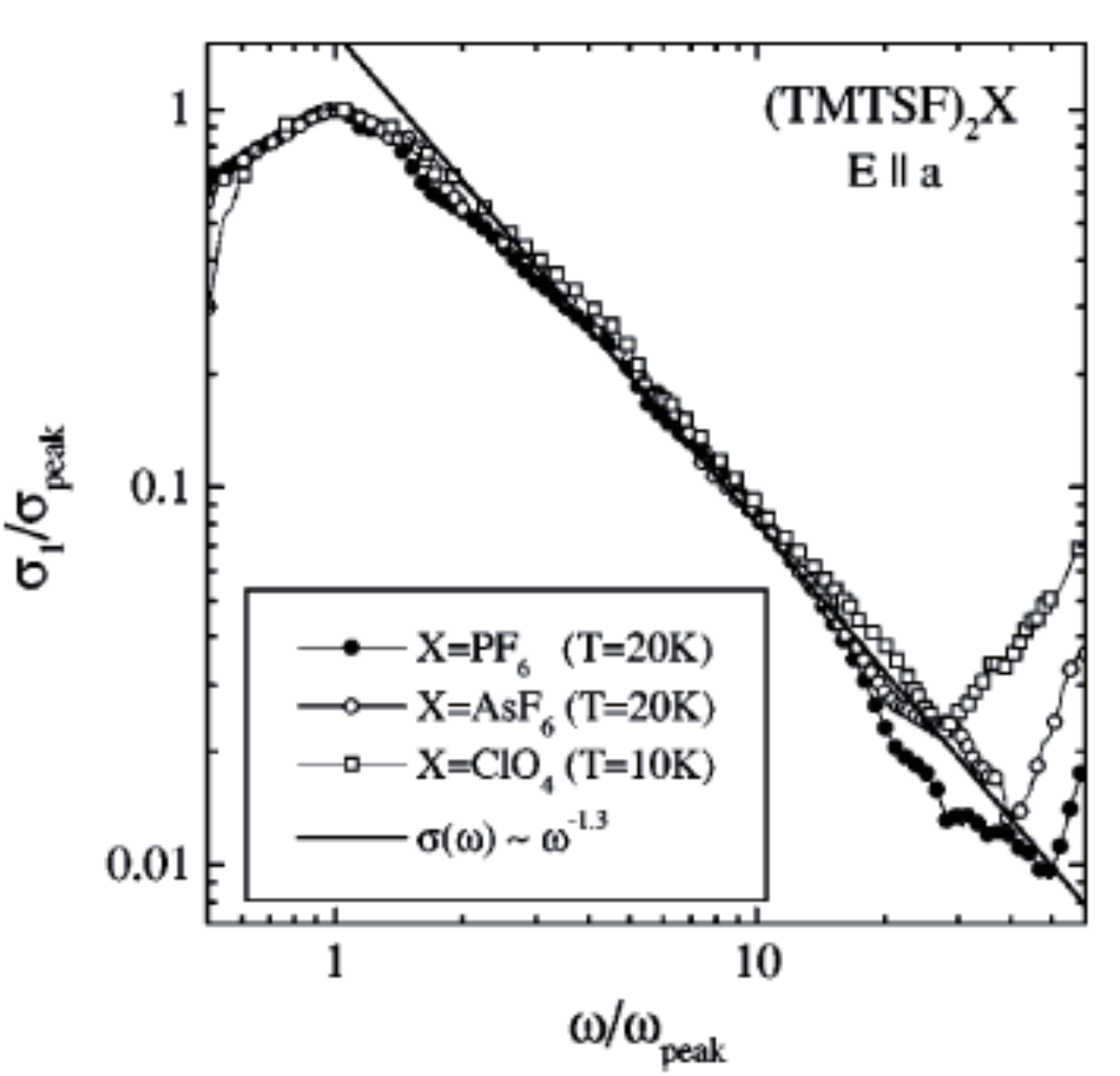}}
\caption{Optical conductivity above the Mott-Hubbard gap in several selenide conductors analyzed in terms of a power law  $\omega^{-1.3}$. After Ref. \cite{Schwartz98}}
\label{loidepuissanceenoptique.eps}
\end{figure}
%%%%%%%%%%%%%%%%%%%%%%

According to the optical experimental data  (see Fig.\ref{loidepuissanceenoptique.eps} \cite{Schwartz98}), $\sigma_{1,a} (\omega ) \propto \omega^{-1.3}$ at high frequency leading to 
$n^{2}K_{\rho}$=0.93. This value for the correlation coefficient is fairly close to the one  derived above from parallel  transport data but none of
these experiments allow by themselves to discriminate between half or quarter-filled Umklapp scattering. 
 
 \subsection{Pseudogap and zero frequency mode in the metallic phase of \tsx }

It is also most illuminating to have a look at the conductivity in the far
infrared regime. A large gap of order 1000K is observed in the frequency dependence of the FIR
conductivity of sulfur compounds\cite{Jacobsen87}. This is in line with the activation energy of the DC
conductivity in those compounds. However, the surprise arose for 
 selenium compounds which behave apparently like normal metals as far as DC transport is concerned, in spite of  the marked gap
observed in the FIR regime at low temperature. The apparent normal behavior of the resistivity varying quadratically in
temperature for \tmc\ or \tmpf above the SDW transition could lead to the  misleading conclusion of a 2 or 3-D Fermi gas in
which the temperature dependence of the transport is governed by e-e scattering. However, the analysis of the conductivity in
terms of the frequency reveals quite a striking breakdown of the Drude theory for single-particles. The inability of the Drude
theory to  describe the optical conductivity has been noticed by a number of experimentalists working on \tfx\   with
X=\cl, \pf\ or \sb \cite{Timusk87}. When the reflectance of \tmc\  in the near-infrared is analyzed with the Drude model  in the whole range of
temperatures from $300$ down to 30K the electron scattering rate is found to decrease gradually from $2.5\times 10^{14}$\,s$^{-1}$ at room
temperature to
$1.3\times 10^{14}$\,s$^{-1}$ at $30 $K \cite{Kikuchi82}. Even if the RT value is not far from the value from DC conductivity, a drastic
difference emerges at low temperature as $\sigma_{\rm DC}$ increases by a factor about 100 between RT and  30K \cite{Bechgaard81},
as compared to the factor 2 for the optical lifetime. 

An other striking feature of the optical conductivity has been noticed
when the Kramers-Kr\"onig transformation of the reflectance is performed in a broad frequency domain for \tmc\  as well for
all conducting materials at low temperature. Given the usual Drude relation $\sigma_{\rm DC}= \omega_{p}^{2}\tau/4\pi$
between transport lifetime and plasma frequency data (the plasma frequency has been found nearly temperature independent
\cite{Kikuchi82,Jacobsen83}) and the measured resistance ratio for $\rho_a$ of about 800 between RT and  2K obtained in
good quality measurements, the Drude conductivity in the frequency range $\approx 40$~cm$^{-1}$ should amount to at least
$4000\,\Omega$cm$^{-1}$\cite{Ng85,Ng83}. The measured optical conductivity  is at most of the order of $500 \,\Omega$cm$^{-1}$\cite{Ng85}. Consequently, the rise in the conductivity as $\omega\rightarrow  0$ has been taken in \tmc\ as well as in the other
salts with \pf \  or \sb\ as an evidence for a hidden zero frequency mode. This mode is actually so narrow that it escapes a direct determination
from K-K analysis of the reflectance, which is limited to the frequency domain above   10 cm$^{-1}$. Estimates of the mode width have been
obtained using  the DC conductivity and the oscillator strength $\Omega _{p}^{2}$ of the mode with $\sigma_{\rm DC}=
\Omega_{p}^{2}\tau/4\pi$ where $\Omega _{p}$ is measured from the first zero crossing of the dielectric constant.
This procedure gives a damping factor $\Gamma = 0.005$ cm$^{-1}$ and  0.09 cm$^{-1}$ at 2 and 25K respectively, in \tmc\ \cite{Ng83}. 
%%%%%%%%%%%%%%%%%%%%
\begin{figure}[htbp]			
\centerline{\includegraphics[width=0.4\hsize]{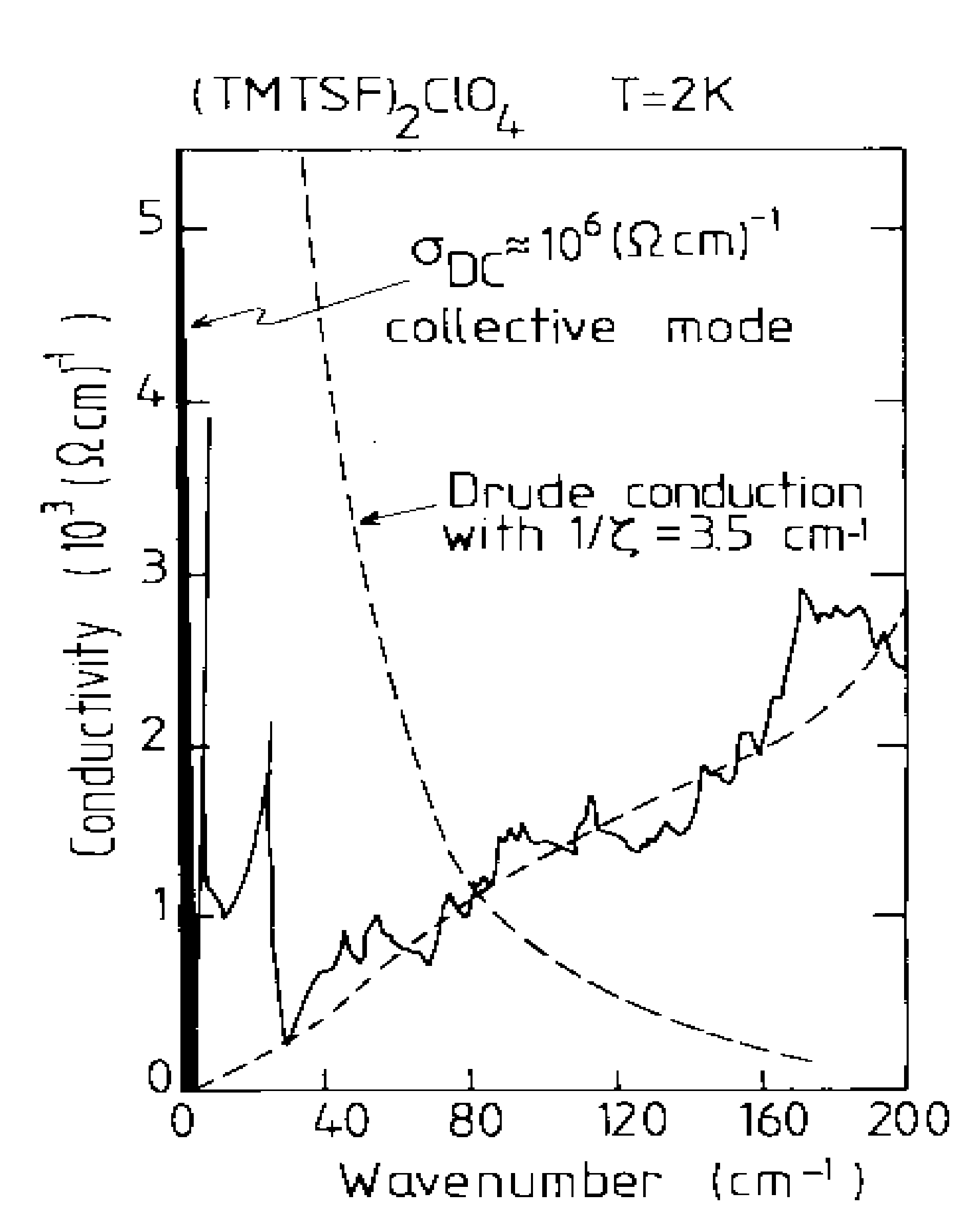}}
\caption{Far infra red data of \tmc. The dashed line is the Drude behaviour with $1/\tau =  3.5$ cm$^{-1}$ and $\omega_{p}
= 10^{4}$ cm$^{-1}$. After Ref.  \cite{Ng83}}
\label{ClO4FIR.eps}
\end{figure}
%%%%%%%%%%%%%%%%%%%%
The confirmation of a very long scattering time for the DC conduction 
 has also been brought  by the rapid suppression of $T_c$ by non-magnetic
defects in the non conventional superconductor \tmc\ leading to $\Gamma_{c} = 0.56$ cm$^{-1}$ at low temperature\cite{Joo04}, meaning that the
electron lifetime at low temperature is actually much longer than the value inferred from a Drude description.  

There is now a wealth of
experimental evidences showing the development of a narrow frequency mode in the Mott gap of
\tmc\  and related conducting compounds. From FIR data in \SePF, it has also been shown that the narrow mode carries only
a small fraction (a few percent) of the total spectral weight\cite{Schwartz98,Jacobsen83}, but it is this mode that explains the very large value
of the DC conduction observed at low temperature.

%%%%%%%%%%%%%%%
 \subsection{Quarter-filled compounds}
 \label{Quarter}
\begin{figure}[htbp]			
\centerline{\includegraphics[width=0.8\hsize]{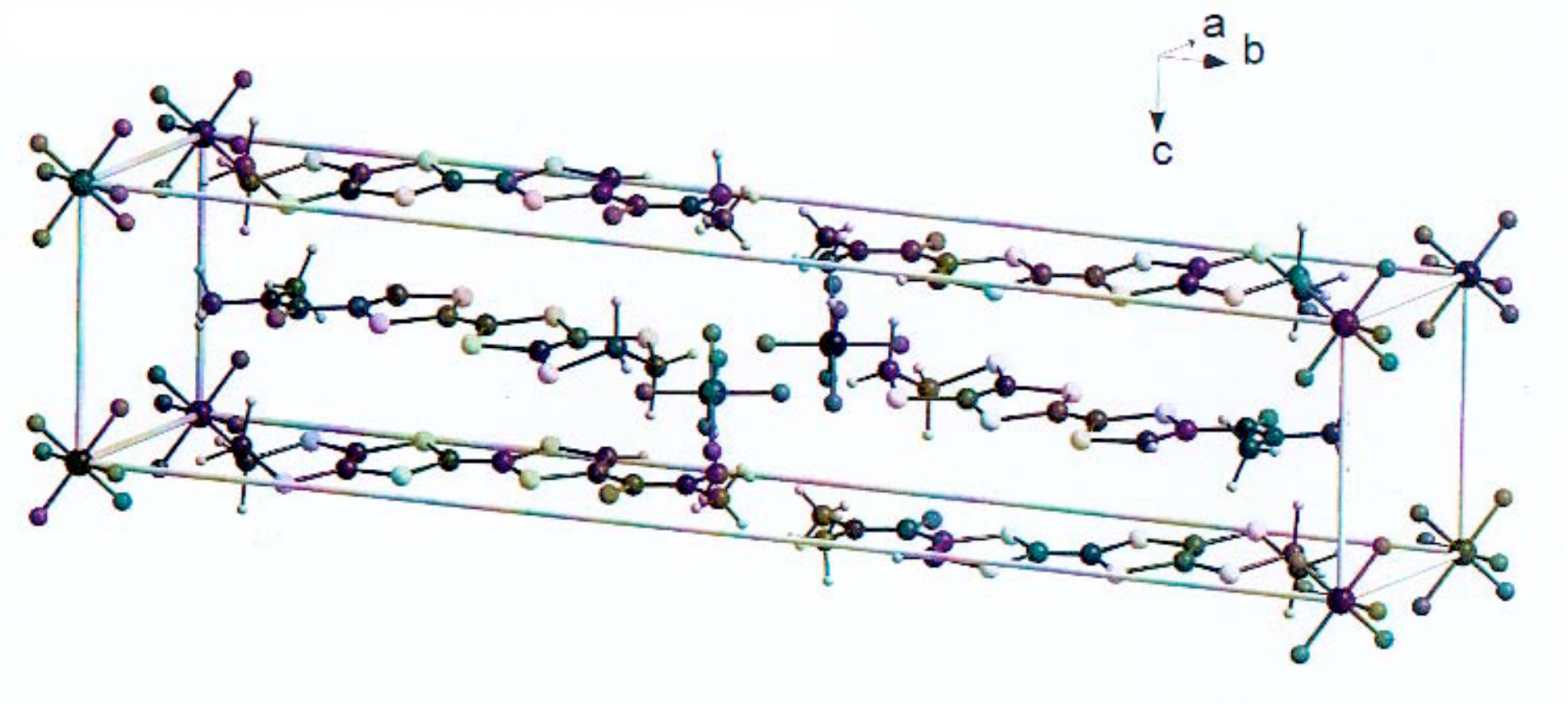}}
\caption{Structure of the quarter-filled compound \edt. After Ref.  \cite{Heuze03} }
\label{structdeEDTAsF6.eps}
\end{figure}

 Even if the  debate between 1/2 and 1/4 fillings may be relevant for \TM, this is no longer the case for new synthesized
compounds in a family whose general  structure precludes any dimerization.
The structural
peculiarity of the salt  \hbox{\edt\,} is the absence of inversion center between adjacent
molecules in stacks and instead the presence of a glide symmetry plane
\cite{Heuze03} (see Fig.\ref{structdeEDTAsF6.eps}). The analysis of the transport data of \hbox{\edt\,} has shown that in
spite of the existence of a glide symmetry  plane, the carriers are localized, and even more localized
than in the most insulating salts of the Fabre series known at present. Since the localization in this compound cannot be ascribed to a 1/2$-$Umklapp
scattering or to the Anderson localization, 1/4-Umklapp scattering seems to be
the only channel left to explain carrier localization in this commensurate 1D
conductor. Under ambient pressure, given the total bandwidth deduced from
quantum chemistry ($W$ = $W_{\rm tot}$($P=1$bar) = 0.350 eV (3850 K)) 
and the experimental Mott gap ($2\Delta_{\rho,a} =2700$ K), the theory
\cite{Giamarchi97} gives, in the case of quarter filling \textit{stricto
sensu}: $2\Delta_{\rho, a}= 2W(U/W)^{3/2(1-4K_{\rho})}$. This leads to the bare value $K_\rho $  = 0.1, with a reasonable $U/W$ = 0.7.

The Mott gap of \hbox{\edt}  is much larger than the expected value
of the bare interstack overlap $t_\perp$, which makes according to Eq. \ref{cross1}  the single
particle hopping between neighboring stacks non-pertinent in the
pressure regime less than 20 kbar since the transverse hopping is renormalized to zero
on account of a strong intrachain electron-hole interaction.

 The interaction over bandwidth ratio $U/W$ $\approx$ 0.7 is also in fair agreement with the result of a crude analysis of the spin susceptibility
of S-salts
\cite{Bourbon99,Wzietek93} and indicates that these compounds lie in the strong coupling sector.
With increasing pressure, the gap decreases steadily up to the pressure of 20 kbar  above which it disappears sharply
due to the competing transverse coupling. Below 15 kbar, the gap of  \edt\  (1350 K) is about equal to
the gap of  \tfx\   measured under ambient pressure \cite{Moser98}. Since the
logarithmic pressure dependences of the gap seem to be similar for both
compounds, we may say that \edt\    can also be considered as
part of the generic Fabre-Bechgaard salt diagram, provided  the origin of the pressure axis
is shifted to the left by 15 kbar. Hence, it is reasonable to expect
that the TL parameter $K_\rho $ increases from left to right in the
\tmx diagram, since both optical and transport data suggest $K_\rho $= 0.23 in the Se
compounds whilst it is only of the order of 0.1 in sulfur compounds.
%%%%%%%%%%%%%%Figure EDT structure
%%%%%%%%%%%%%%
Let us add that the uniformly stacked 1-D conductor $\mathrm{(DI-DCNQI)_{2}Ag}$, an other quarter-filled compound reveals localization properties quite similar to those observed in the \tmx series\cite{Itou04}. The normal state of this system is insulating at low pressure probably due to strong electron correlations but at pressures exceeding 15 kbar,  the longitudinal restivity is metallic  above 100K with a quasi linear temperature dependence leading to $K_\rho $= 0.25 in the quarter-filled band hypothesis. This result shows once more that the conductor lies at the border with  the  quarter-filled Mott localized insulator.

 \subsection{ A robust 1-D compound:\TTdm} 

For the sake of completeness we shall mention the behavior  of an  interesting organic  salt,  which unlike all \tmx compounds, has failed to reveal the usual suppression of the insulating phase under pressure\cite{Lopes02}. The \TTdm \ system, in spite of a strong structural analogy with the \tmx materials exhibiting stacks of donors arranged in layers with short interchain contacts, shows a unique extreme 1-D character together with a strong bond dimerization. According to extended Huckel calculations, the dimerization gap at the middle of the HOMO band amounts to 0.027 eV \textit{i.e} 13\%  of the upper  band dispersion and the interchain coupling within the layers is practically zero. As the smallness of this  contact is due to terminal  sulfur atoms in intermolecular contacts not participating to the HOMO of the molecule, we can expect a survival of the 1-D character under pressure and no pressure-induced dimensionality crossover. Consequently, contrasting with the members of the Fabre-Bechgaard series,  the Mott-Hubbard insulating nature persists up to a pressure of 25 kbar\cite{Lopes02}, which is usually large enough to severely decrease (if not suppress) the localization in the latter family as shown in the present article.
 
 %%%%%%%%%%%%%%%%%%%%%%%%%%%%%%%%%
  \subsection{The spin-density-wave phase} 
 \label{SDW}
 At the core of the unity shown by  the phase diagram of Fig. \ref{fig:1}  is the shift on the pressure scale when selenium is substituted for the sulfur atom and yields the  Bechgaard salts  series \Bech \cite{Jerome80,Bechgaard80}. The selenium series was at the start considered more promising compared to previous organic compounds, mainly because  the  metal insulator transition at ambient pressure only occurs below 20K  after high metallic conductivity have been attained\cite{Bechgaard80}.

 In the two  compounds of the series \SePF\  and \SeAsF, the transition occurs at $T_c\approx 12$K at 1 bar\cite{Bechgaard80,Jerome80,Brusetti82}. The transition early showed all the characteristics of SDW long-range ordering \cite{Mortensen81,Andrieux81}.  Similar SDW is also found in  \Fabre\ but at much higher values on the pressure scale. In   \SPF\  for example, about 40~kbar of pressure is needed to reach a $T_c\sim 10$K (Fig. \ref{PF6generic.eps})\cite{Jaccard01,Adachi00}, whereas for \SBr\  \cite{Balicas94,Klemme95}  and \SBF\cite{Auban03} about  10~kbar and 27~kbar must be applied, respectively. 
 
The gradual emergence of a plasma edge in the $b$ direction below  100K\cite{Jacobsen81,Vescoli98}, and the recovery of transverse metallic and longitudinal Fermi liquid ($\sim T^2$) resistivity below $T^\star$  for \SePF\ indicate that the  transverse electron  band motion has developed some coherence  at the onset of  the SDW transition.  The mechanism of the instability will then naturally follow from the property of nesting of the Fermi surface based on the property $E^*_{-p}({\bf k})=-E^*_{p}({\bf k} +{\bf q}_0)$ of the spectrum (\ref{3Dspect}) for a special wave vector $\mathbf{q}_0$, called the nesting vector. The relevance of the  Fermi surface for the transition has been confirmed by the determination of the modulation wave vector of SDW by NMR\cite{Takahashi86}, which coincides with the best nesting vector obtained by band calculations \cite{Ducasse86}. In the simplified model spectrum   Eq. (\ref{3Dspect}) for an approximate orthorhombic lattice,  perfect nesting is found at $\mathbf{q}_0=(2k_F,\pi,\pi)$.  Deviations with respect to this ideal situation, however,  are likely to be found in practice. This amounts to use the spectra
 \begin{equation}
\label{spectrumb}
E^*_p({\bf k}) = v(pk-k_F) -2t^*_{\perp b}\cos k_{\perp b} -2t^*_{\perp c}\cos k_{\perp c} -2t'_{\perp  b}\cos 2 k_{\perp b},
\end{equation}
that contains small next-to-nearest-neighbor hopping $t'_{\perp  b}$ along the $b$ direction ($t'_{\perp b}\ll t_{\perp b}$). This leads to the modified nesting condition 
 \begin{equation}
\label{nesting}
E^*_p({\bf k})=-E^*_{-p}({\bf k} +{\bf q}_0) + 4t'_{\perp  b}\cos 2 k_{\perp b},
\end{equation}       
with $k_{\perp b}$ dependent nesting frustration. 

The determination of the temperature scale for the SDW instability follows   the analysis given in \S\ \ref{interchain}, where the ladder result Eq. (\ref{ladder}) becomes
\begin{equation}
\label{Stoner}
\tilde{J}(T) = {\tilde{J}^*\over 1- {1\over 2}J^*\chi^0(\mathbf{q}_0,T)},
\end{equation}  
where
\begin{equation}
\label{barechi}
 \pi v_F \chi^0(\mathbf{q}_0,T) = \ln {T_{x^1}\over T} + \psi\Big( {1\over 2}\Big) - \Re{\rm e} \Big\langle \psi\Big({1\over 2} + i{t'_{\perp b} \cos 2k_{\perp b}\over \pi T}\Big)\Big\rangle_{k_{\perp b}},
\end{equation}
is the bare susceptibility in the presence of nesting frustration $t'_{\perp b}$ ($\psi(x)$ is the Digamma function and $\langle \ldots \rangle_{k_{\perp b}}$ is an average over $k_{\perp b}$) \cite{Montambaux88}.  As mentioned previously, the above expressions differ slightly from previous mean-field approaches \cite{Horovitz82,Yamaji82,Montambaux88,Gorkov96}, in that the contribution of intermediate electron-hole excitations  to $\chi^0$ has been here restricted to an energy shell $\pm T_{x^1}$ (instead of $\pm E_F$) around the Fermi level. More energetic excitations extending up to the Fermi energy are one-dimensional in character and are governed by Eqs. (\ref{flow1D})\cite{Bourbon86,Bourbon91,Emery82,Seidel83}. The above Stoner form also neglects  the finite coupling between electron-hole and electron-electron pairings, an interference that persists even  below $T_{x^1}$. This approximate weak coupling description of SDW remains qualitatively correct, however, as long as $t'_{\perp b}$ does not reach too large values, that is where the interference between density-wave and superconductivity   can change the nature of the ground state (see \S\  \ref{Supra}). 

The Stoner form (\ref{Stoner}) develops a singularity at the critical temperature $T_c$ that depends on $t'_{\perp b}$, the main parameter that is standardly used to mimic the actual effect of pressure on the SDW state \cite{Horovitz82,Yamaji82,Montambaux88,Duprat01}. A finite $t'_{\perp b}$ will then reduce $T_c$ with respect to the BCS limiting value $T^0_c= T_{x^1} \,e^{-2/\tilde{J}^*}$  at perfect nesting (Eq. (\ref{BCSTc})), a feature of the model that was soon linked  with experiments done under pressure \cite{Jerome80,Jerome82,Brusetti82,Vuletic02} (inset of  Fig. \ref{tmpf6supraetsdwsousP.eps}). Besides the monotonic increase of $T_{x^1}$  and the decrease of $\tilde{J}^*$ under pressure, the  detrimental influence of $t_{\perp b}'$ on $T_c$ remains the most dominant effect.  

The expression (\ref{Stoner}) predicts that at the approach of the critical value $t'^{\rm cr}_{\perp b}  \sim 0.7 T_c^0$, $T_c$ rapidly goes down to zero --  though the possibility of SDW at incommensurate $\mathbf{q}_0$ and very low temperature may introduce a change in the slope of $T_c$ near $P_c$ \cite{Hasegawa86b}.  To  $t'^{\rm cr}_{\perp b}$ will then correspond a critical pressure $P_c$ for the suppression of SDW; this variation of $T_c$ agrees with the characteristic pressure profile generally observed for \Bech\ at moderate pressure (Fig. \ref{tmpf6supraetsdwsousP.eps}) and for \Fabre\ at higher pressure (Figs. \ref{PF6generic.eps} and \ref{fig:1}). The fitness of the model to describe the $T_c$ of the SDW state in \Bech\ can be  further assessed if one considers  the influence of a transverse magnetic field on $T_c$. A perpendicular magnetic field  $H\| {\rm c}^*$ tends to confine the motion of electrons along the  chain direction and gradually restores  better nesting conditions. This  in turn increases   $T_c$ with $H$ \cite{Montambaux88}, consistently with early field dependent measurements of $T_c$ in \SePF\ \cite{Kwak86}.

\subsubsection{Intrusion of  charge-density-wave order}
\label{CDW}
 The reexamination of  diffuse scattering X-ray patterns of (TMTSF)$_2$PF$_6$ by Pouget and Ravy \cite{Pouget96,Pouget97} revealed the emergence, besides SDW,   of a charge-density-wave superstructure (CDW) at  $T_c$; both having the same modulation vector $\mathbf{q}_0$. These results were subsequently confirmed  by Kagoshima {\it et al.} \cite{Kagoshima99}, who also found a similar superstructure in \SeAsF, but with a weaker amplitude.  These results came as a surprise since at variance with ordinary Peierls phenomena, the CDW order is not preceded by  any lattice softening in the normal state (the  $2k_F$ diffuse scattering lines do exist at high temperature but their amplitudes become  vanishingly small in the vicinity of  $T_c$ in the normal state\cite{Pouget82,Pouget96}). The CDW superstructure would then be entirely electronic in character with no lattice displacement involved.  
 
 In connection with these X-ray results, it is worth mentioning the  earlier  optical conductivity measurements of  Ng {\it et al.}\cite{Ng84}, on the isostructural member of the series (TMTSF)$_2$SbF$_6$.  The results show the growth in the infrared of new phonon lines at $T_c$, precisely those  usually expected for the excitations of a CDW superstructure; their  temperature dependent intensity  follows roughly  the one of the SDW order parameter below $T_c$. CDW phonon lines in the far infrared conductivity have also been found in the metallic phase of \SeClO\ at low temperature \cite{Cao96}, indicating that $2k_F$-CDW  and SDW correlations apparently  coexist in the normal phase\cite{Bourbon84,Wzietek93}.
 
 On theoretical grounds the possibility for  SDW and CDW to coexist has been analyzed recently by considering the redistribution of charge and spin  in the unit cell, a possibility that emerges when its internal -- two-molecules -- structure and the long-range Coulomb interaction are taken into account. In the framework of extended Hubbard model,  numerical and mean-field approaches show that  charge and spin can  be so rearranged in the  unit cell  that   SDW, BOW and CDW can coexist  \cite{Riera00,Mazumdar99,Kobayashi98,Seo97}.  Moreover, it was shown recently  that when interchain Coulomb interaction is included, this  can favor -- even for small amplitude --  BOW and CDW correlations while not affecting SDW\cite{Nickel05}.

% \subsection{Organic superconductivity}
 %\label{OrgSuper}
 \subsection{Some features of the superconducting state}
\subsubsection{The superconducting transition}

\begin{figure}[htbp]			
\centerline{\includegraphics[width=0.38\hsize]{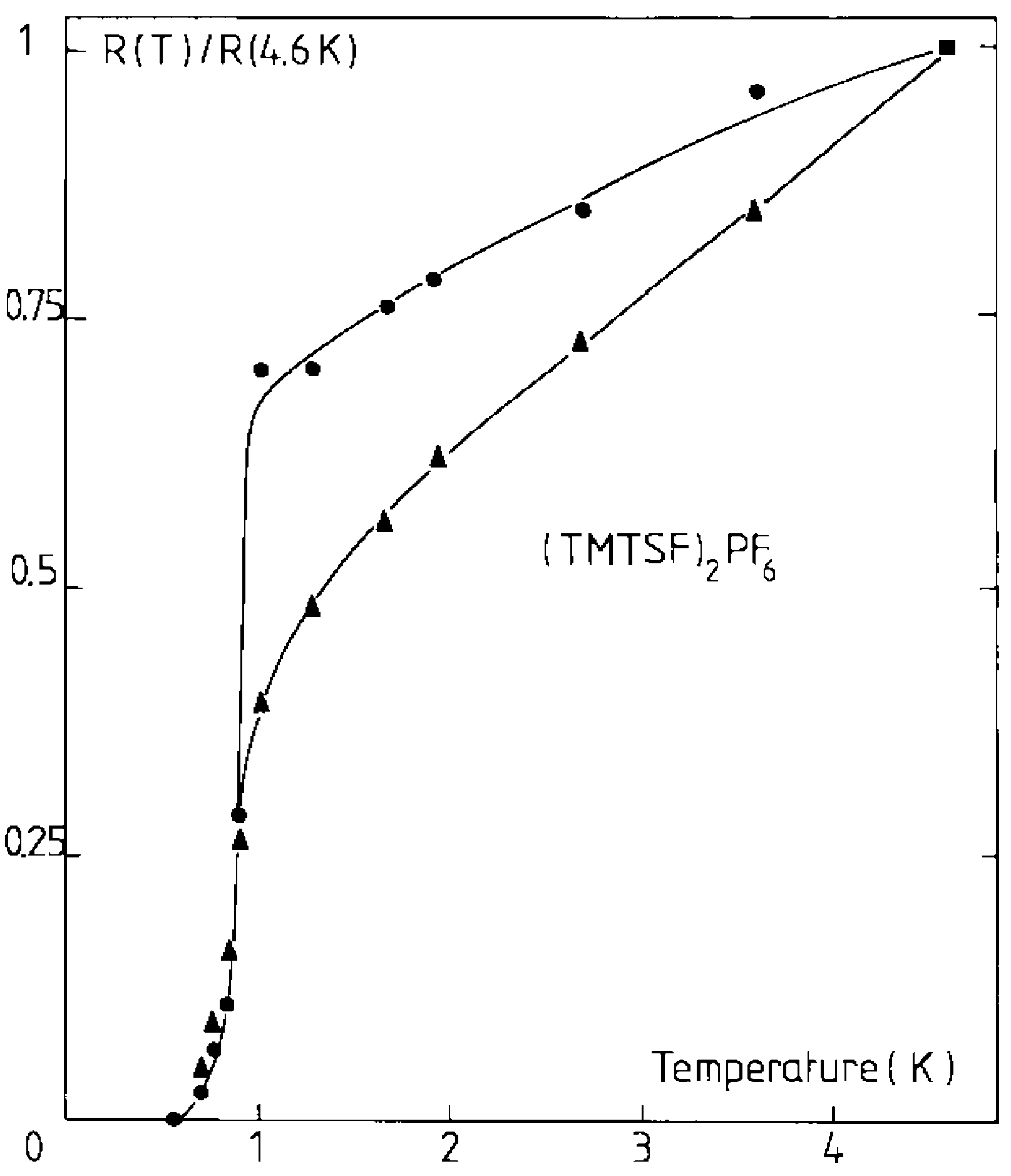}}
\caption{$\mathrm{(TMTSF)_{2}PF_{6}}$, first observation of organic superconductivity under pressure. After Ref. \cite{Jerome80}}
\label{tm2pf6supra.eps}
\end{figure}

For all cases of superconductivity in the \tmx\  series, the first evidence has been provided by a drop of the resistivity below the critical temperature and the suppression
of this drop under magnetic field. We shall focus the presentation on the two members of the \tsx series, which have attracted most attention: 
\begin{itemize}
\item
i) \tmpf,
because this has been the first organic superconductor to be found by transport measurements \cite{Jerome80}, and subsequently confirmed by magnetic shielding
\cite{Ribault80,Andres80}, and also because the electronic properties   of the 1-D electron gas on the organic stacks  are only weakly (if at all) affected by
the centrosymmetrical  anions \pf. The finding of a very small and still non-saturating resistivity  under ambient pressure reaching the value of
$10^{-5}\Omega^{-1}$ cm$^{-1}$  at 12K  triggered  further pressure studies at a pressure of 9 kbar
 in a dilution refrigerator, which  led to the discovery of  a zero resistance state 
below 1K (see Fig.~\ref{tm2pf6supra.eps}). The non saturation of the resistivity had been taken as a signature of superconducting precursor effects. We shall come again to this important question later.
\end{itemize}
%%%%%%%%%%%%%%Figure supra TMTSF2PF6
%%%%%%%%%%%%%%
\begin{itemize}
\item
ii) \tmc,  because it is the only member of the \tmx\ series displaying  superconductivity at ambient pressure. 
%%%%%%%%%%%%%%%Figure Supra de TMTSF2ClO4
\begin{figure}[htbp]			
\centerline{\includegraphics[width=0.5\hsize]{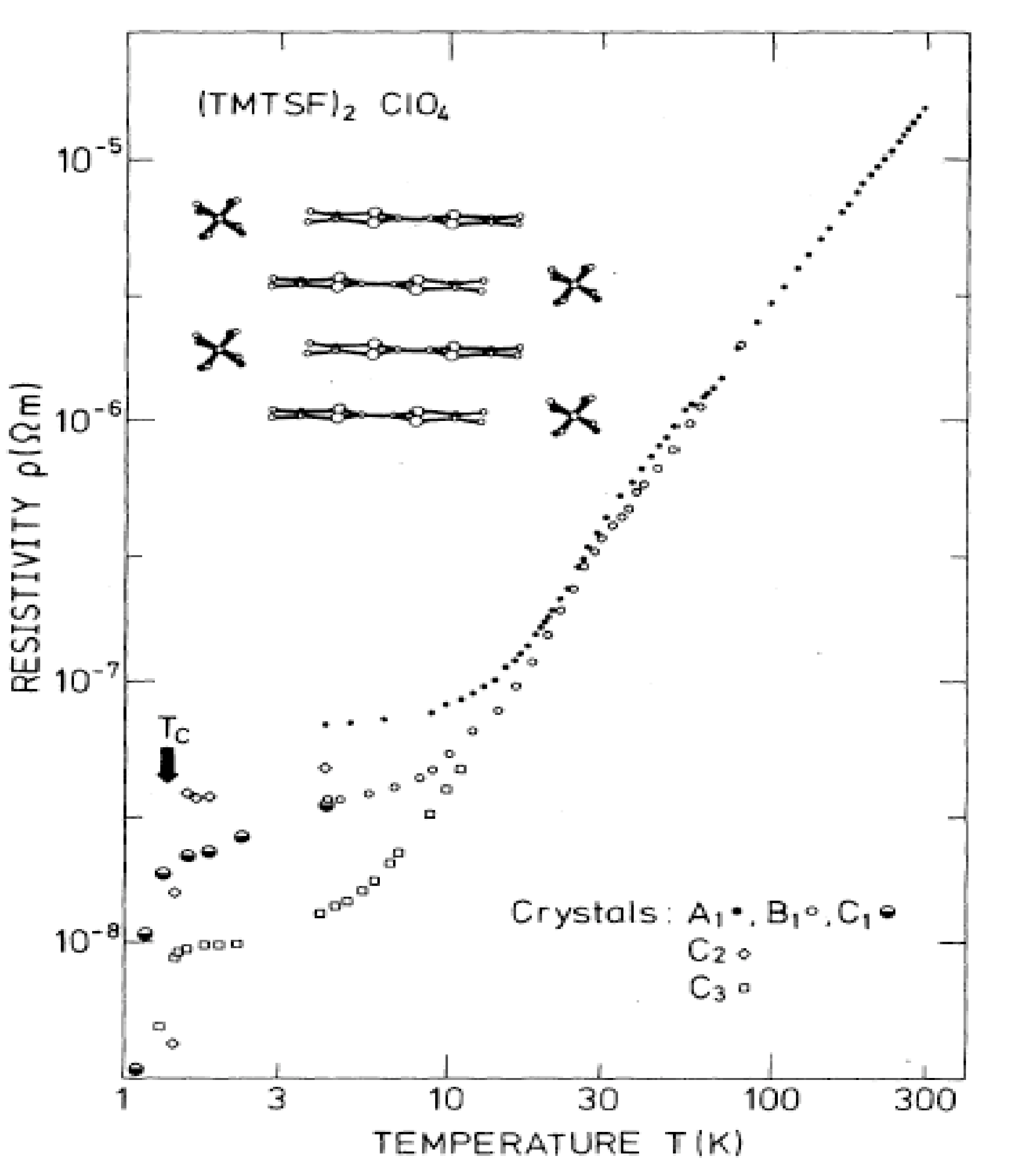}}
\caption{$\mathrm{(TMTSF)_{2}ClO_{4}}$, first observation of organic superconductivity at ambient pressure. After Ref. \cite{Bechgaard81}}
\label{tm2clo4supra.eps}
\end{figure}
%%%%%%%%%%%%%%%
However, the study of the superconducting state in  \tmc\  is meeting the problem of the \cl \, anions
ordering at  24K, doubling the periodicity along the
$b$-axis \cite{Pouget87}. Consequently, great care must be taken to cool the sample slowly enough in order to reach a well anion-ordered state (R-state) at low temperature; otherwise superconductivity is faced to its great sensitivity to disorder, a very  important feature for organic superconductors that will
be discussed more extensively below. 
\end{itemize}

\begin{figure}[htbp]			
\centerline{\includegraphics[width=0.65\hsize]{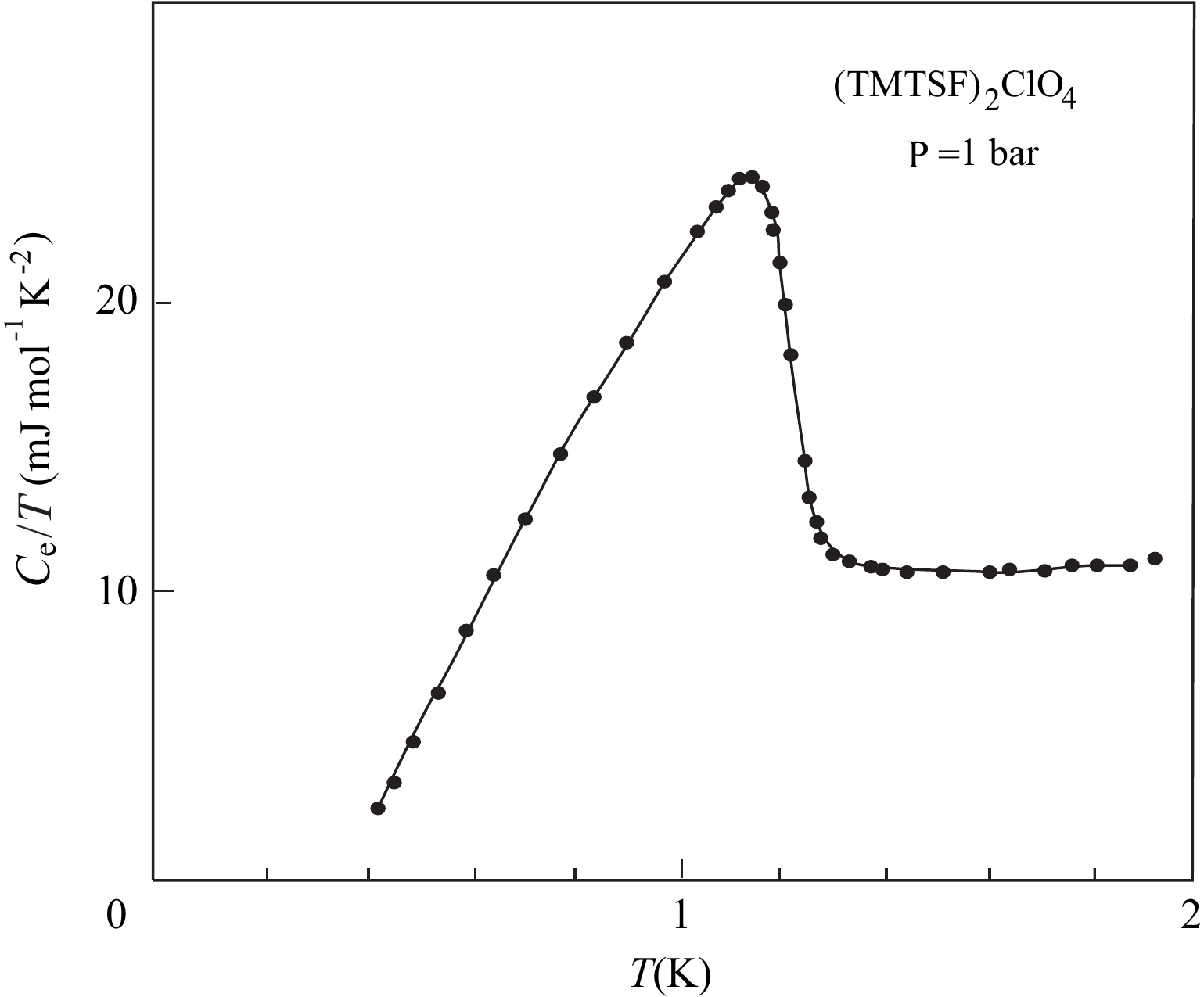}}
\caption{Electronic contribution to the specific heat of \tmc, plotted as $C_e/T$ versus $T$. After \cite{Garoche82}}
\label{Cvtm2clo4.eps}
\end{figure}

Since \tmc\ is an ambient pressure superconductor the thermodynamic evidence of the phase transition  has been obtained from specific heat on single crystals\cite{Garoche82}.
The electronic contribution to the specific heat of \tmc\ in a $C_e/T$ {\it vs} $T$ plot (Fig.~\ref{Cvtm2clo4.eps}), dipslays a very large anomaly around $1.2$K\cite{Garoche82}.
Above $1.22$K, the total specific heat obeys the classical relation in metals $C/T	= \gamma + \beta T^2$,	where the Sommerfeld constant for electrons $\gamma$ = 10.5
mJ mol$^{-1}$K$^{-2}$, corresponding to a density of states at the Fermi level $N(E_F)$ = 2.1 states eV$^{-1}$ mol$^{-1}$ for the two spin directions\cite{Garoche82}. The specific
heat jump at the transition amounts then to $\Delta C_e /\gamma T_c = 1.67$, \textit{i.e.} only slightly larger than the BCS ratio for a {\it s}-wave superconductor.
The behavior of $C_e(T)$ in the superconducting state leads to the determination of the thermodynamical critical field $H_c =
 44\pm 2$ Oe and the quasi-particle gap $2\Delta = 4$K. $T_c$ is depressed  at a rate of 1.1 mK/Oe$^{-1}$, when a magnetic
field is applied along the $c^*$ axis\cite{Brusetti83}. Comparing the value of the density of states derived from the specific heat and the value
of the Pauli susceptibility\cite{Miljak83}, lends support to a weak coupling Fermi liquid picture (at least in the low temperature
range)\cite{Bourbon99}. 
%%%%%Figure specific heat
%%%%%%%%%%%%

\begin{figure}[htbp]			
\centerline{\includegraphics[width=0.65\hsize]{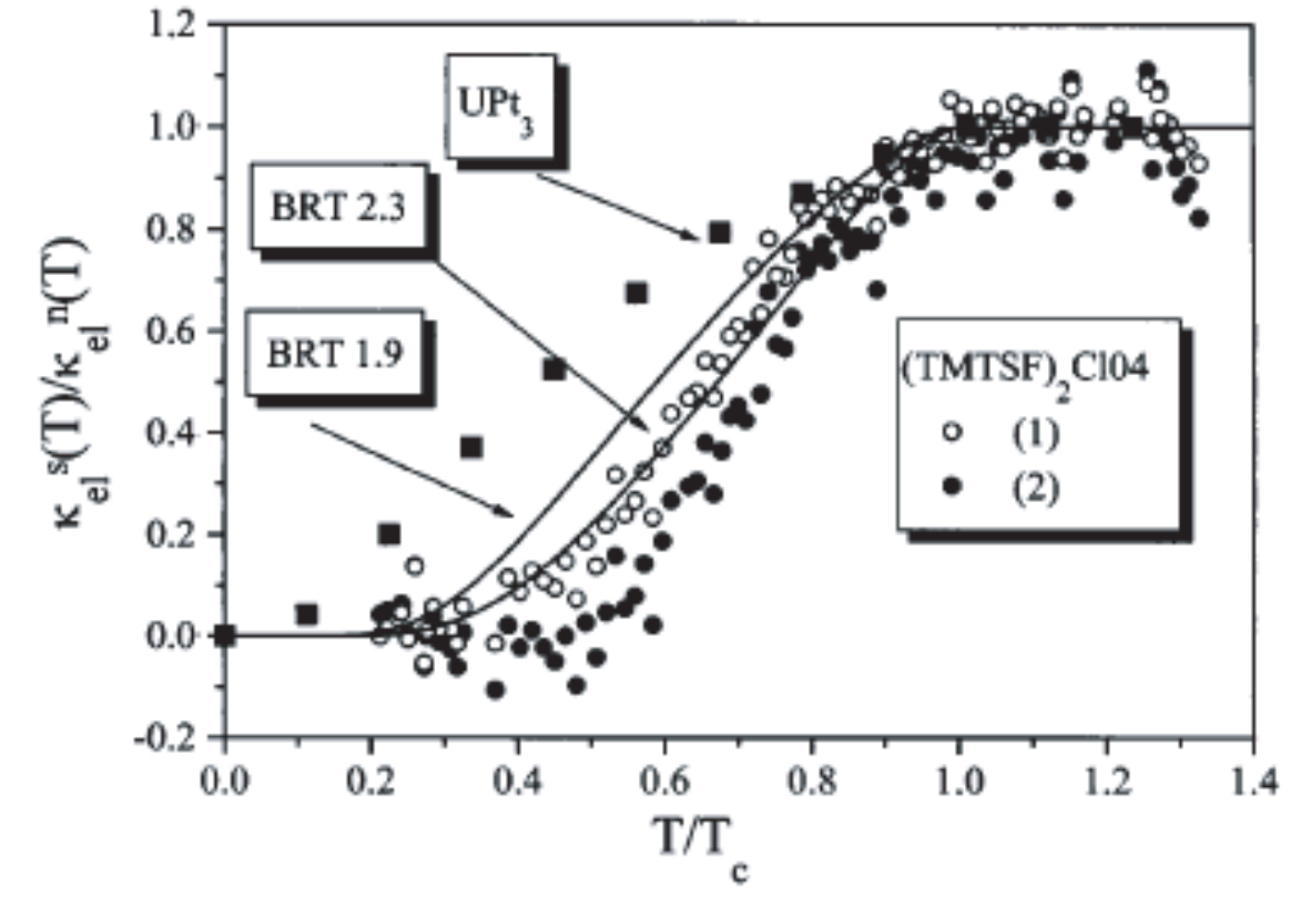}}
\caption{\SeClO, normalized electronic thermal conductivity compared to the BRT theory and the data in the unconventional heavy fermion superconductor. After Ref. \cite{Belin97} }
\label{Thermalcond.eps}
\end{figure}

Another confirmation of organic superconductivity has been provided by the measurement of thermal conduction in \tmc\ \cite{Belin97}. From the difference between thermal conductivity in magnetic field (larger than the critical field) and in zero applied field, the authors of Ref. \cite{Belin97} have been able to extract the electronic contribution of the thermal conductivity below $T_c$ down to about $T_c /5$, Fig. (\ref{Thermalcond.eps}). These data  lead to a ratio $\Delta (0) /k_{B}T_{c}=2$ within the Bardeen  Rickaysen and   Teword (BRT) theory of the thermal conduction in the superconducting state\cite{Bardeen59}. Such a ratio is in fair agreement with the specific heat jump data mentioned above. The saturation of the electronic thermal conduction observed at low temperature on Fig.(\ref{Thermalcond.eps}) is in favor of a well defined gap in the quasi-particle spectrum (possibly due to the interplay with  the anion gap of this particular compound) but does not necessarily implies $s$-pairing for the  orbital symmetry of the superconducting wave function (see the discussion in \S \ref{Supra}). The results of thermal conductivity contrast with those  NMR spin-lattice relaxation rate measurements obtained earlier by Takigawa {\it et al.,} on \SeClO\cite{Takigawa87}, which show the absence of an Hebel Slichter peak at $T_c$ and  a power law dependence $T_1^{-1} \propto T^3$  for protons NMR ($^1$H)-- these two features  being compatible with the existence of nodes for the superconducting gap \cite{Hasegawa87}. A similar algebraic dependence on temperature for $T_1^{-1}$   has been found for  $^{77}$Se nuclei in \SePF\  above the critical pressure $P_c$ \cite{Lee02}.\par
%%%%%%%%%%%%%Figure thermal conduction
The onset of superconducting order has also been detected by a shift of the muon precession frequency entering the superconducting state below $T_c$ under a field of 180 G \cite{Le93}. However, the supeconducting state is not accompanied by any enhancement of the muon relaxation rate according to the data of reference \cite{Le93},  and the recent zero field data of \cite{Luke03}. Such a behavior is \textit{at variance} with the behaviour of $\mathrm{Sr_{2}Ru0_{4}}$ \cite{Luke98}, in which the increase of the relaxation rate in a zero applied field suggests the development of spontaneous magnetic fields and is taken as a possible (but not unique) evidence for time-reversal symmetry breaking and triplet superconductivity in this oxide material. 

\subsubsection{Critical fields}
The anisotropic character of the electronic structure already known from the anisotropy of the
optical data in the normal phase, is reflected in a severe anisotropy of the critical fields $H_{c2}$ measured along the three
principal directions in \tmc\ \cite{Gubser82,Greene82,Murata82,Murata87}. Early data in
\tmc\ \cite{Murata87} are not in contradiction with the picture of singlet pairing  but no data were given below $0.5$K, the
temperature domain  where it would be most rewarding to see how $H_{c2}$ compares with the Pauli limit, when $H$ is
 perfectly aligned along the
$a$ and $b^\prime$ axes. This study has been revisited  quite recently in perfectly aligned magnetic fields down to 0.2K\cite{Joo06}. 

%%%%%%%%%%%%

The linearity of the critical fields with temperature in the vicinity of
the
$T_c$ suggests an orbital limitation in the Ginzburg-Landau formalism for the critical field and rules out a Pauli limitation,
which would favour a $(1-T/T_{c})^{1/2}$ dependence\cite{Greene82, Gorkov85}. 

Furthermore, in support to an early suggestion\cite{Gorkov85}, the  band structure parameters of \tmc\  can explain the values and the anisotropies of the critical fields assuming the existence of nodes of the superconducting order parameter \cite{Joo06}. These results  imply  that critical fields values calculated without any contribution from the spin-orbit coupling  can overcome the Pauli limit at low 
temperature by factors of two or more\cite{Lee00,Oh04}. The situation for \tmpf may, however, be quite different as  the possibility of a non homogenous superconducting phase in the vicinity of the critical pressure opens an other possibility for an enhancement of the superconducting critical fields, \textit{vide infra}.
%%%%%%%%%%%%%Figure critical fields in TMTTF2PF6 under pressure
\begin{figure}[htbp]			
\centerline{\includegraphics[width=0.65\hsize]{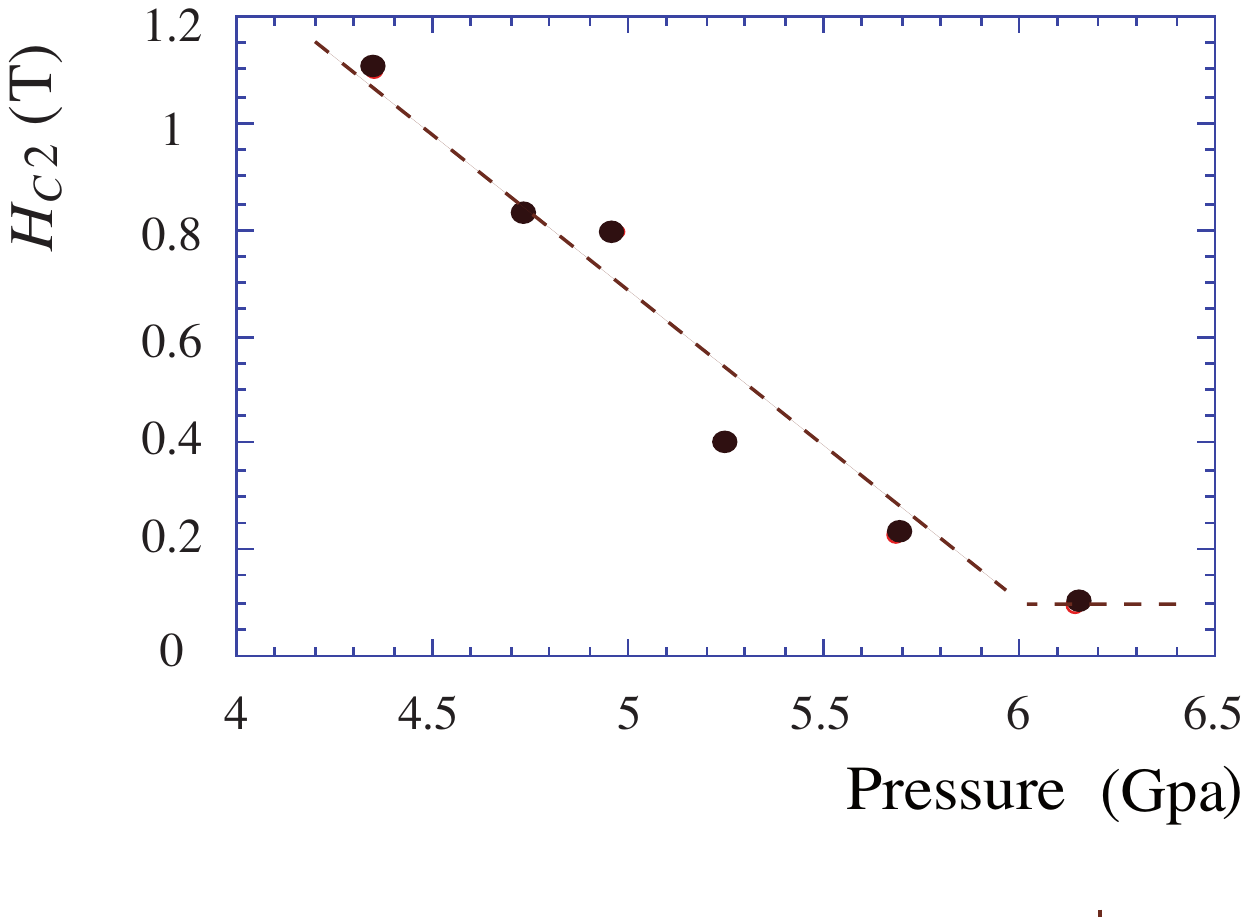}}
\caption{\tms, critical fields along  $c^\star $ under pressure in the SDW/SC coexistence regime. The lines are guides to the eyes. After Ref. \cite{Wilhelm04} }
\label{Hc2TMTTF2PF6.eps}
\end{figure}
\subsubsection{Superconductivity and Pressure}
The pressure dependence of $T_c$ is admittedly a remarkable feature for the \tmsx\  compounds since it is the pressure parameter that
enabled organic superconductivity to be discovered. However, pressure suppresses also the superconducting phase very quickly. As far as \tmpf is concerned, in the vicinity of the critical pressure  the pressure coefficient amounts  to  $\delta \ln T_c / \delta P$  =
11\%  kbar$^{-1}$ leading in turn to a Gruneisen constant  for superconductivity $\delta \ln T_c / \delta \ln V  =
18$ at 9~kbar\cite{Schulz81}, with the compressibility data measured under pressure (at 16~kbar  $\delta \ln V / \delta  P = 0.7\%$
kbar$^{-1}$ )\cite{Gallois87} in the same compound.  This value  is indeed sizably larger than 7, the value which is obtained in tin,  the elemental superconductor exhibiting 
the strongest sensitivity to pressure\cite{Jennings58}. A look at the  \tmpf phase diagram shows that the strong pressure dependence   of $T_c$ is however restricted to the close vicinity of the border with the SDW phase.

The pressure coefficient  of superconductivity in \tmc\ is even more dramatic since  then $\delta \ln T_c /\delta \ln V  = 36$\cite{Mailly83a} using the compressibility of 1\% kbar$^{-1}$ (this is the value measured for  \tmpf at ambient
pressure \cite{Gallois87}, since to the best of our knowledge compressibility data for \tmc\ are still missing). However, this remarkable sensitivity of $T_c$ in \tmc\ might actually be related to the very specific
problem of anion ordering in this compound as it has been suggested from the recent study on the sensitivity of $T_c$
against the presence of non-magnetic disorder\cite{Joo04}. Anion ordering reveals an uprise of the ordering temperature under
pressure\cite{Creuzet85b,Murata85,Mailly83a}, which can be derived from the pressure dependence of a small kink in the temperature dependence of the
resistivity, the signature of the ordering,  moving from 24 up to 26.5K under 1.5~kbar and corroborated by  studies at even higher pressures\cite{Guo98}. Together with this uprise, there
exists a slowing down in the dynamics of the anions needed for the ordering. Hence, high pressure studies require a special attention to the cooling
rate which must be kept low enough to allow anion ordering  at low temperature. This may  be the explanation for the discrepancy  between
high pressure data showing the signature of anion ordering up to 8~kbar \cite{Murata85}, and  the absence of ordering claimed from the
interpretation of magneto-angular oscillations\cite{Kang93}.

\subsubsection{SDW-SC coexistence}
\begin{figure}[htbp]			
\centerline{\includegraphics[width=0.6\hsize]{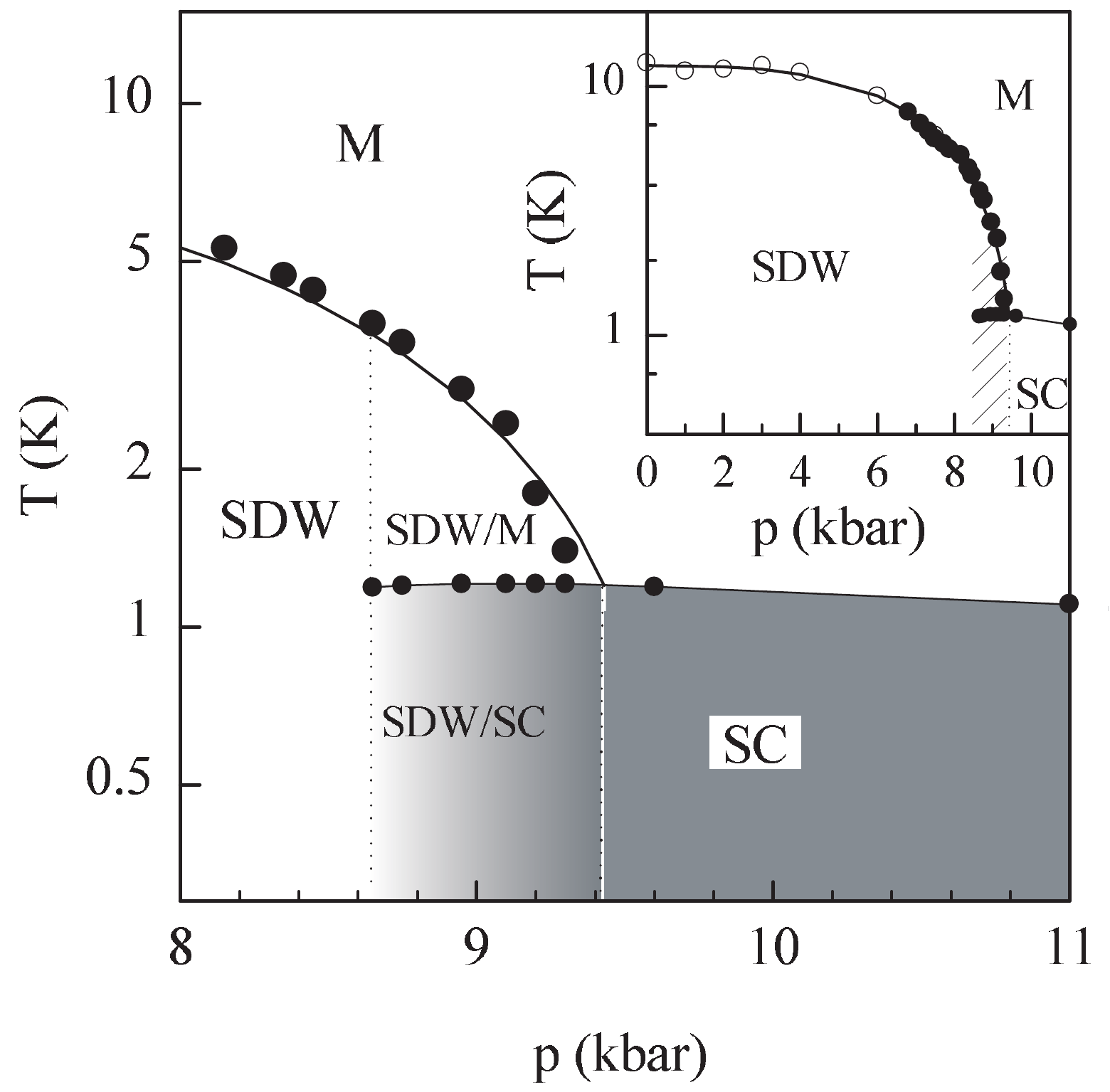}}
\caption{Coexistence between SDW and Superconductivity in \tms in  the  vicinity of the criticial pressure for suppression of the SDW ground
state. After Ref. \cite{Vuletic02}}
\label{tmpf6supraetsdwsousP.eps}
\end{figure}

A situation of non-homogenous superconductivity have been clearly identified near the border between SDW and
Superconductivity in the \tsx\  phase diagram\cite{Vuletic02}. At high pressure $(P>$ 9.4~kbar ), the superconducting phase emerges from a
metallic state and can be reasonably thought of as homogeneous with a critical current density along the $a$-axis $J_c = 200$~A cm$^{-2}$. Below this critical pressure, there exists a pressure domain for \tmpf ($\approx$ 1~kbar  wide) in which a
superconducting signature is observed  at a nearly pressure independent temperature below the onset of a SDW instability, where the
critical current density is greatly reduced, $J_c \approx $ 10~A cm$^{-2}$.  This feature points in favor of a  coexistence of SDW and SC macroscopic domains consisting of metallic (SC) tubes parallel to the $a$ axis (Fig. \ref{tmpf6supraetsdwsousP.eps}).
%%%%%%%%%%%%%%%%%%%%FigureSDW/SC coexistence experimental
%%%%%%%%%%%%%%%%%%%%
The existence of coexisting macroscopic regions of SDW and SC order is also supported by a recent NMR investigation performed at a pressure slightly lower than the critical pressure for the establishment of the homogenous state\cite{Lee05}. This latter study has enabled a quantification of the relative volume fractions in the SDW-metallic regime using the proton NMR linewidth as the local probe.
 A related consequence of the existence of macroscopic insulating  domains in the superconducting phase allowing a channeling of the
lines of force in the material is
the large increase of the upper critical field $H_{c2}$\cite {Lee02}, which had already been reported long time ago
in
\tmpf
 with the second confirmation  of organic superconductivity \cite{Greene80},  and also later in
\tmas
\cite{Brusetti82}. 

%%%%%%%%%%%%Figure Coexistence theory
\begin{figure}[htbp]			
\centerline{\includegraphics[width=0.7\hsize]{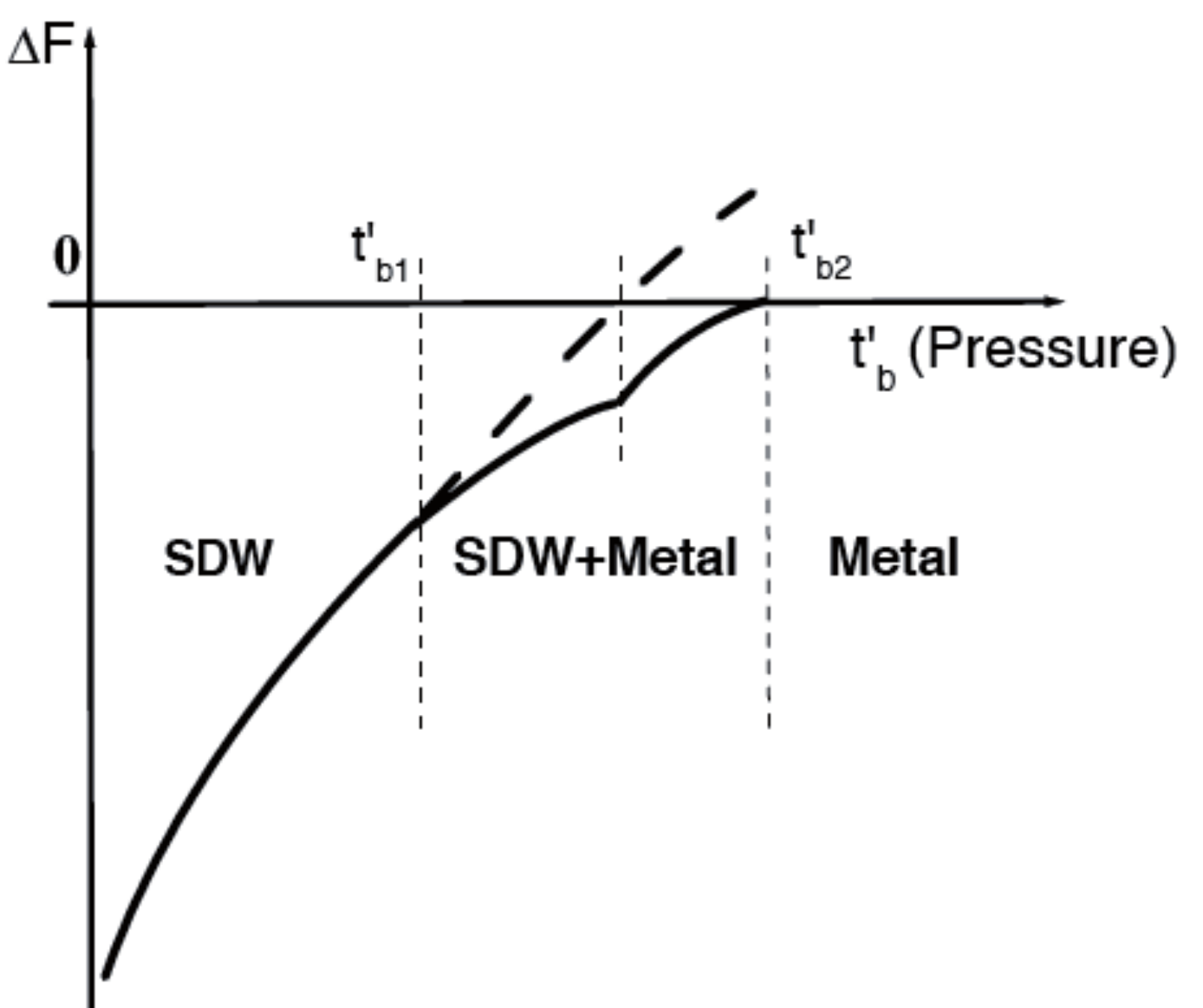}}
\caption{Sketch of the theory showing from the free energy {\it vs } the parameter $t'_b$, the region of coexistence between SDW and Superconductivity in   the  vicinity of the criticial pressure\cite{Vuletic02}. $t_{b}^{\prime} (\equiv t'_{\perp b} $ in the text) parametrizes the deviation from the perfect nesting of the Q-1-D Fermi surface.  After Ref. \cite{Vuletic02}}
\label{Coexistence.eps}
\end{figure}
%%%%%%%%%%%%%%%
It must be kept in mind that the slab formation  in the vicinity of the SDW state with the formation of insulating domains is not related to the penetration of the magnetic field in a type II superconductor but the result from a competition between  insulating and conducting phases. The claim for the existence of tubular domains  $\| a$ is based on an analysis of the resistance in the vicinity of the critical pressure at low temperature  above $T_c$, and of the critical currents in the superconducting state\cite{Vuletic02}. The theoretical approach relies on a variational model leading to an inhomogenous phase with an energy lower than the energy of the homogenous states (Metallic or Insulating SDW). Since it is the transverse $b$ parameter that is expected to govern the respective stability of the SDW and metallic phases at a pressure lower than the critical one but close enough to the homogenous critical line the formation of macroscopic metallic domains with a smaller $b$ is energetically favored in-between $b$ expanded SDW domains \cite{Heritier84}. Similarly, at pressures larger than  the critical pressure, insulating SDW domains should be present between metallic regions (see Fig. \ref{Coexistence.eps}). 
This  interpretation is at variance with a model based on similar experimental data for the critical fields in the coexistence regime in which  the formation of thin superconducting slabs perpendicular to $a$ sandwiched between SDW insulating domains is the result of a self organization process taking advantage of the largest field penetration length perpendicular to the direction of the field\cite{Lee02}.

  An other approach has been taken   by Gorkov and Grigoriev for the interpretation of the SDW/SC coexistence regime in \tmpf \cite{Gorkov05}. This model predicts the existence of  soliton domain walls in the SDW phase close to the critical pressure between the uniformly gapped SDW phase and the SC phase. As these domain walls should be perpendicular to the molecular stacking axis, a strong anomaly of the transport anisotropy could be anticipated related to the formation of conducting slabs perpendicular to the $a$ axis in the coexistence pressure regime. However, no such anomaly has been detected experimentally\cite{Auban06}.

Recent data obtained with  \SePF\ under very high pressure, where the coexistence regime is much broader than for \tmpf, have clearly shown that the critical field for $H \| c^*$
is  enhanced by a factor 10 at the border with the SDW phase. In this pressure domain, it can reach 1T, (Fig. \ref{Hc2TMTTF2PF6.eps}), while it amounts to about 0.1 T  at
very high pressure when the superconducting state is homogenous\cite{Wilhelm04}, a value similar to the observation in the R-state of \tmc. 

 Finally, an experimental study performed in (TMTSF)$_2$ReO$_4$ under pressure has revealed the existence of conducting filaments parallel to $a$ in the pressure domain close to 10 kbar  when metallic and insulating domains coexist at low temperature as a consequence of two coexisting anion orders in this material\cite{Colin06}.

\subsubsection{Superconductivity and non magnetic defects}

It is  the  remarkable sensitivity  of organic
superconductivity to irradiation\cite{Bouffard82,Choi82} that led Abrikosov to suggest the possibility of
triplet pairing in these materials\cite{Abrikosov83}.
Although irradiation was recognized to be an excellent method for the introduction of defects in a
controlled way\cite{Zuppirolli87}, defects thus created can be magnetic \cite{Sanquer85}, and
the suppression of superconductivity by irradiation induced defects  as a signature of
non-conventional pairing must be taken with `a grain of salt'  since local magnetic moments can also  act  as
strong pair-breakers on \textit{s}-wave superconductors. Several routes have been followed to introduce an 
intrinsically non-magnetic perturbation modulating the potential seen by the carriers on the organic stacks.
 Non magnetic disorder has been achieved  substituting \tst\ for \tsm\  
on the cationic stacks of \tmx salts with \pf  \cite{Coulon82} and \cl\  salts  \cite{Johannsen85}. However,  in both
situations cationic alloying induces drastic modifications of the normal state electronic properties
since the SDW transition of \tmpf \, is quickly broadened and pushed towards higher temperature upon
alloying\cite{Mortensen84}. 

\begin{figure}[htbp]			
\centerline{\includegraphics[width=0.65\hsize]{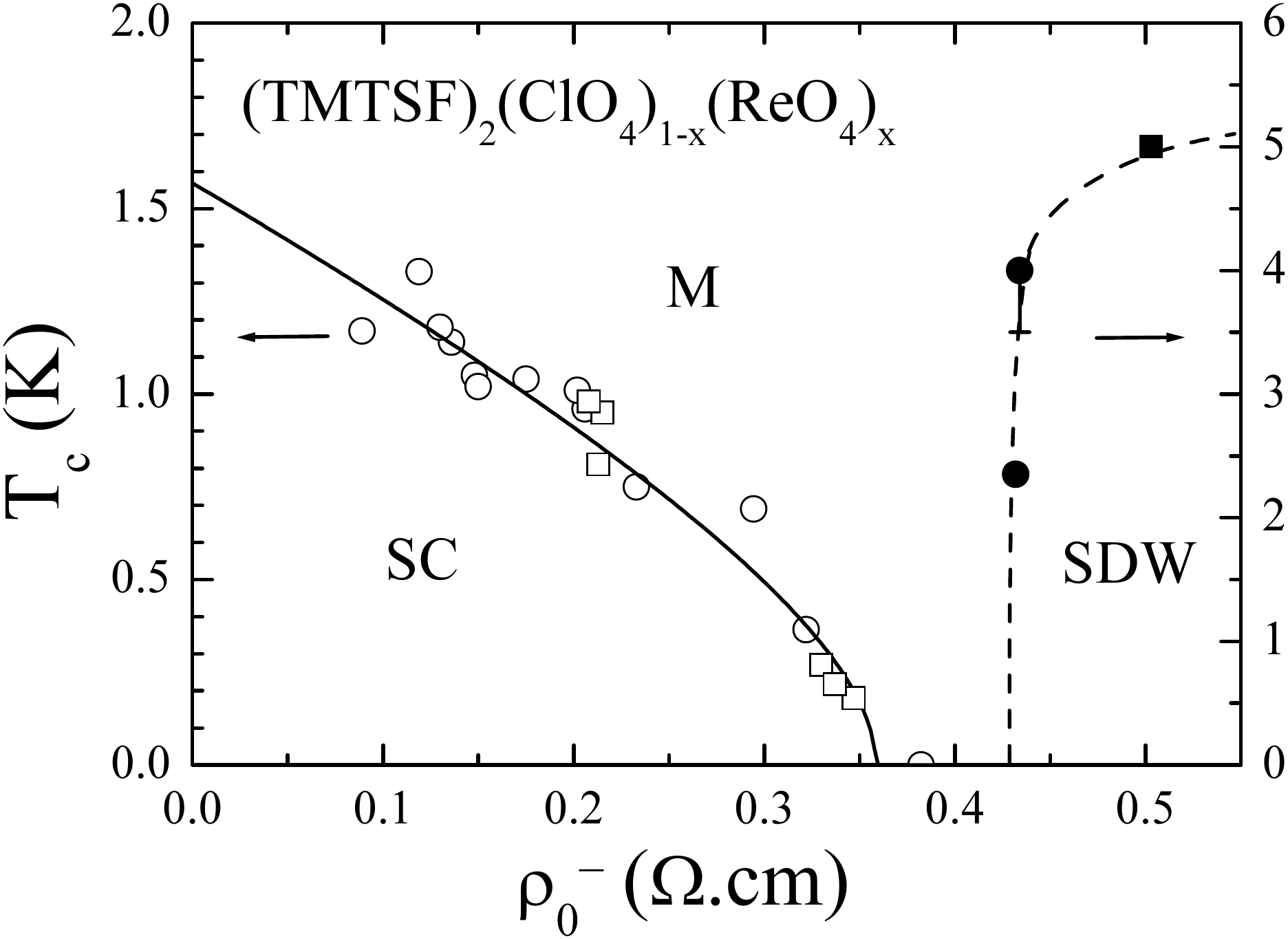}}
\caption{Phase diagram of \tmx \, governed by  non magnetic disorder. All open circles refer to the very slowly cooled samples in the
R-state with different \R contents. Open squares are  data from the same samples corresponding to slightly larger cooling rates although
keeping a metallic behavior above $T_c$. A 10\% sample with $\rho_0^- \approx 0.32 \Omega$.cm  has provided four different $T_c$
depending on the cooling rate. One sample (8\%) did not reveal any ordering down to 0.1K. These data show that the residual
resistivity is a better characterization for the disorder than the nominal \R concentration.  Full dots (15 and 17\%) are relaxed
samples exhibiting a SDW ground state. The vertical bar is the error bar for a sample in which a maximum of the logarithmic derivative
could not be clearly identified and therefore the actual  SDW temperature should lie below 4K, the temperature of minimum
resistivity. The full square is the Q-state of a 6\% sample. The continuous line at the Metal-SC transition  is the  best fit of the data
 with the diGamma function model providing
$T_{c}^0$ = 1.57K. The dashed line at the Metal-SDW transition is only a guide for the eye. After Ref. \cite{Joo05}}
\label{Tcvsrho.eps}
\end{figure}

Leaving the cation stack  uniform, scattering centers can also be created on the anion stacks with the solid
solution  \tmcrx,  where Tomi\'c  \textit{et al.,}  first mentioned the suppression of superconductivity upon alloying with a very
small concentration of \re\  anions\cite{Tomic82,Hamzic88}. 
In the case of a solid solution with tetrahedral anions such as
\cl\ or \re, one is confronted to two potential sources of non-magnetic disorder that  act additively on the
elastic electronic lifetime according to the Mathiessen's law. First the modulation due to the different chemical natures of the anions and second a
disorder due to  a progressive loss of long-range ordering at $T_{\rm AO}$ in the \tmx\, solid solution, although X-ray investigations have revealed that long-range order is preserved up to 3\%  \re\ with a correlation length $\xi_a$  $>$ 200\AA\cite{Ravy86}. Studies of superconductivity in
\tmc\ performed under extremely slow cooling conditions have shown that
$T_c$ is a fast decreasing function of the non-magnetic disorder\cite{Joo04}, where the residual resistivity along the
$c^*$ axis has been used for the measure of the disorder in the  alloys with different concentrations (Fig. \ref{Tcvsrho.eps}). It must be emphasized that the residual resistivity is derived from a fit of the temperature dependence of the normal state resistivity according to a Fermi liquid model below the anion ordering temperature of 24K, namely $\rho(T) = \rho_0^- + AT^2
$. This treatment of the resistivity below 10K or so, allows us to remove the influence of superconducting precursor effects above the ordering temperature.

%%%%%%%%%%%%%%%%%%%%
%%%%%%%%%%%%%%%%%%%%
The suppression of $T_c$
is clearly related to the enhancement of the scattering rate in the solid solution. 
Since the additional scattering cannot be ascribed to magnetic scattering according to the  EPR checks
 showing no additional traces of localized spins in  the solid solution, the data in figure (\ref{Tcvsrho.eps}) 
 cannot be reconciled with the picture of a superconducting gap keeping a constant sign over the whole
$(\pm k_F)$ Fermi surface. They require a picture of pair breaking in a superconductor with an unconventional gap
symmetry. The conventional pair breaking theory for magnetic impurities in usual superconductors has been generalized to
the case of non-magnetic impurities in unconventional materials and the correction to $T_c$ obeys the following relation \cite{Maki04,Larkin65},
\begin{eqnarray}
\ln\Big(\frac{T_{c}^0}{T_{c}}\Big)=\psi\Big(\frac{1}{2}+\frac{\alpha T_{c}^0}{2\pi T_{c}}\Big)-\psi\Big(\frac{1}{2}\Big),
\end{eqnarray}
with $\psi(x)$ being the Digamma function, $\alpha = \hbar /2 \tau k_{B}T_{c}^0$ the depairing parameter, $\tau$
the elastic scattering time 
and $T_{c}^0$ the limit of $T_c$ in the absence of any
scattering. From the data in figure (\ref{Tcvsrho.eps}),  the best fit leads to $T_{c}^0= $1.57K  and  a critical scattering for
the suppression of superconductivity of $1/\tau _{\rm cr} = 1.85$~cm$^{-1}$. Accordingly, $1/\tau$ amounts to 0.56 cm$^{-1}$  in the pristine
\tmc\, sample. Such a value for the inverse carrier  life time is admittedly significantly smaller than the predicted  width at half height namely, $1/\tau \approx 2$ cm$^{-1}$ assuming a classical Drude behaviour involving the  temperature dependence of the DC conductivity and the longitudinal plasma frequency\cite{Jacobsen83}. The present derivation  of the electron life time compares fairly well with far infrared optical measurements leading to a zero frequency conductivity peak   with a  width less than 2-4 cm$^{-1}$\cite{Ng85,Timusk87}. Our results support the existence of a very narrow zero frequency peak carrying a minor fraction of the total spectral weight, which is probably the signature of a correlated low dimensional Fermionic gas.

The sensitivity of $T_c$ to non-magnetic disorder  cannot be reconciled with a model of
conventional superconductors. The gap must show regions of positive and negative signs on the Fermi surface, which can be averaged
out by a finite electron lifetime due to elastic scattering. As  these defects are local, the scattering momentum of order
$2k_F$ can mix $ + k_F $ and $-  k_F$ states and therefore the sensitivity to non-magnetic scattering is still unable to tell the difference
between \textit{p}$_x$ and  \textit{d} orbital symmetries for the superconducting wave function.
A noticeable
progress could be achieved paying attention to the spin part of the wave function. In the close vicinity of
$T_c$, orbital limitation for the critical field is expected to prevail and therefore the analysis of the critical fields close to
$T_c$ does not  imply a triplet pairing \cite{Gorkov85}. When the magnetic field is oriented along the
intermediate $b$-axis violations of the  Pauli limitation have been claimed in  \tmpf  \cite {Lee00} and recently in  \tmc\
superconductors\cite{Oh04}. However, it must be kept in mind that in all these experiments under transverse magnetic
field along the $b^{\prime}$ axis, the electronic structure is profoundly affected by  the application of the field which tends to localize the electrons as shown by
the normal state crossing over from a metallic to an insulating state when investigated with a current along $c^*$\cite{Joo06}. Furthermore it is still unclear
whether the superconducting phase remains uniform under very strong transverse field\cite{Lebed86,Dupuis93}.  

The nature of the superconducting coupling in \tmx conductors is still intensively studied and debated.
The absence of temperature dependence of the $^{13}$C Knight shift  through the critical temperature at a pressure where \tmpf is
superconducting implies  a triplet pairing \cite{Lee02}.  However, the question of sample thermalization during the time of the NMR
experiment has been questioned\cite{noteNMRtriplet} and this result will have to be reconfirmed. The nature of the  coupling in \tmx
superconductors has not yet reached  a consensus. This is due in part  to the lack of unambiguous experimental data for samples
exhibiting superconductivity in the very low temperature region. This is \textit{at variance} with the  singlet coupling found in 2D
organic superconductors with a $T_c$ in the  10K  range as clearly indicated by the Knight shifts measurements in the
superconducting state\cite{Mayaffre95,Kanoda96}. It can be noticed that in spite of the established singlet coupling, the critical fields
$H_{c2}$ of 2D superconductors can also greatly exceed the paramagnetic limit in the parallel geometry\cite{Kamiya02,Zuo00,Jerome05}.

\begin{figure}[t]
\centering
\includegraphics*[width=.7\textwidth]{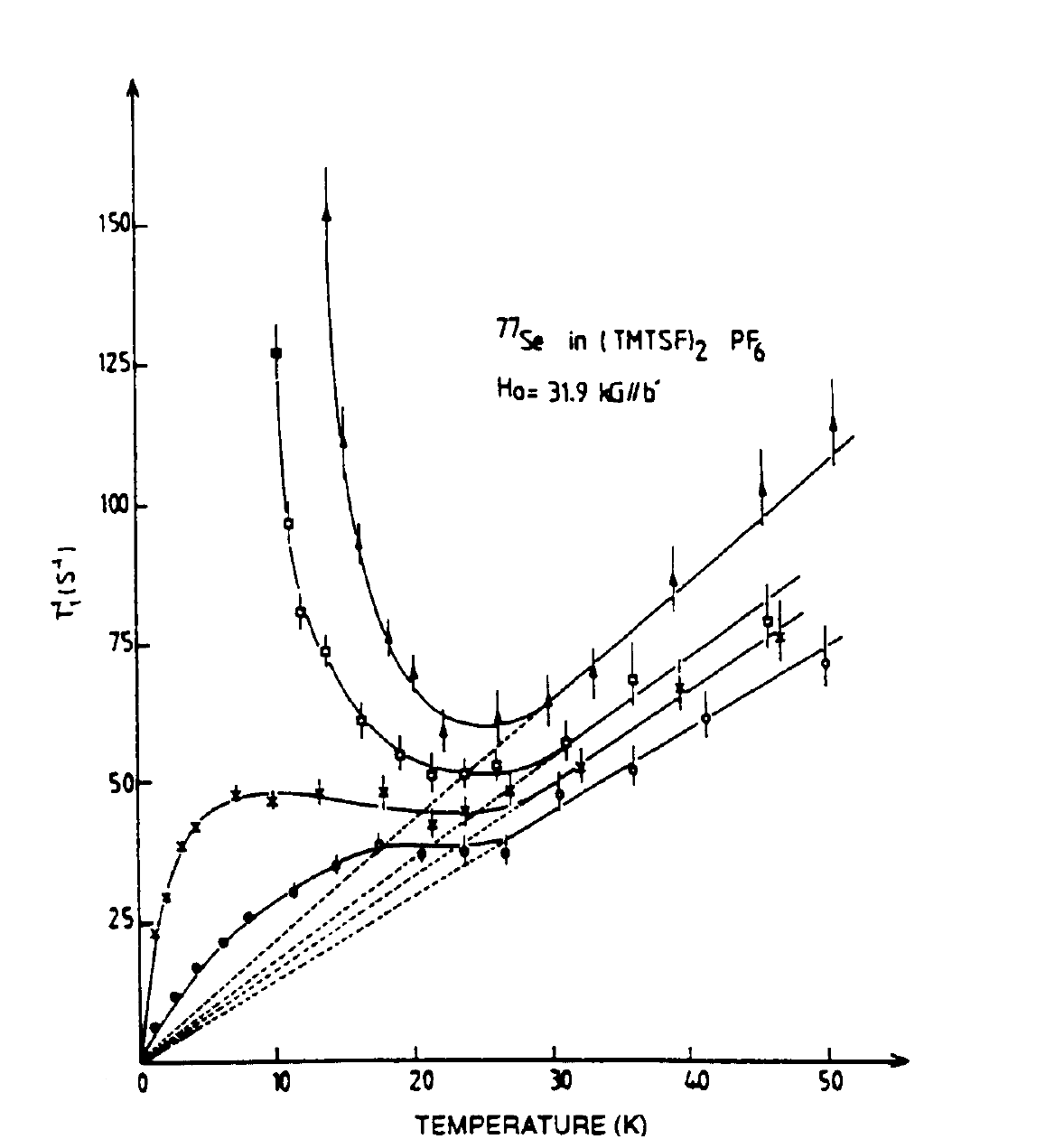}
\caption[]{Temperature dependence of the nuclear spin relaxation rate $T_1^{-1}$ of (TMTSF)$_2$PF$_6$ at  P=1bar (triangles), 5.5kbar (squares), 8kbar (crosses) and 10 kbar (full circles). Deviations with respect to a linear temperature behavior result from antiferromagnetic fluctuations. After Ref.\cite{Creuzet87} }
\label{T1Se}       % Give a unique label
\end{figure}

 \subsubsection{An approach to the mechanism of superconductivity}
 \label{Supra}
Given the experimental uncertainty about the nature of superconductivity in the Bechgaard and Fabre salts, it is therefore still premature to privilege  the triplet scenario for pairing over the    singlet one.  In any case, however, superconductivity   certainly differs from what is commonly seen in ordinary metals and this raises the question  of the  possible causes  of unconventional pairing in these materials \cite{Dupuis06}.  This problem perplexed almost   every one in the field from the start, since the requirements for a traditional  phonon-mediated mechanism for pairing are apparently not met \cite{Emery83}.  Antiferromagnetism that completely surrounds superconductivity in the phase diagram represents the main obstacle for an effective attraction mediated by phonons to  take  place. Superconductivity is indeed  invariably replaced by an SDW instability whether one moves backward on the pressure scale or whether at fixed pressure $P>P_c$, one moves along the magnetic field axis ${\bf H}(\| c^*)$,  where a cascade of field-induced SDW states is  found\cite{Wang93}.  (TMTSF)$_2$ClO$_4$ is another example that illustrates how close \TM\ are to the threshold of a SDW instability at $P>P_c$. This compound presents an anion ordering on slow cooling and is already a superconductor at ambient pressure ($P_c < 1$ bar, Fig. \ref{tm2clo4supra.eps}); it develops an SDW instability  either beyond a critical alloying\cite{Joo05,Tomic82} (Fig. \ref{Tcvsrho.eps}), or by just cranking up the cooling rate\cite{Takahashi82,Tomic82,Schwenk84}. Additional experimental weight supporting underlying coupling conditions for SDW comes from the properties of the normal state  at $P>P_c$.   NMR experiments show indeed a strong enhancement of the spin-lattice relaxation rate $T_1^{-1}$ as a function of temperature, revealing the existence of strong antiferromagnetic spin fluctuations in a very broad temperature domain of the metallic  state above the superconducting $T_c$ (Fig. \ref{T1Se})\cite{Bourbon84,Creuzet87,Wzietek93,Wu05}.  Therefore all this goes to show that even in the presence of superconductivity, the interactions in \TM\ at $P>P_c$ remain repulsive and   favorable to a SDW state,   only  nesting conditions are apparently  changing in each case.

According to the model described in \S \ref{SDW},  these fluctuations are made at the microscopic level of  electron-hole pairs  at $(\mathbf{k}, \mathbf{k}\pm \mathbf{q}_0)$. These will
then coexist with electron-electron (and hole-hole) pairing  at $(\mathbf{k}, -\mathbf{k})$, namely those  responsible for the superconducting instability of the normal state at $P>P_c$. Since these two different pairings refer to the same electronic excitations around the Fermi surface, there will be some intrinsic dynamics or interference between them.     We already came up against the problem of interfering  pairing instabilities in the one-dimensional case (\S\ \ref{theory}). For repulsive interactions and perfect nesting at $2k_F$, we have seen that interference between electron-hole and electron-electron  pairing is maximum for a 1D -- two points -- Fermi surface and enters as a key ingredient in the formation of  either a Luttinger liquid or  a Mott insulating phase at commensurate filling \cite{Bychkov66,Dzyaloshinskii72,Solyom79}. At finite $t_{\perp b}$ and for temperature well below $T^\star $, however,  the outcome differs and may provide a logical link between SDW and  superconductivity.   

The connection between superconductivity and density-wave correlations in isotropic systems goes back to the work of  Kohn and Luttinger in the mid sixties\cite{Kohn65}. They showed that the coupling between electron-hole (density-wave) and electron-electron correlations, albeit very small, is still present for a  spherical  Fermi surface. In  this isotropic limit, $2k_F$ Friedel (charge) fluctuations act as a  oscillating pairing potential for electrons giving rise to a purely electronic mechanism for superconductivity  at large angular momentum.  Emery suggested that this non-phonon mechanism should be  working in the spin sector as well, being boosted by the proximity of a SDW state in the quasi-1D geometry in order to  yield experimentally reachable $T_c$ \cite{Emery86} -- an effect that was early on confirmed in the framework of renormalized mean-field theory\cite{BealMonod86,Caron86,Bourbon88} and various RPA approaches \cite{Scalapino86,Shimahara89,Kino99}. However, these approaches amount to extract  an effective superconducting coupling from short-range density-wave correlations, which in turn serves as   the input  of a ladder diagrammatic summation. It turns out that the ladder theory, as  a single channel approximation, neglects the quantum interference between the different kinds of pairings, and as such it cannot capture the dynamical emergence of superconductivity.

Because of the finite value of $t_{\perp b}$, interference becomes non uniform along the Fermi surface. This introduces a momentum dependence in the scattering amplitudes, which  can be parametrized by  the set of transverse wave vectors  for in going $(k_{\perp1}k_{\perp2})$ and outgoing $(k'_{\perp1}k'_{\perp2})$ electrons (here $k_\perp\equiv k_{\perp b}$). The generalization of the 1D scaling equations Eqs. (\ref{flow1D}) to now $k_{\perp}$-dependent interactions  $g_{i=1,2,3}(k'_{\perp1}k'_{\perp2}k_{\perp2}k_{\perp1})$ in  the quasi-1D case, where both $t_{\perp b}$ and $t'_{\perp b}$ are present,  has been worked out recently \cite{Nickel05,Nickel06,Duprat01,Bourbon04}. The results can be put in the following schematic form:
\begin{eqnarray}
\label{DLG}
\partial_\ell g_i(k'_{\perp1}k'_{\perp2}k_{\perp2}k_{\perp1}) && = \sum_{k_\perp}\sum_{n,n'=1}^3\Big\{\epsilon^{nn'}_{C,i}\,g_n(\{k_\perp\})\,g_{n'}(\{k_\perp\}){\cal L}'_C(k_\perp, q_{C\perp}) \cr 
&& +\, \epsilon^{nn'}_{P,i}\,g_n(\{k_\perp\})\,g_{n'}(\{k_\perp\}){\cal L}'_P(k_\perp, q_{P\perp},t'_{\perp b}) \Big\},
\end{eqnarray} 
Here ${\cal L}'_{C,P}=\partial_\ell{\cal L}_{C,P} $ where  ${\cal L}_{C,P} $ are the Cooper (electron-electron)  and Peierls (electron-hole) loops, with $q_{C,P\perp}$ as their respective $\{k_\perp\}$-dependent transverse momentum variables, and $\epsilon^{nn'}_{C,P,i}=\pm 1$, or 0.  By integrating  these flow equations, the singularities shown by interactions signal instabilities  of the normal state at a critical temperature $T_c$. The nature  of ordering is determined  by the profile of interactions in $\{k_\perp\}$ space, which in turn  corresponds to a divergence of a given order parameter susceptibility $\chi_\mu$.      Feeding these equations with a realistic set of bare parameters for the  repulsive intrachain interactions $g_i$ and the band parameters $t_{\perp b}$ and $E_F$ in  \TM, it is possible to follow the instabilities of the normal state as a function of the nesting deviations parameter  $t'_{\perp b}$, which simulates the main influence  of pressure in the model\cite{Nickel05,Nickel06,Duprat01,Bourbon04,Fuseya05}.

%%%%%%
\begin{figure}[t]
\centering
\includegraphics*[width=.6\textwidth]{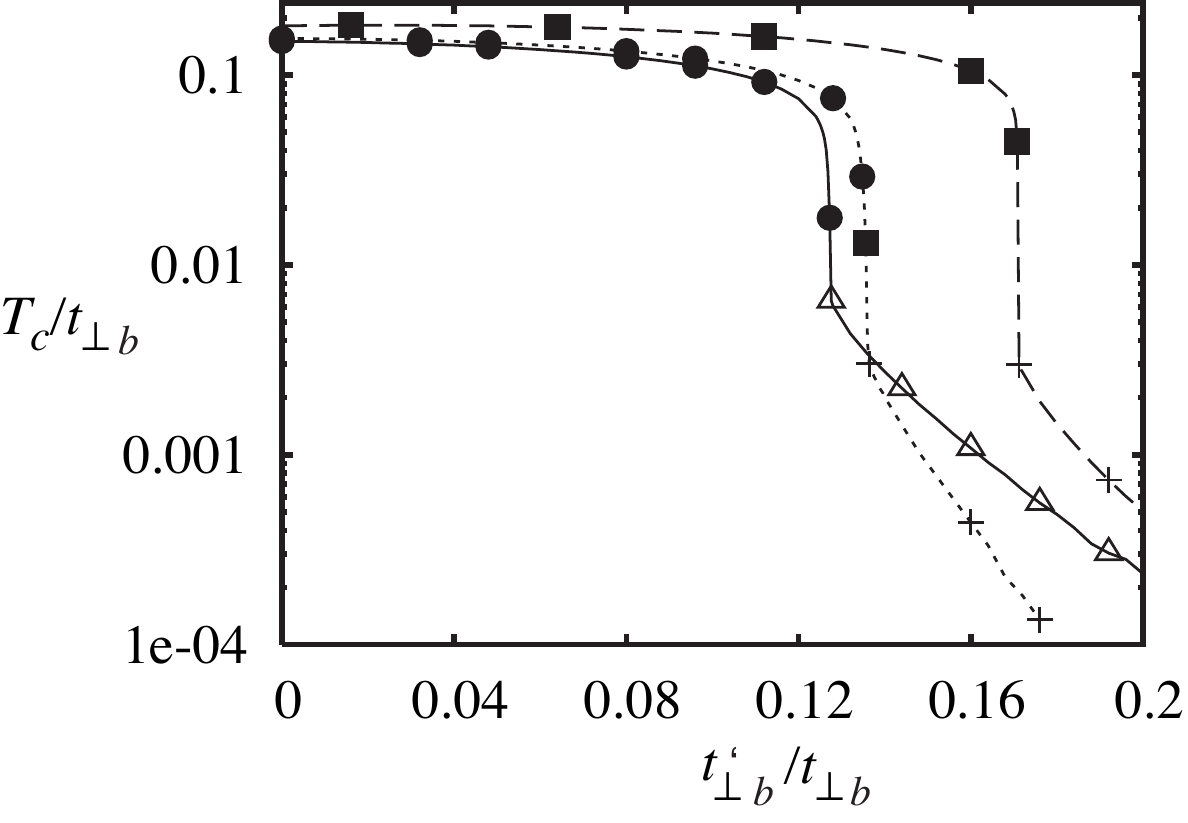}
\caption[]{Calculated phase transition temperature of  the quasi-one-dimensional electron gas model as a function of the nesting deviation parameter $t'_{\perp b}$ for repulsive intrachain interactions $g_{1,2,3}$. Continuous line ($\tilde{g}^\perp_{1}=0$), dotted line ($\tilde{g}^\perp_{1}=0.11$) and dashed line ($\tilde{g}^\perp_{1}=0.14.$). After Nickel {\it et al.,} Ref. \cite{Nickel05} }
\label{Tc}      
\end{figure}
%%%%%%%

Thus at perfect nesting, when $t'_{\perp b}=0$, the normal state develops an SDW instability at $T^0_c\sim 20$K, which for small Umklapp scattering corresponds to the range of  $T_c$ expected in most of \Bech\ at ambient  pressure and  \Fabre\ at relatively high pressure. The range  of $T_c$ roughly squares with the one obtained in the single channel approximation with no interference below $T_{x^1}$ (\S \ref{SDW}). As $t'_{\perp b}$ increases, $T_c$ is gradually decreasing until  the critical range  $t'^{\rm cr}_{\perp b}\simeq 0.8 T_c^0$ is reached where the SDW is suppressed (full circles, Fig.\ \ref{Tc}). The metallic phase remains unstable at finite temperature, however, but the instability now takes place in the superconducting  channel (open triangles,  Fig.\ \ref{Tc}). The  order parameter is of the form $\Delta(k_\perp) = \Delta \cos k_\perp$ and has nodes at $k_\perp \pm \pi/2$; it corresponds to a interstack singlet or $d_{x^2-y^2}$-wave pairing. Therefore for  repulsive intrachain interactions,  an attraction between electrons can be dynamically generated from the interference between Cooper and Peierls scattering channels. The attraction between carriers on neighboring  chains  can be seen as being mediated by   spin fluctuations. The fact that SC{\it d} and  SDW instability lines meet at the maximum of the superconducting $T_c\sim 1$K  and that the ratio $T_c^0({\rm SC}{\it d})/T_c^0({\rm SDW}) \sim 1/20$, together with  their respective $t'_{\perp b} $ dependence are worth noticing features in regard to the  experimental phase diagram (Fig. \ref{fig:1}).        

Regarding the possible symmetries of the superconducting order parameter, an analysis of the momentum dependence of the scattering amplitude $g_i(\{k_\perp\})$ reveals that for the electron gas model defined with only  {\it intra}chain repulsive interactions, the strongest superconducting instability is invariably found in the singlet SC{\it d}-wave channel\cite{Duprat01,Fuseya05}. Triplet superconductivity in the {\it p}$_x$ channel, which has been proposed on phenomenological grounds as a possible candidate to describe superconductivity in the Bechgaard salts \cite{Lebed00}, is strongly suppressed. In effect, the triplet SC{\it p}$_x $ superconductivity, which has a gap order parameter $\Delta_r = r \Delta$ with $r={\rm sign }\,k_x$,  is an intrachain pairing that is  subjected to the strongest repulsive part of the oscillating potential produced by SDW correlations \cite{BealMonod86}. More favorable conditions for triplet pairing do exist but they take place at higher angular momentum, in the interchain {\it f}-wave channel  with a order parameter $\Delta_r(k_\perp)= r\Delta \cos k_\perp$, a possibility that was shown to come  out from the  mean-field analysis\cite{Kuroki01}. However, for  intrachain repulsive couplings alone the renormalization group analysis show that the amplitude of  triplet correlations are always subordinate to those of the SC{\it d}-wave channel,  which yields the highest superconducting $T_c$ \cite{Nickel05,Fuseya05}. 

\begin{figure}[t]
\centering
\includegraphics*[width=.6\textwidth]{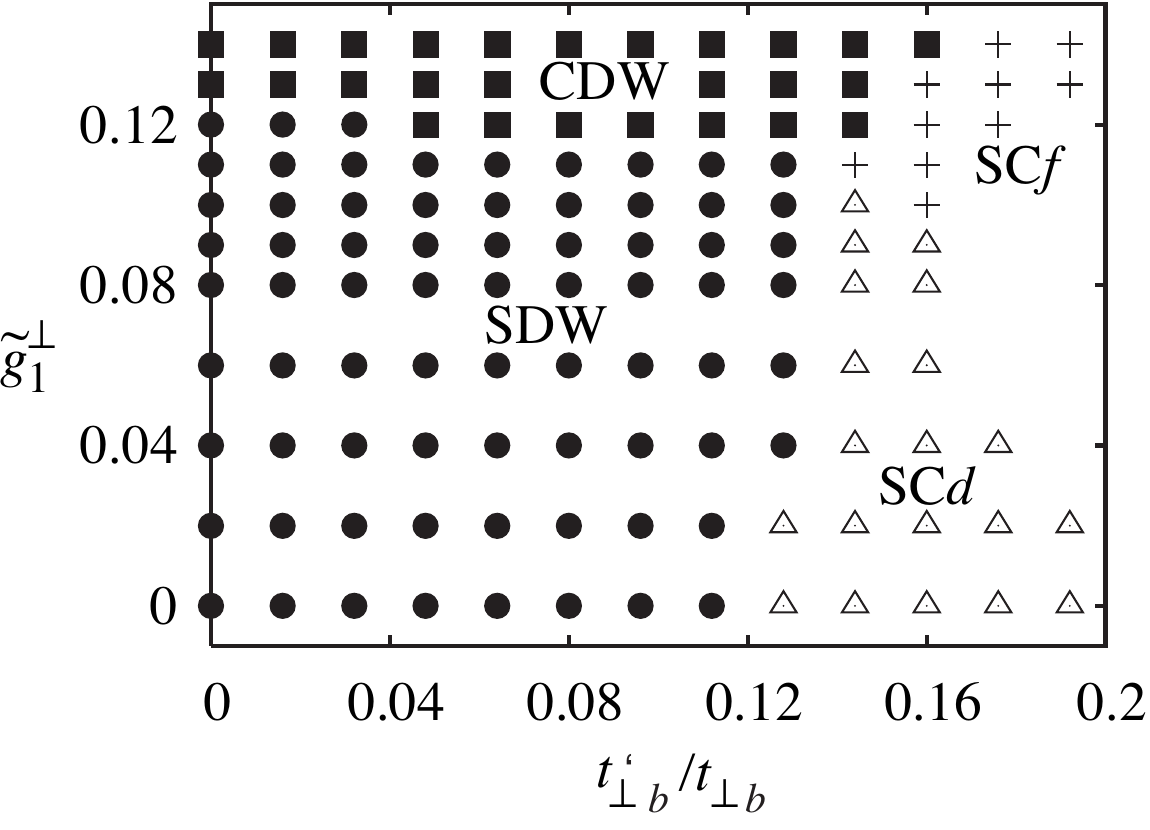}
\caption[]{Calculated phase diagram of the extended quasi-one-dimensional electron gas model for repulsive couplings. After Nickel {\it et al.,} Ref. \cite{Nickel05} }
\label{PhaseT}       % Give a unique label
\end{figure}

Following the Kohn-Luttinger picture, triplet pairing at high angular momentum is actually  connected to charge-density-wave fluctuations\cite{Kohn65}. The presence of a CDW superstructure that coexists with a SDW state in the Bechgaard salts (see \S\ \ref{CDW}) has in this respect stressed their importance in these salts close to $P_c$. Following the example of most quasi-one-dimensional systems in which  a CDW superstructure is found\cite{Pouget89,Barisic85b},  {\it inter}chain Coulomb interaction  is a physically relevant  coupling  that must be taken into account in the presence of charge correlations. By including, besides the $g_i$, {\it inter}chain backward ($g_1^\perp$), forward ($g_2^\perp$) and  Umklapp ($g_3^\perp$) scattering amplitudes, one defines the so-called  extended electron gas model\cite{Gorkov74,Lee77}. For realistic repulsive couplings, this model allows us to expand  the range of  possibilities of both superconducting and density-wave long-range orders. The  RG solution of Eqs.\ (\ref{DLG}) in the  $T-t'_{\perp b}$ phase diagram shows that  $g_1^\perp$ plays a key role on the one hand, in  the stability of  SDW and SC{\it d} orders, and on the other hand in  the emergence of triplet superconductivity and CDW order  \cite{Nickel05,Nickel06}. The RG  results depicted in Fig. \ref{PhaseT} indeed show  that beyond a relatively small threshold in  $g_1^\perp$,   SC{\it d} long-range order is no longer stable and  the instability of the normal state is rather found  in the interchain triplet $f-$wave channel above a critical $t'_{\perp b}$. The SC{\it f} $T_c$ are  comparable to SC{\it d} but show stronger `pressure' coefficient along the  $t'_{\perp b}$  axis (Fig. \ref{Tc}). The results also show that the strength of SDW correlations remain essentially unaffected and can still dominate the normal phase in addition to CDW correlations, whose amplitude grows with the strength interchain coupling. Although  SDW order  can  precede the triplet SC{\it f} instability along the $t'_{\perp b}$ `pressure' scale near the interchain coupling threshold, it is very close in stability with a CDW superstructure (Figs. \ref{Tc} and \ref{PhaseT}). 

Given the observation of a close proximity between SDW and CDW in the phase diagram of the  Bechgaard salts, it follows that not only singlet SC{\it d} but also SC{\it f} are serious candidates for the description of superconductivity in these materials. Since many parameters of the model are likely to change under pressure besides $t_{\perp b}$, this puts some haziness about how one actually moves in the phase diagram of Fig. \ref{PhaseT}, and in turn on the most stable type of superconductivity in these compounds.  

\section{Conclusion and outlook}

In this brief account we went through the main  physical properties of the Bechgaard and Fabre salts series of organic superconductors.   The global phase  diagram that was gradually built over the years around these two series of compounds under either hydrostatic or  chemical pressure stands out as a model of unity for the physics of low dimensional correlated systems. Much effort went to  explain  the  multifaceted phase diagram of \TM\ as a whole, an attempt that also proved to be an active quest of unity for the theory.

In this respect, while  the theoretical description of the  spin-Peierls and antiferromagnetic instabilities in the phase diagram of \TM \ does not meet any serious conceptual difficulty, their strong competition  at the putative quantum critical point on the pressure scale  is to our knowledge without precedent in the field of low dimensional compounds. Although the quantum criticality that is behind this competition would certainly gain to be further clarified on experimental ground,  its comprehension clearly challenges the traditional framework of critical phenomena and fosters some new conceptual focus in unifying antiferromagnetism and a lattice distorted spin liquid phase.   In the last few years, the phenomenon of charge ordering has also played an important part in improving and even expanding the structure of the  phase diagram of \TM.  Though its observation has been so far restricted to members of the Fabre salts series, this phenomenon  raises important  questions about the influence of charge disproportionation  in  the relative stability of the spin-Peierls and N\'eel states in this  series of compounds.

The interplay of different types of commensurability in weakly dimerized  quarter-filled compounds like the \TM, is  another issue that is at the heart of  a better understanding of  strong electronic correlations that characterize the properties normal phase in \TM.  This problem is linked to the persistent issue of the dimensionality crossover or  about how the restoration of  a Fermi liquid is achieved in quasi-one-dimensional conductors like the \TM.

As the  starting point  of the study of the Bechgaard salts more than twenty-five years ago, superconductivity is certainly  one of the hardest part of  the phase diagram to both explain and  characterize.    In spite of  the recent experimental advances, which again  confirm  the non conventional nature of superconductivity in \TM,  the  problem of the symmetry of the superconducting order parameter, as well as the  issue of the presence and location of nodes for the gap, are all enduring questions for which  a consensus of views has   yet to be reached. Given the    experimental constraints and difficulties tied to the use of extremely low temperature and high  pressure conditions  in  \TM, these questions will certainly continue to consume major experimental efforts in the next few years. 

The  mechanism of organic superconductivity in quasi-one-dimensional molecular crystals is  a related  key issue in want of a satisfactory explanation. The extensive experimental evidence  in favor of  the  systematic emergence of superconductivity in \TM \ just below their  stability threshold for antiferromagnetism  has shown the need for a unified description of electronic excitations that at the core of  both density-wave and superconducting correlations. In this matter, the  recent progress achieved by the renormalization group method have resulted  in definite predictions about the possible symmetries of the superconducting order parameter when a purely electronic mechanism is involved -- predictions that often differ from  phenomenologically based   approaches to superconductivity.   The results for the electron  gas model, albeit appealing when confronted  to existing data,  remain only  indicative, however,  of what may be the actual origin of superconductivity in these complex materials. In this respect, the future progress on the experimental side
 will  be certainly decisive for the theory.

\subsubsection{Acknowledgments}
A large part of the results presented in this review are  based on a long term activity in the domain of low dimensional organic superconductors with a large number of contributors at Orsay and Sherbrooke. We acknowledge in particular the recent contributions of P. Auban-Senzier, J. C.  Nickel, R. Duprat and N. Dupuis. This extended activity would not have been possible without the fruitful cooperation of our colleagues in Chemistry J. M. Fabre (Montpellier), K. Bechgaard (Copenhagen), and P. Batail (Angers).

%
% BibTeX users please use
% \bibliographystyle{}
% \bibliography{}
%
% Non-BibTeX users please follow the syntax
% the syntax of "referenc.tex" for your own citations

\bibliographystyle{prsty}
\bibliography{articlesII}

%\input{referenc}
%%%%%%%%%%%%%%%%%%%%%%%%%%%%%%%%%%%%%%%%%%%%%%%%%%%%%%%%%%%%%%%%%%%%%%  }

%%%%%%%%%%%%%%%%%%%%%%%%%%%%%%%%%%%%%%%%%%%%%%%%%%%%%%%%%%%%%%%%%%%%%%

%\printindex
\end{document}